\documentclass[sigconf,10pt]{acmart}

\usepackage{amsmath}
\usepackage{multirow, makecell}
\usepackage{pifont}
\usepackage{subfig}
\usepackage{graphicx}
\usepackage{colortbl}
\usepackage{balance}
\usepackage{xcolor}
\usepackage[space]{grffile}
\usepackage{lscape}
\usepackage{caption}
\usepackage{commath}
\usepackage{tablefootnote}
\usepackage{tikz}

\AtBeginDocument{%
  }

\newcommand{\ourmethod}{\textit{DALTON}}

\setcitestyle{sort&compress}
\setcopyright{none}
\renewcommand\footnotetextcopyrightpermission[1]{}

\usepackage{tcolorbox}

\tcbuselibrary{theorems}
\newtcbtheorem
  []% init options
  {takeaway}% name
  {Takeaway}% title
  {%
    colback=white,
    colframe=black!90,
    fonttitle=\bfseries,
  }% options
  {tay}% prefix

\begin{document}
%\title{Unveiling Indoor Well-being: An In-the-Wild Study on Indoor Air Pollution in a Developing Nation}
%\title{Unmasking Indoor Health: An In-the-Wild Study on Indoor Air Pollution in a Developing Nation}
\title{Exploring Indoor Health: An In-depth Field Study on the Indoor Air Quality Dynamics}
%\title{\ourmethod: An IoT platform to rank Indoors by pollution exposure}
% \title{\ourmethod{}: Towards Distributed Indoor Air Quality Monitoring}

%Authors
\author{Prasenjit Karmakar}
\email{prasenjitkarmakar52282@gmail.com}
\affiliation{%
  \institution{IIT Kharagpur}
  \country{India}
}

\author{Swadhin Pradhan}
\email{swadhinjeet88@gmail.com}
\affiliation{%
	\institution{Cisco Meraki}
	\country{USA}
}

\author{Sandip Chakraborty}
\email{sandipchkraborty@gmail.com}
\affiliation{%
  \institution{IIT Kharagpur}
%  \city{kgp}
%  \state{WB}
  \country{India}
}

\renewcommand{\shortauthors}{Trovato et al.}

%Abstract
\begin{abstract}

    Indoor air pollution, a significant driver of respiratory and cardiovascular diseases, claims 3.2 million lives yearly, according to the World Health Organization, highlighting the pressing need to address this global crisis. In contrast to unconstrained outdoor environments, room structures, floor plans, ventilation systems, and occupant activities all impact the accumulation and spread of pollutants. Yet, comprehensive in-the-wild empirical studies exploring these unique indoor air pollution patterns and scope are lacking. To address this, we conducted a three-month-long field study involving over 28 indoor spaces to delve into the complexities of indoor air pollution. Our study was conducted using our custom-built DALTON air quality sensor and monitoring system, an innovative IoT air quality monitoring solution that considers cost, sensor type, accuracy, network connectivity, power, and usability. Our study also revealed that conventional measures, such as the Indoor Air Quality Index (IAQI), don't fully capture complex indoor air quality dynamics. Hence, we proposed the Healthy Home Index (HHI), a new metric considering the context and household activities, offering a more comprehensive understanding of indoor air quality. Our findings suggest that HHI provides a more accurate air quality assessment, underscoring the potential for wide-scale deployment of our indoor air quality monitoring platform.
    
\end{abstract}

%\begin{CCSXML}
%<ccs2012>
%  <concept_id>10010520.10010553.10010562</concept_id>
%  <concept_desc>Computer systems organization~Embedded systems</concept_desc>
%  <concept_significance>500</concept_significance>
%  <concept_id>10010520.10010575.10010755</concept_id>
%  <concept_desc>Computer systems organization~Redundancy</concept_desc>
%  <concept_significance>300</concept_significance>
%  <concept_id>10010520.10010553.10010554</concept_id>
%  <concept_desc>Computer systems organization~Robotics</concept_desc>
%  <concept_significance>100</concept_significance>
%  <concept_id>10003033.10003083.10003095</concept_id>
%  <concept_desc>Networks~Network reliability</concept_desc>
%  <concept_significance>100</concept_significance>
%</ccs2012>
%\end{CCSXML}
%
%\ccsdesc[500]{Computer systems organization~Embedded systems}
%\keywords{datasets, neural networks, indoor air quality, sensor networks}

\maketitle

%Sections
\section{Introduction}
\label{sec:intro}
%In the post-COVID era, the world has witnessed a shift towards online education, doorstep grocery delivery services, hybrid workplace model, and work-from-home culture~\cite{}, which have become the prevailing norms these days. Consequently, both children and adults are devoting significant amount of time~\cite{} doing their day-to-day activities from indoors. Studies~\cite{} have found that unlike outdoors, indoors are susceptible to accumulation of pollutants due to poor ventilation and often more seriously polluted~\cite{} in even the largest and most industrialized cities, leading to long-term pollution exposure to its inhabitants. For example, carbon-dioxide ($\geq$ 2000 ppm) alone causes headaches~\cite{}, lack of concentration~\cite{}, and increased heart-rates~\cite{} etc. Yet, long-term exposure to pollutants such as volatile organic compounds (VOCs), nitrogen oxides (NO\textsubscript{x}), carbon oxides (CO\textsubscript{x}), particulate matters (PM\textsubscript{x}), etc., can be very detrimental to our physical~\cite{} as well as mental~\cite{} well being. Recent research~\cite{} indicates that we spend almost 90\% of our time indoors; thus ensuring healthy indoor air quality is essential for the sustainable development of our society.

Amid the transformative post-COVID era, where online education, doorstep grocery delivery services, and remote work have become the prevailing norms, we must pay attention to the air we breathe in our indoor spaces. Extensive studies have revealed that indoor environments, due to inadequate ventilation, are susceptible to the accumulation of pollutants, even in the largest and most industrialized cities~\cite{hsu2020smell,adhikary2020we,sakhnini2018mycitymeter,wu2020sharing,patel2022samachar}. This long-term exposure to pollutants such as carbon dioxide, volatile organic compounds (VOCs), nitrogen oxides (NO\textsubscript{x}), carbon oxides (CO\textsubscript{x}), and particulate matters (PM\textsubscript{x}) poses significant health risks. For instance, elevated carbon dioxide levels ($\geq$ 2000 Parts per million (ppm)) can lead to symptoms like headaches, decreased concentration, and increased heart rates. As we spend approximately 90\% of our time indoors~\cite{epaspend,klepeis2001national}, ensuring healthy indoor air quality becomes crucial to create an environment supporting our physical well-being and promoting mental clarity and happiness~\cite{kim2013inair,kim2020awareness,zhong2021complexity,moore2018managing}. 

\begin{figure}
	\captionsetup[subfigure]{}
	\begin{center}
		\subfloat[Household-1 (H1)\label{fig:hm_house_marked}]{
			\includegraphics[width=0.48\columnwidth,keepaspectratio]{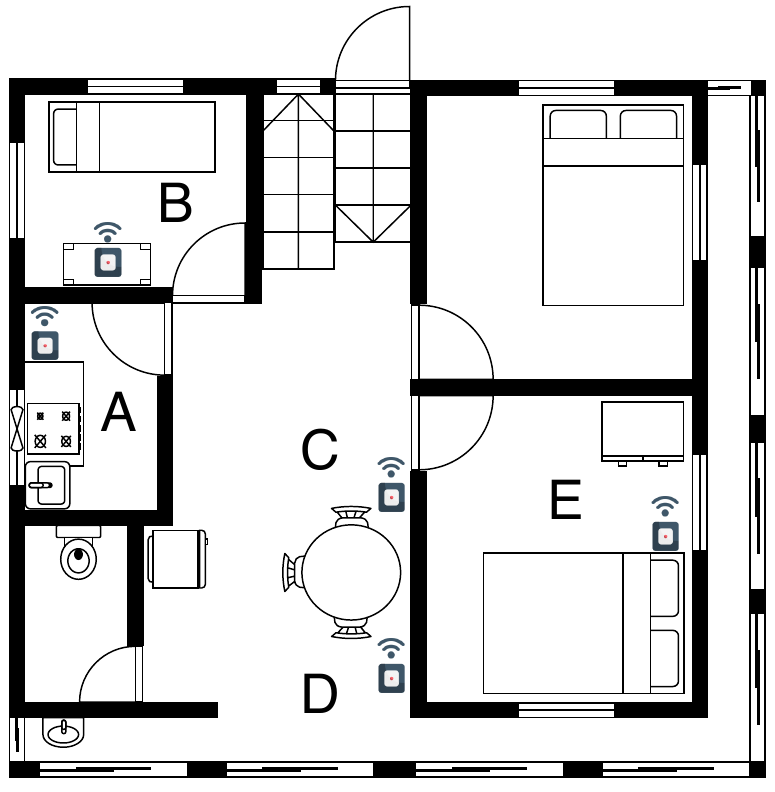}
		}
		\subfloat[VOC Distribution at H1 \label{fig:cross_corr_hm}]{
			\includegraphics[width=0.48\columnwidth,keepaspectratio]{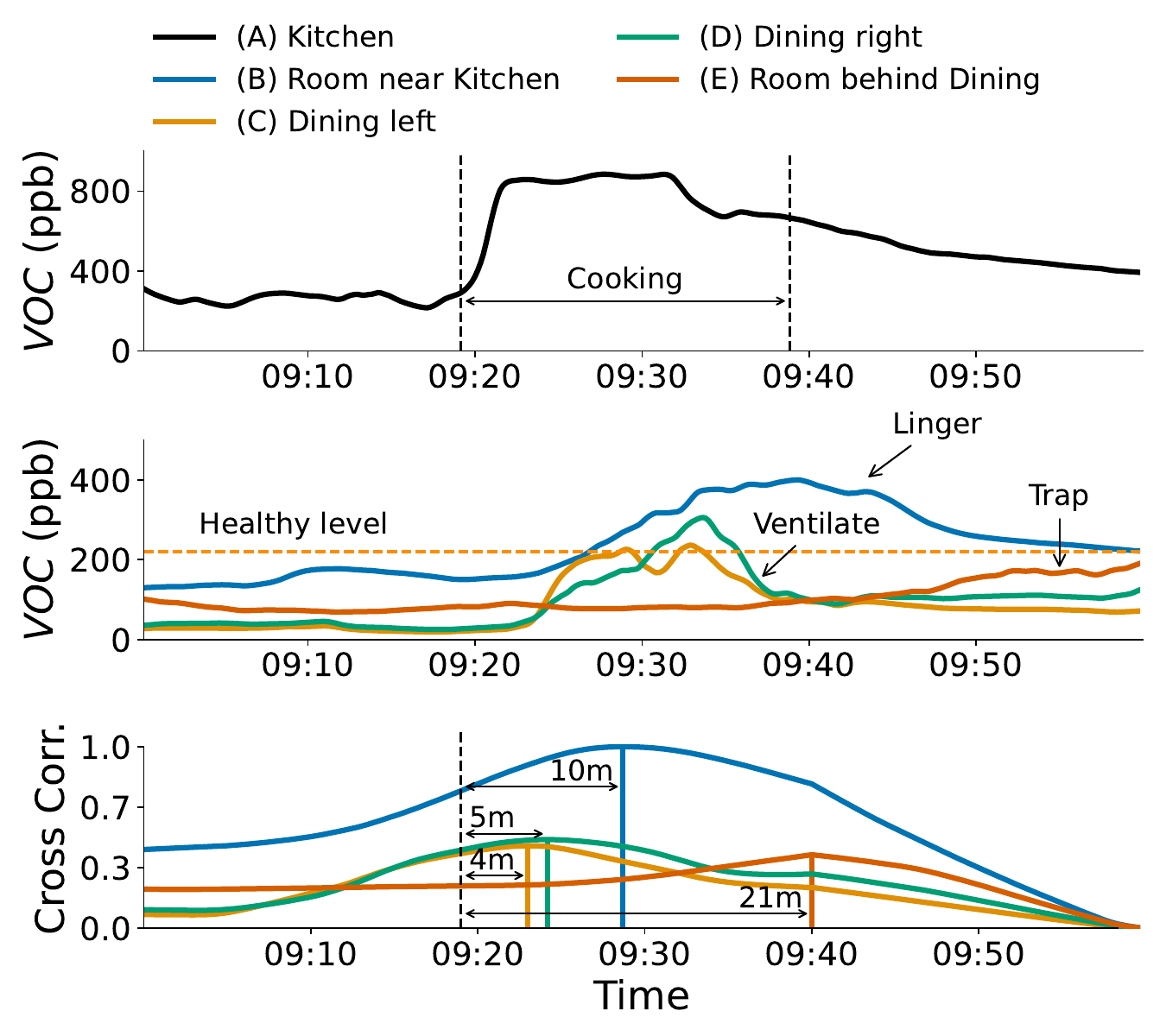}
		}
	\end{center}
	\caption{Pattern and Cross Correlation of VOCs at H1.}
	\label{fig:into_plot1}
\end{figure}

\begin{comment}
\begin{figure*}
	\centering
		\subfloat[Household-1 (H1)\label{fig:hm_house_marked}]{
		\includegraphics[width=0.25\textwidth]{Figures/Diagrams/PK_marked.pdf}
		}
            \hfill 
            \subfloat[Household-2 (H2)\label{fig:sc_house_marked}]{
		\includegraphics[width=0.32\textwidth]{Figures/Diagrams/SC_marked.pdf}
		}
            \hfill 
            \subfloat[Household-3 (H3)\label{fig:gb_house_marked}]{
		\includegraphics[width=0.28\textwidth]{Figures/Diagrams/GB_marked.pdf}
		}\
            \subfloat[VOC Distribution at H1\label{fig:cross_corr_hm}]{
			\includegraphics[width=0.28\textwidth]{Figures/Motiv/cross_corr_hm_house.pdf}
		}
            \hfill 
            \subfloat[VOC Distribution at H2\label{fig:cross_corr_sc}]{
			\includegraphics[width=0.28\textwidth]{Figures/Motiv/cross_corr_sc_house.pdf}
		}
            \hfill 
            \subfloat[VOC Distribution at H3\label{fig:cross_corr_gb}]{
			\includegraphics[width=0.28\textwidth]{Figures/Motiv/cross_corr_gb_house.pdf}
		}
	\caption{Cross Correlation of VOC with Kitchen}
	\label{fig:into_plot}
\end{figure*}
\end{comment}

\begin{figure}
	\captionsetup[subfigure]{}
	\begin{center}
		\subfloat[Household-2 (H2)\label{fig:sc_house_marked}]{
			\includegraphics[width=0.46\columnwidth,keepaspectratio]{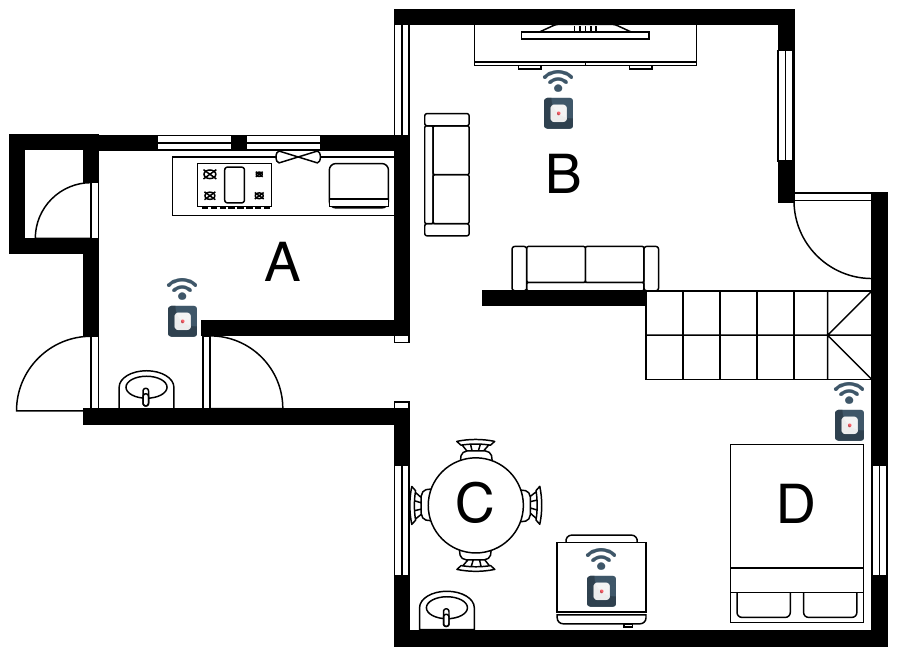}
		}
		\subfloat[VOC Distribution at H2 \label{fig:cross_corr_sc}]{
			\includegraphics[width=0.46\columnwidth,keepaspectratio]{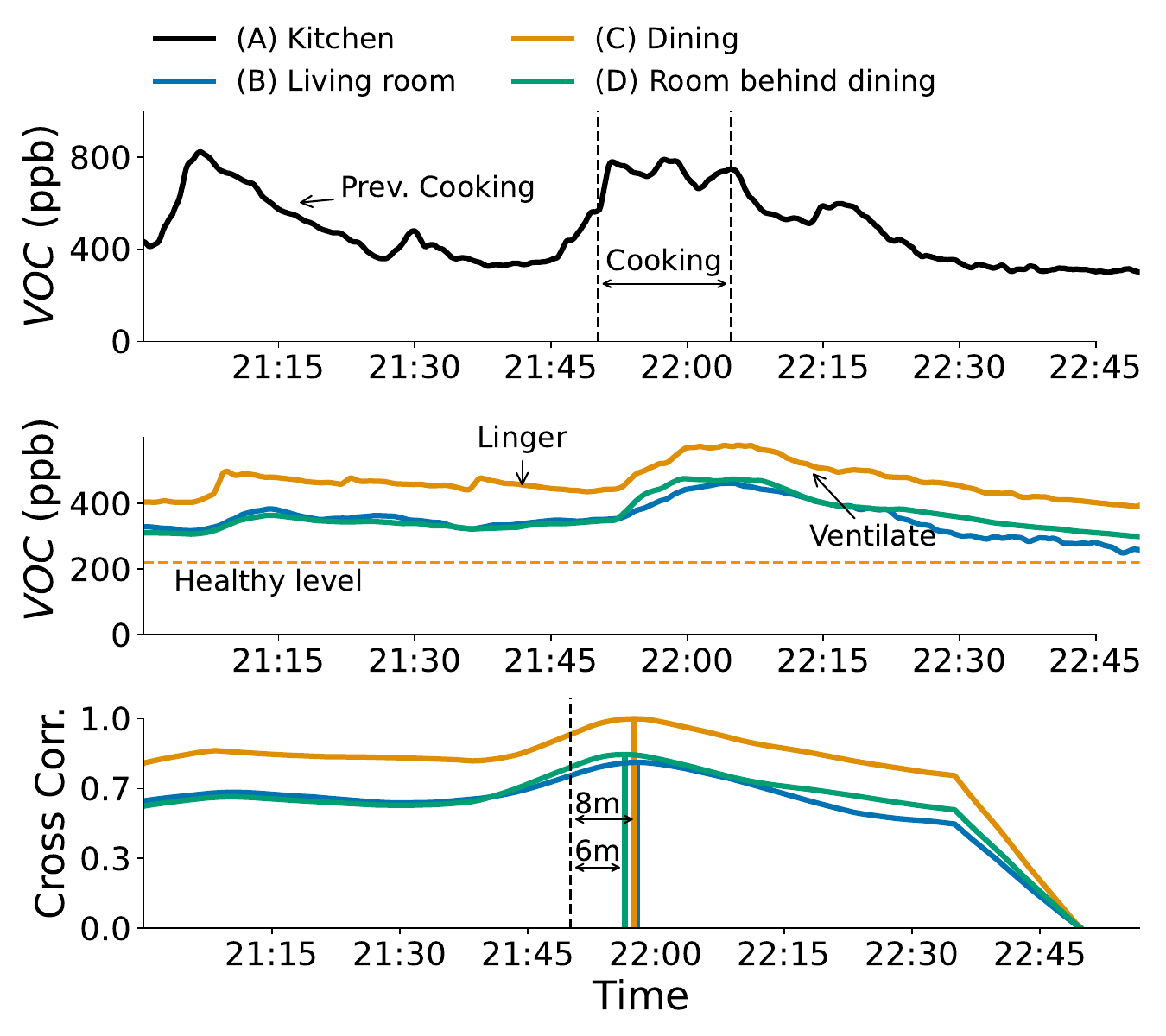}
		}
	\end{center}
	\caption{Pattern and Cross Correlation of VOCs at H2.}
	\label{fig:into_plot2}
\end{figure}

Unlike the air quality monitoring stations (AQMS)~\cite{hsu2017community} that are installed at strategic locations to sense outdoor air pollution and compute metrics such as the \textit{Air Quality Index} (AQI) that represent the overall air quality in a region, continuous monitoring of indoor air quality poses several challenges. \textbf{First}, the indoor pollution sources are complex, localized, and activity-driven. Cooking, for example, contributes significantly to indoor pollution, but how it affects the room depends on the type of food being cooked~\cite{pratiti2020health}. Furthermore, activities such as cutting fruit (produces Ethyl Alcohol)~\cite{moore2018managing,dudley2004ethanol}, cleaning rooms (increases VOCs)~\cite{epavoc}, or charging mobile phones (produces CO\textsubscript{2})~\cite{phoneco2} produce pollutants at different levels. In addition, the level of pollution in a room depends on the number of people present and their activities. \textbf{Second}, in contrast to outdoor pollutants that diffuse quickly because of open space, indoor pollution can undergo significant and rapid changes depending on the dynamics of the room, such as the people in the room and their movements, ventilation, switching on or off a fan or air conditioner, and opening and closing a door~\cite{chhaglani2022flowsense,qin2023system,adhikary2020we,moore2018managing}. As a result of such dynamics, different pollutants can dominate at different times. For example, as the variation of VOC for multiple sensors shown in \figurename~\ref{fig:into_plot1}, for H1, we can observe the degree of spread as the kitchen (A) exhaust is off, and the dining ceiling fan pulls the pollutant. The dining left (C) and dining right (D) sensors situate below the ceiling fan; thus, we observe a rapid increase in the VOC.  \textbf{Third}, the spatial spread of pollutants indoors depends on the rooms' structure, connectivity, ventilation, and many other factors. Unlike some indoor pollutants (such as CO\textsubscript{2}), VOCs can linger in a room for a considerable period of time. Additionally, they can contaminate neighboring rooms even if they spread from one room to another. For example, as shown in \figurename~\ref{fig:into_plot1}, for H1, VOC \textit{lingers} in the room (B) for a while due to poor ventilation (only one open window). The room behind the dining (E) \textit{traps} some amount of VOC over time (significant after $21$min from the start of cooking) coming from the dining. In contrast, we observe different behavior for H2, as illustrated in \figurename~\ref{fig:into_plot2} where the VOC level always remains excessive since this household is located near a busy construction site and most of the windows are left wide open. In H2, the VOC lingers within the rooms connected to the kitchen (A). %Therefore, monitoring pollutants only in a few rooms might not indicate how healthy an apartment or home is overall.

To analyze such indoor pollution dynamics, in this paper, we perform a large-scale study for $3$ months over $28$ deployment sites by sensing different indoor pollutants and correlating them with the ground truth information (room structure, events, activities, etc., details in Section~\ref{sec:pilot}). For this study, we have developed a custom-made indoor air quality solution called \ourmethod{} (\textbf{D}istributed \textbf{A}ir Qua\textbf{L}i\textbf{T}y M\textbf{ON}itor) that considers the trade-off in cost, sensor type, accuracy, network connectivity, power and usability (details in Section~\ref{sec:datacollection}), and deployed multiple such platforms at every deployment site. In the last decade, several works have conducted field studies to understand the distribution of indoor pollutants~\cite{mujan2021development, gao2022understanding, temprano2020indoor, kumar2016real,chartier2017comparative,putri2022spatial, cao2020sensor} and explored the challenges in sensor placement. However, these works are either very small-scale (<10 measurement sites) or done in a very controlled manner within the Lab setup. Recently developed IoT frameworks~\cite{verma2021sensert,hsu2017community} have employed real-time data streaming over the wireless sensor networks to compute pollution overlays~\cite{ji2019indoor,brazauskas2021real,brazauskas2021data} for buildings; but, these works are tuned again for a single environment with a specific task. So, To the best of our knowledge, this is the first horizontal study to focus on sensing infrastructure design, sensor placement, and pollution dynamics analysis on a large scale in the wild.

Notably, AQI is extensively used to characterize the outdoor air quality~\cite{pramanik2023aquamoho}. The metric has consistently performed well outdoors because air quality is impacted similarly over a wide area by weather and pollution outlets (e.g., manufacturing). Indoors, however, exhibit various pollutants with spatio-temporal dominance due to household activities, room structures, floor plans, ventilation systems, etc. So, we believe that AQI is not perfectly suitable for indoor environments. In this vein, there have been a few studies that have proposed metrics such as Indoor Air Quality Index (IAQI)~\cite{mujan2021development}, and Actionable Indoor Air Quality Index~\cite{breeze}. However, such metrics only consider the average exposure for the major pollutant, resulting biased estimate of the effective air quality. Realizing this research gap, in this paper, we propose a novel metric \textit{Healthy Home Index} (HHI) that quantifies indoor air quality, embedding indoor activities, and spatial spread of pollutants. Furthermore, the metric adjusts to the floor plan and ventilation of the interior. To our knowledge, this is the first work that attempts to formulate a robust metric to measure the holistic \textit
{healthiness} of indoor spaces. Our contributions to this paper are the following. 
%To facilitate real-time data analytics and fault recovery, we have further developed an end-to-end IP-agnostic IoT backbone that processes the data stream from multiple sensing modules distributed across several indoor environments.
%who have actively participated in the study by providing indoor activity and event annotation in real-time via the VocalAnnot app
%at two cities in India and one city in the United States 
%Based on the observations of the field study and realizing the limitations of the existing quantifiers for indoor air quality, w
\begin{enumerate}
  \item \textbf{Development of \ourmethod{} Platform}:
    We have designed a lunch-box-size sensing module that can be placed at any indoor location to sample pollution data with minimal user intervention. Moreover, a simple speech-to-text Android application VocalAnnot\footnote{VocalAnnot app source: \url{https://anonymous.4open.science/r/VocalAnnot}} is also developed for the participants to annotate indoor events and activities easily. % which perturb the indoor air quality.
    
    \item \textbf{Large-scale Field Study}:
    We have conducted a large-scale in-the-wild field study across different indoor environments such as households, research Labs, canteens, classrooms, studio apartments, etc., for a period of $3$ months, expanding $28$ indoor deployment sites with a total of $42$ occupants. A segment of the collected data is open-sourced\footnote{HHI Implementation and Sample Dataset available \url{https://anonymous.4open.science/r/DALTON}} for the research community.
    
    \item \textbf{Formulation of Healthy Home Index (HHI)}:
   	We formulated a novel adaptive metric named the Healthy Home Index (HHI) that better captures long-term pollution exposure, the impact of indoor activities, ventilation, and the spatio-temporal spread of the pollutants. \ourmethod{} platform also provides the capacity for distributed multi-point pollution sampling to compute the HHI  for any indoor setting. We open-sourced the implementation of HHI with the sample dataset\footnotemark[2].
    
    \item \textbf{Scalability of HHI}:
	We have thoroughly evaluated the HHI metric on diverse indoor setups. Our evaluation shows that HHI more efficiently represents the effective pollution exposure to other baseline metrics. Moreover, HHI is categorized into Healthy, Alert, and Actionable thresholds considering the well-being of the occupants.
 
 % triggers \textbf{X} \% less false ventilation alarms and adapts to the specific indoor environment within \textbf{H} hours of deploying the sensing modules.
 
\end{enumerate}

%paper organisation We have further highlighted the benefits of HHI over existing metrics like IAQI. to ensure its scalability and applicability in the real world
%\input{Sections/platform}
\section{Data Collection}
\label{sec:datacollection}
This section briefly overviews \ourmethod{} and the detailed data collection procedure. We also highlight the patterns and anomalies observed over the entire period ($3$ months) of sensor deployment at $28$ different sites.
\begin{figure}[!ht]
	\centering
	\includegraphics[width=0.95\columnwidth]{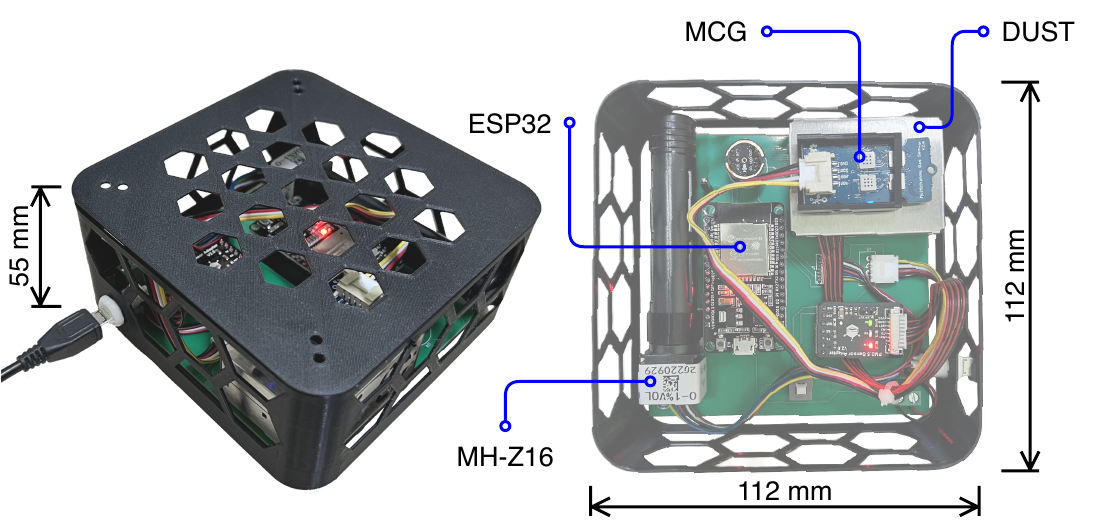}
	\caption{\ourmethod{} Sensing Module.}
	\label{fig:sensors}
\end{figure}

\subsection{\ourmethod{} Sensing Platform}
\label{sec:dalton}
To perform the large-scale analysis of indoor pollutants, we developed an IoT sensing platform called \ourmethod{}, as shown in \figurename~\ref{fig:sensors}. However, before building this customized platform, we researched various consumer indoor air quality measurement platforms, such as Prana Air Monitor~\cite{pranaair}, SmileDrive Portable Monitor~\cite{smiledrive}, YVELINES Monitor~\cite{yvelines}, Pallipartners Monitor~\cite{pallipartners}, Airthings Monitor ~\cite{airthings}, etc. However, they either lacked all pollution sensors with desirable resolution (SmileDrive, YVELINES, PalliPartners, etc.), or they lacked WiFi interfaces for remote communication (Airthings, PalliPartners, Smiledrive, etc.), or API access to data or proper open data streaming (except Prana Air). Therefore, we constructed this module while also considering the cost (around \$250) of a large-scale measurement exercise. Furthermore, the size of the module is of a lunchbox size (112 mm $\times$ 112 mm $\times$ 55 mm), equipped with multiple research-grade sensors that together measure the concentration of pollutants, such as \textit{Particulate matter} (PM\textsubscript{x}), \textit{Nitrogen dioxide} (NO\textsubscript{2}), \textit{Ethanol} (C\textsubscript{2}H\textsubscript{5}OH), \textit{Vola\textit{tile organic compounds} (VOCs), \textit{Carbon monoxide} (CO), }Carbon dioxide (CO\textsubscript{2}), etc., with \textit{Temperature} (T) and \textit{Relative humidity} (RH). We utilize the ESP-WROOM-32 chip as the on-device processing unit that packs a dual-core Xtensa 32-bit LX6 MCU with WiFi $2.4$GHz HT40 capabilities. The connectivity board is a two-layer printed circuit board (FR4 material). The outer shell of the module is a 3D printed (PLA+ material) hollow structure with honeycomb holes so that the air within the module is the same as outside, resulting in unbiased measurement of pollutants (at a sampling frequency of $1$Hz). Section~\ref{sec:eval} details the IoT backbone implementation of \ourmethod{} along with its detailed specification and operational bounds. 

\begin{figure}[!t]
	\captionsetup[subfigure]{}
	\begin{center}
            \subfloat[Indoor types\label{fig:deploy_pie}]{
			\includegraphics[width=0.42\columnwidth,keepaspectratio]{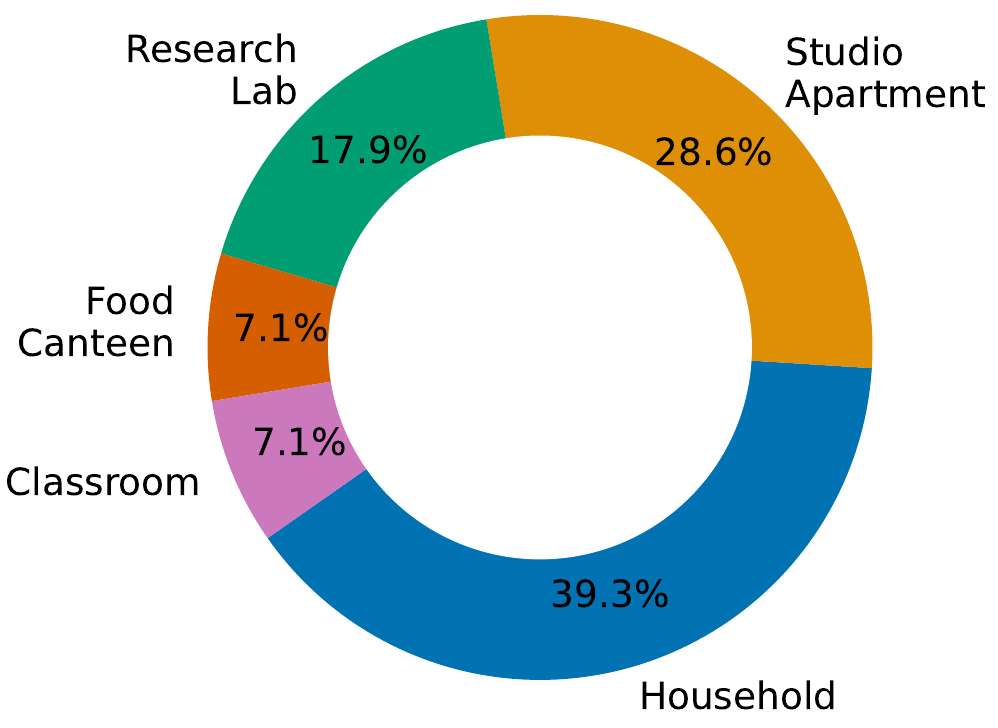}
		}
            \subfloat[Site Area\label{fig:area}]{
			\includegraphics[width=0.26\columnwidth,keepaspectratio]{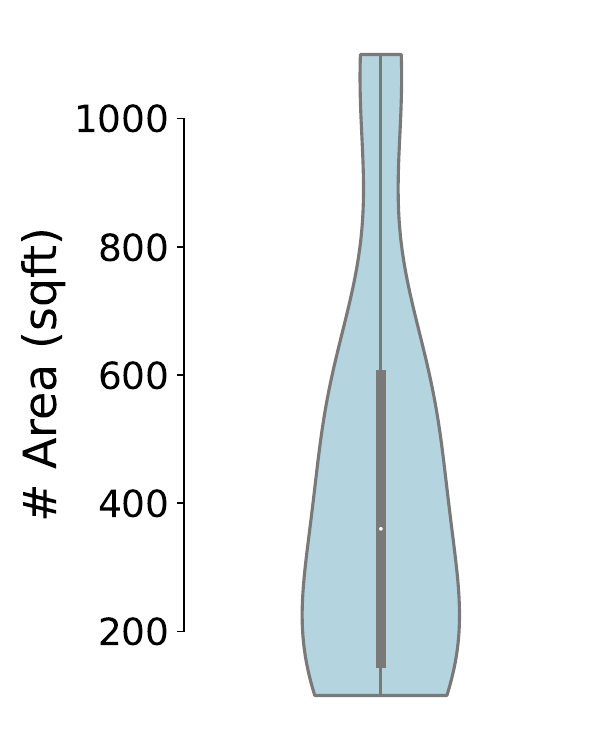}
		}
           % \subfloat[Sensors per site\label{fig:sens_site}]{
		%	\includegraphics[width=0.25\columnwidth,keepaspectratio]{Figures/data/sens_site_dist.pdf}
		%}\
            \subfloat[Sensors per 500sqft\label{fig:sens_500}]{
			\includegraphics[width=0.30\columnwidth,keepaspectratio]{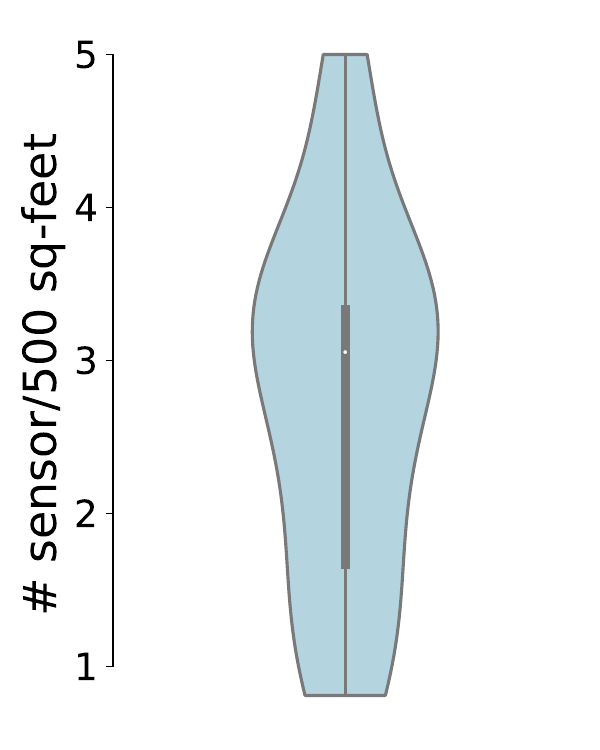}
		}\
            \subfloat[Occupants per site\label{fig:occu_site}]{
			\includegraphics[width=0.32\columnwidth,keepaspectratio]{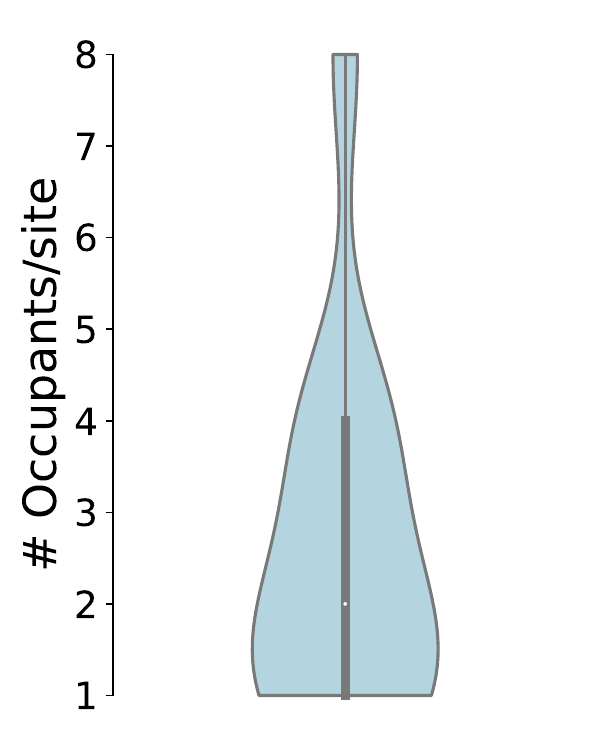}
		}
            \subfloat[Occupants' Age\label{fig:age_dist}]{
			\includegraphics[width=0.32\columnwidth,keepaspectratio]{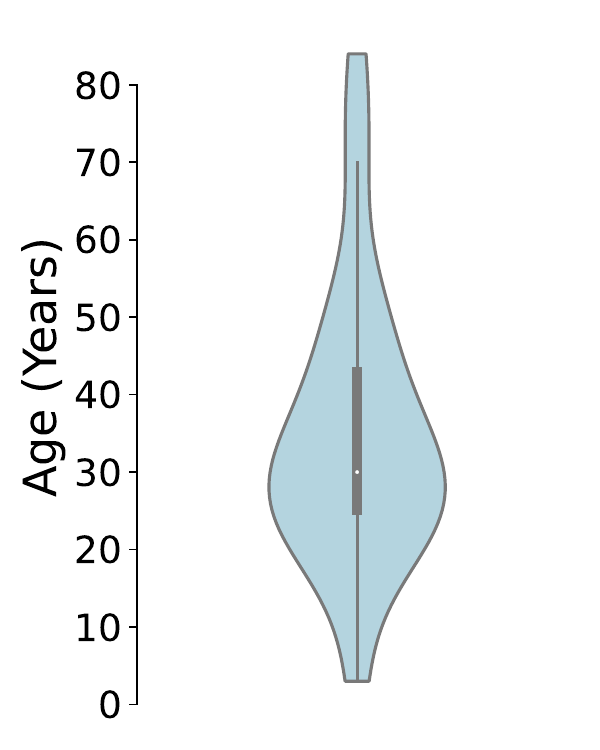}
		}
		\subfloat[Occupants' Gender\label{fig:gender}]{
			\includegraphics[width=0.32\columnwidth,keepaspectratio]{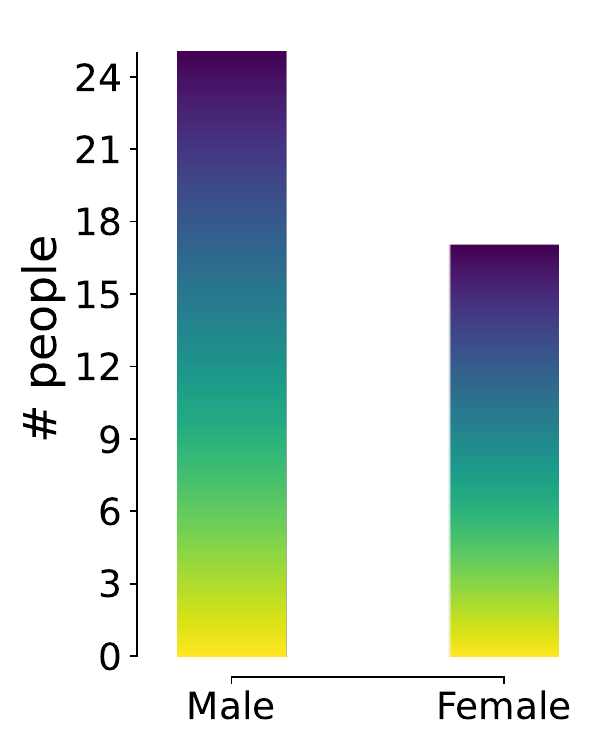}
		}\
	\end{center}
	\caption{Demographics and Details of the field-study.} %: (a) Indoor types, (b) Area of the sites, (c) Sensors deployed per site, (d) Sensors deployed per 500 sqft area, (e) Occupants per site, (f) Age, and (g) Gender of the occupants \figurename~\ref{fig:sens_site} shows the distribution of the number of deployed sensors per site during the data collection.
	\label{fig:demog}
\end{figure}

\subsection{Demographics of the Field Study}
We have collected data for $28$ measurement sites on primarily five types of indoor environments, namely, households, studio apartments, research labs, food canteens, and classrooms (\figurename~\ref{fig:deploy_pie}). The area of the deployment sites varies from $150$ sqft (typically, the studio apartments) to $1100$ sqft (typically, the households), as shown in \figurename~\ref{fig:area}. We deployed $1$-$2$ sensors per studio apartment, $3$-$4$ sensors per classroom and lab, and $3$-$6$ sensors per household.   \figurename~\ref{fig:sens_500} depicts the distribution of number of deployed sensors per $500$ sqft area. In total, $42$ occupants participated in the study and annotated their indoor activities throughout the day for about $3$ months. \figurename~\ref{fig:occu_site} and~\ref{fig:age_dist} show the average number of occupants per site and their age distribution, where \figurename~\ref{fig:gender} shows the degree of participation from both genders.

\subsection{Annotation and Selected Observations}
\begin{figure}[!ht]
	\captionsetup[subfigure]{}
	\begin{center}
            \begin{minipage}{0.48\columnwidth}
                \subfloat[VocalAnnot app\label{fig:app}]{
			\includegraphics[width=\columnwidth,keepaspectratio]{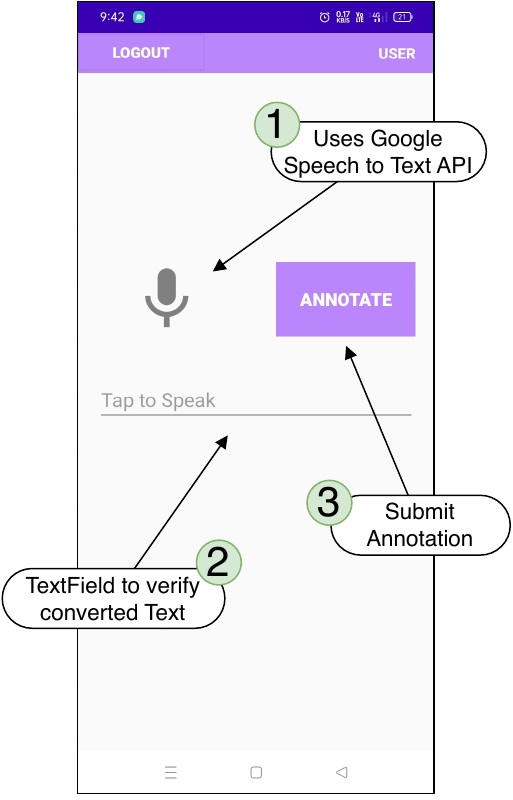}
		}
            \end{minipage}\hfil
            \begin{minipage}{0.38\columnwidth}
                \subfloat[Mosquito Repellent\label{fig:mosq_rep}]{
			\includegraphics[width=\columnwidth,keepaspectratio]{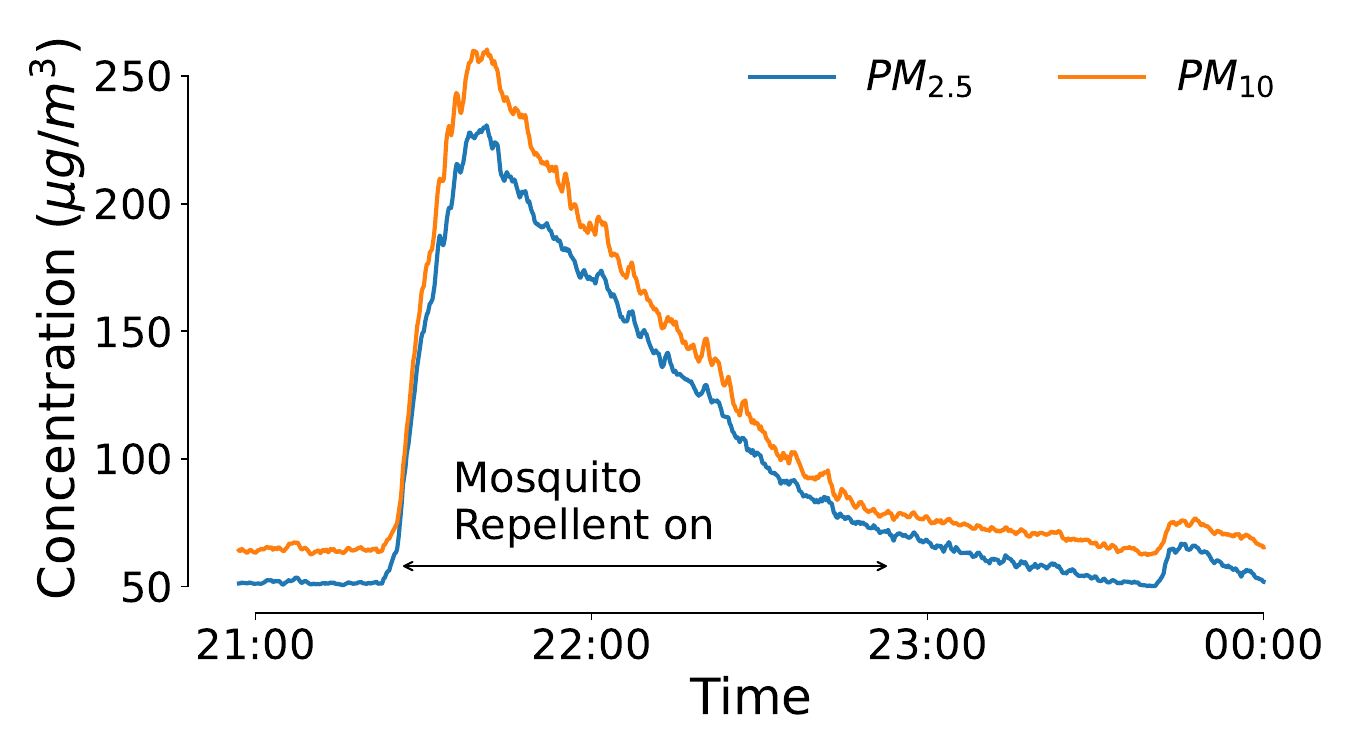}
		}\
		\subfloat[Leftover Food\label{fig:leftover}]{
			\includegraphics[width=\columnwidth,keepaspectratio]{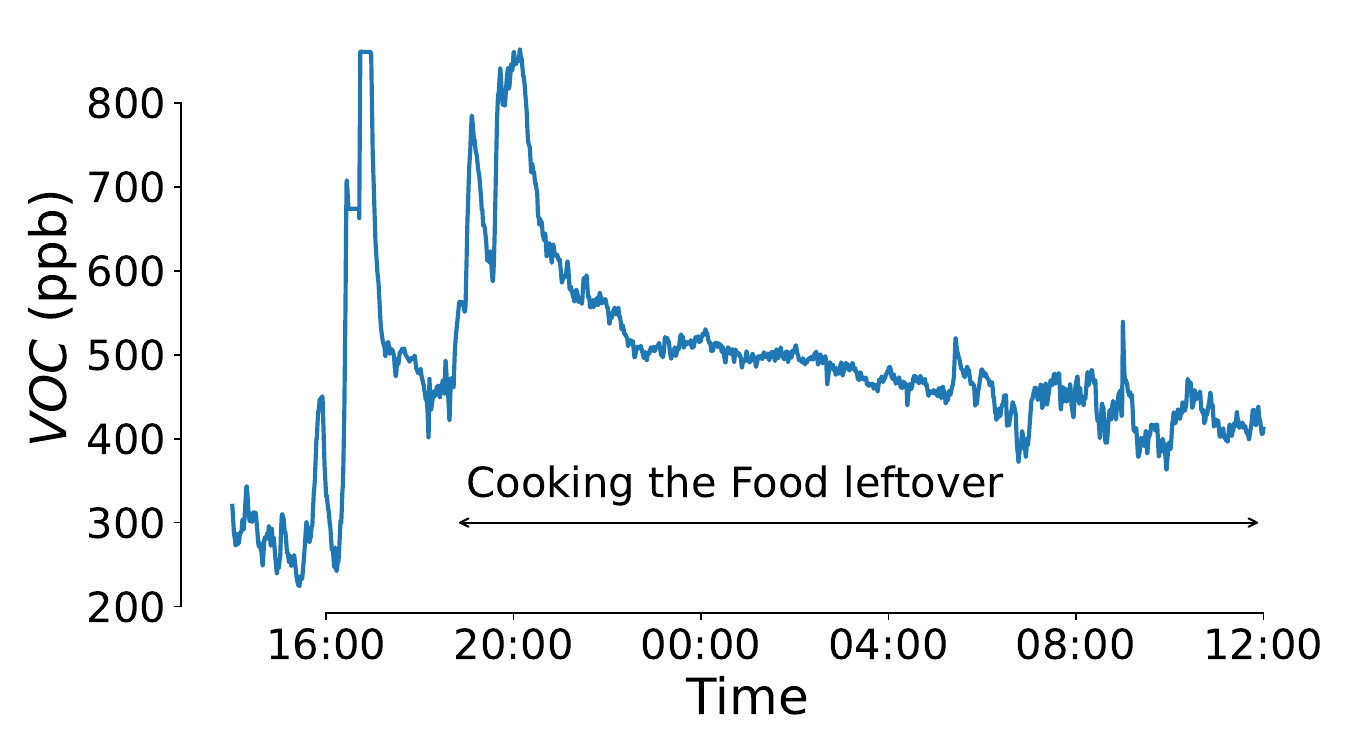}
		}\
            \subfloat[Charging Electronics\label{fig:ph_charge}]{
			\includegraphics[width=\columnwidth,keepaspectratio]{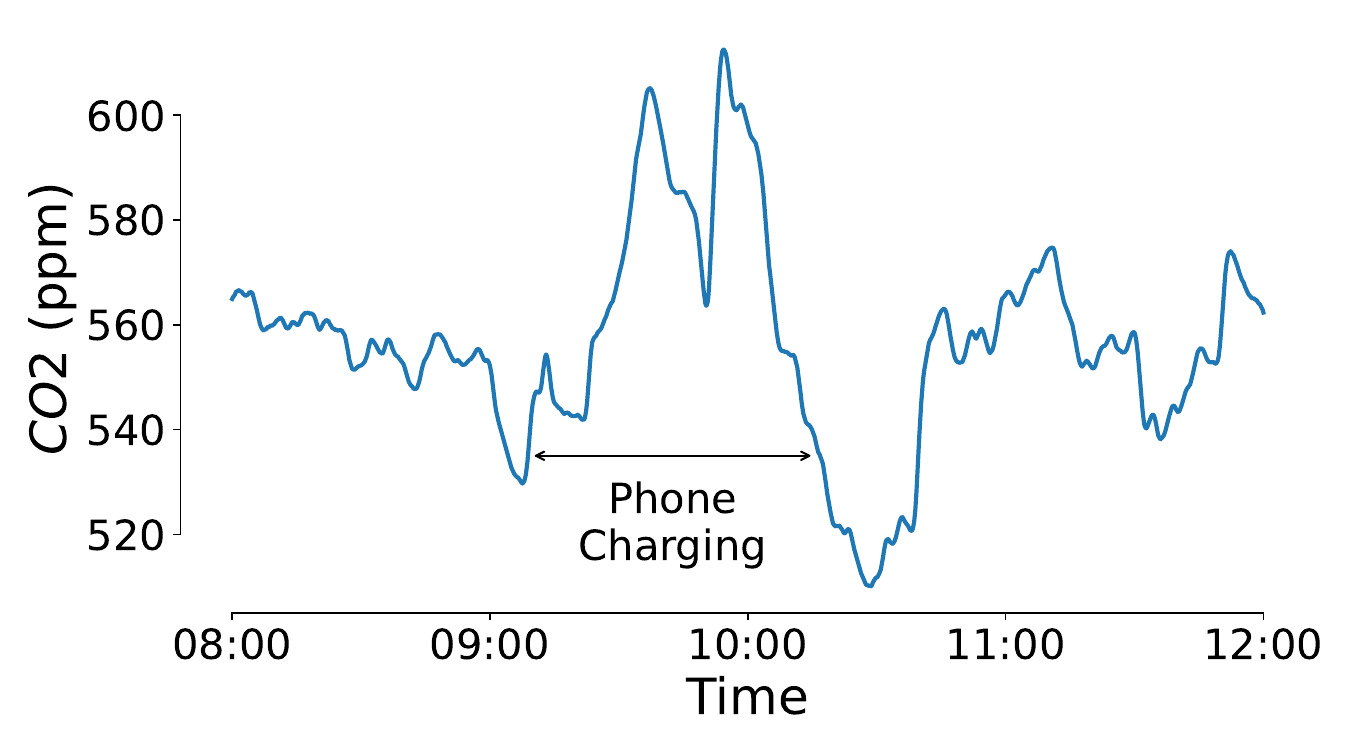}
		}
            \end{minipage}
	\end{center}
	\caption{Annotation app and some pollution sources observed from the annotation.}
	\label{fig:minor_obs}
\end{figure}

The participants are asked to install the VocalAnnot android application (shown in \figurename~\ref{fig:app}) to easily annotate their indoor activities, specifically starting and ending of activities using the \textit{Google Speech to Text API}\footnote{\url{https://cloud.google.com/speech-to-text} (Accessed: \today)}. The API submits the annotation using \ourmethod{} Annotation micro-service. In case of any inconsistency, the participants can manually edit the text by clicking on the text-field. 

Due to the precise timestamp-tagged annotations, we observe several interesting pollution dynamics concerning indoor activities over the data collection period. For example, turning on a mosquito repellent initially hugely impacts the particulate-matter levels due to the burning of the repellent cartridge, as shown in \figurename~\ref{fig:mosq_rep}. Such abrupt exposure to high particulate matter concentrations harms our respiratory system~\cite{patra2016emissions,yang2013source}. Next, we observe that the leftover foods in the kitchen emit VOC overnight that may linger till the following day (\figurename~\ref{fig:leftover}). Such long-term accumulation of VOC impacts the overall healthiness of the household as it gets spread to other adjacent rooms~\cite{huang2011characteristics,song2019source}. Finally, we observe that even charging electronic gadgets like smartphones, tablets, etc., emits a significant amount of CO\textsubscript{2} (\figurename~\ref{fig:ph_charge}). Such observations indicate that there are unconventional indoor pollution sources that a typical user may not perceive; however, they significantly impact the overall pollution dynamics of the room, as we see later in detail.

\begin{figure}
	\captionsetup[subfigure]{}
	\begin{center}
		\subfloat[Kitchen VOC\label{fig:kitchen_voc}]{
			\includegraphics[width=0.45\columnwidth,keepaspectratio]{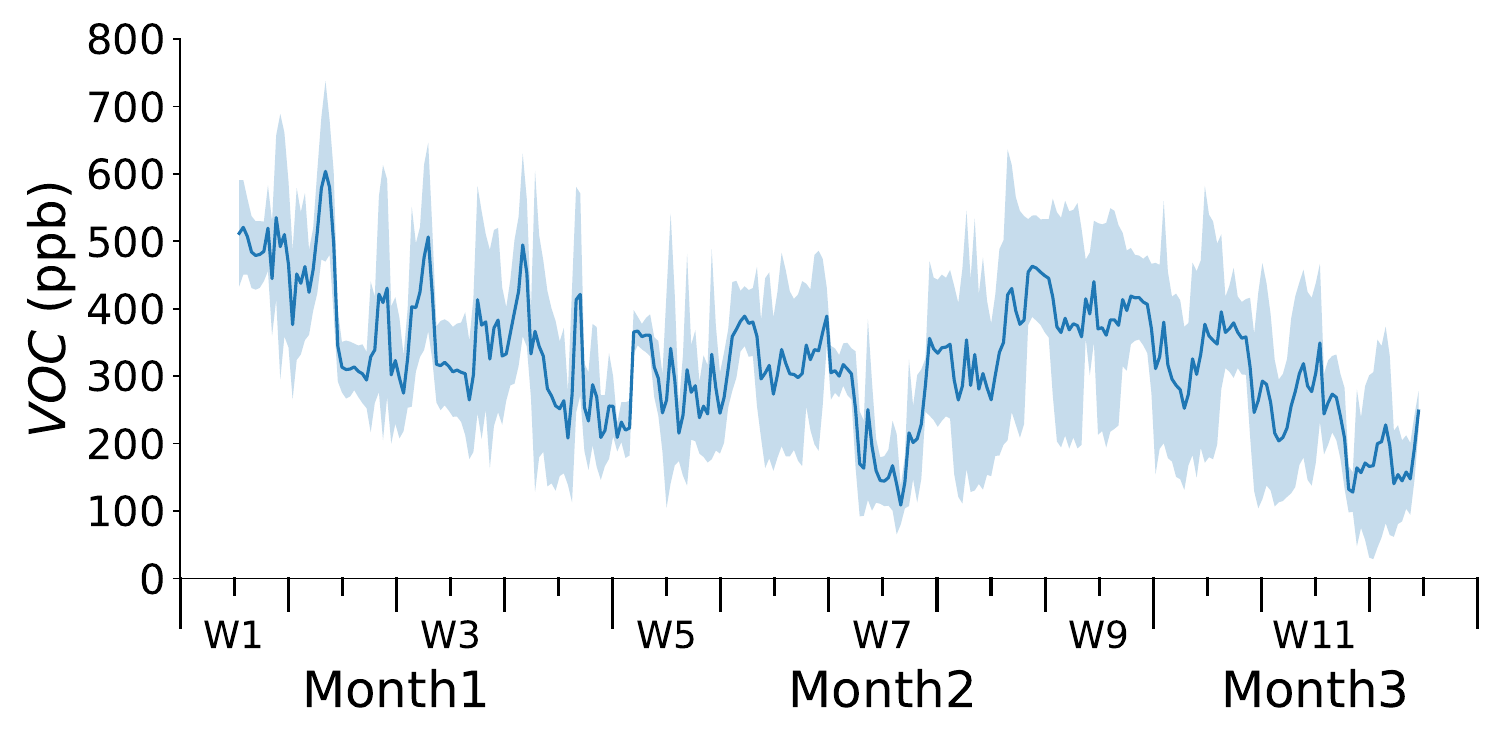}
		}
	    \subfloat[Bedroom VOC\label{fig:bedroom_voc}]{
	    	\includegraphics[width=0.45\columnwidth,keepaspectratio]{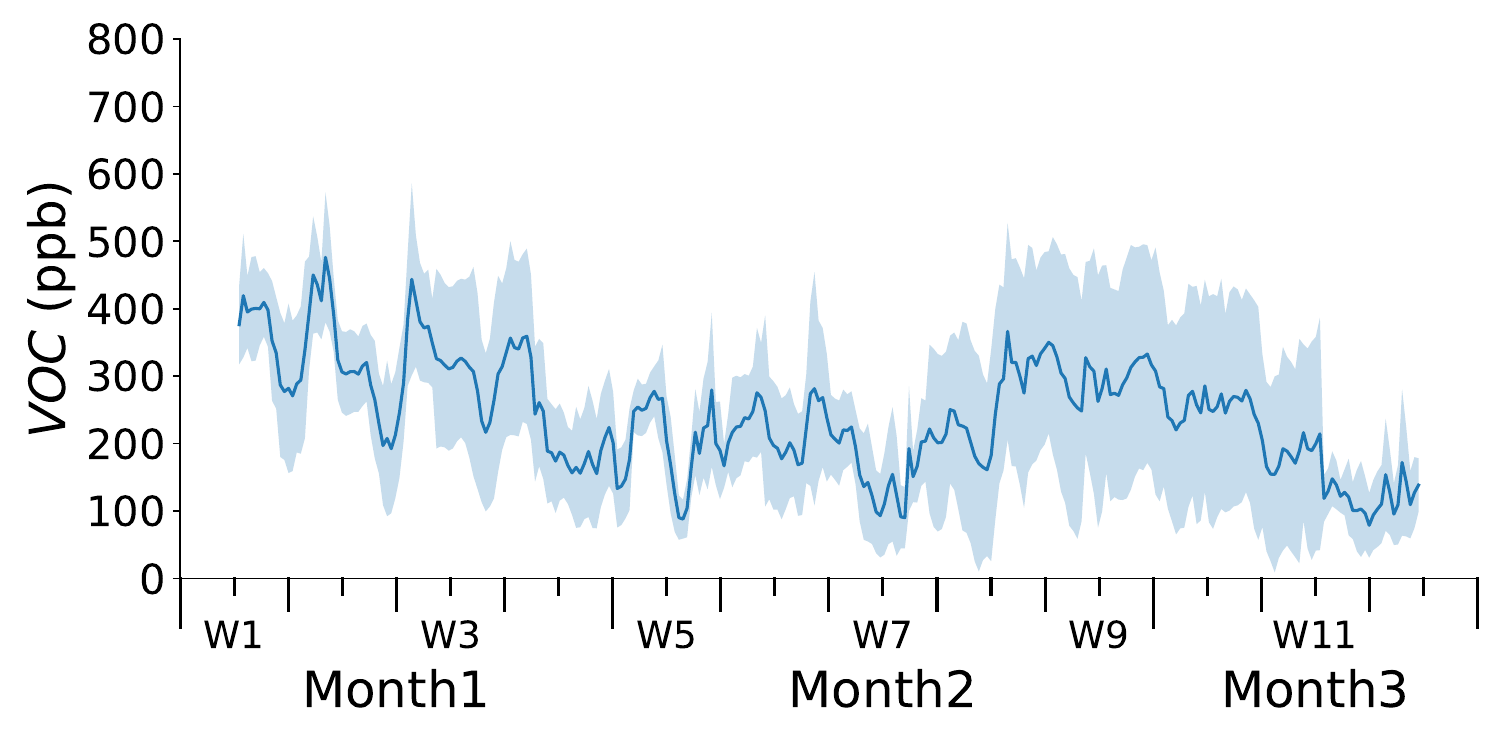}
	    }\
		\subfloat[Kitchen CO\textsubscript{2}\label{fig:kitchen_co2}]{
			\includegraphics[width=0.45\columnwidth,keepaspectratio]{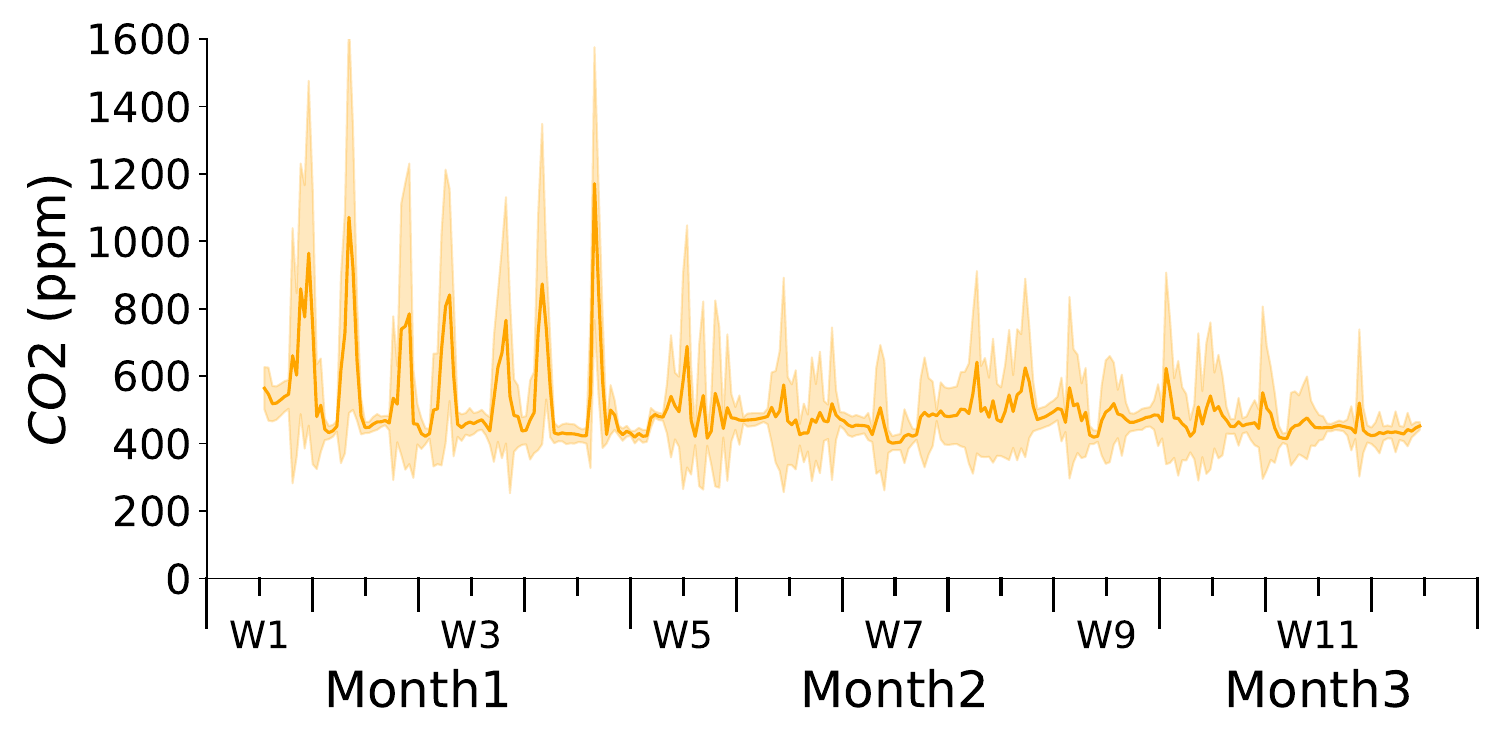}
		}
	    \subfloat[Bedroom CO\textsubscript{2}\label{fig:bedroom_co2}]{
	    	\includegraphics[width=0.45\columnwidth,keepaspectratio]{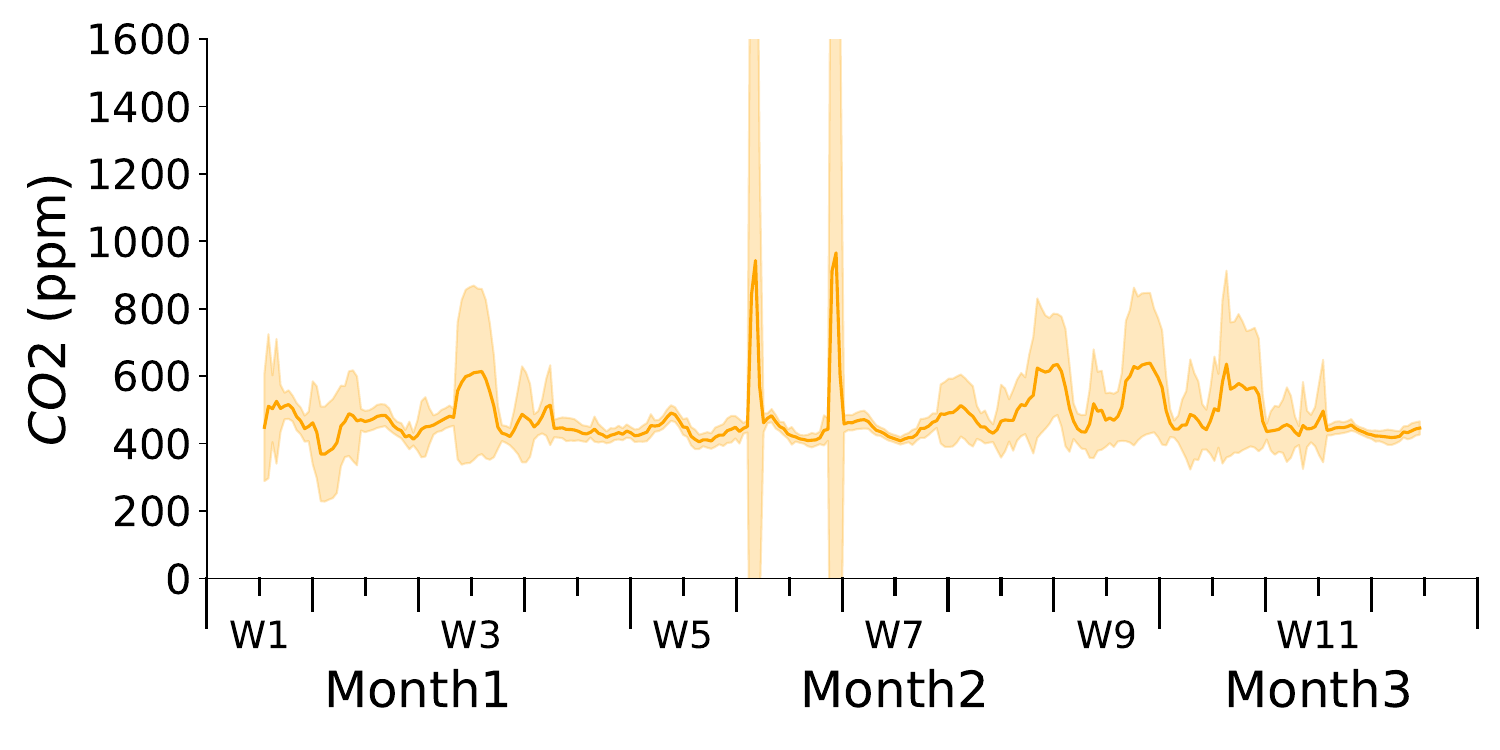}
	    }\
            \subfloat[Temperature\label{fig:overall_t}]{
			\includegraphics[width=0.45\columnwidth,keepaspectratio]{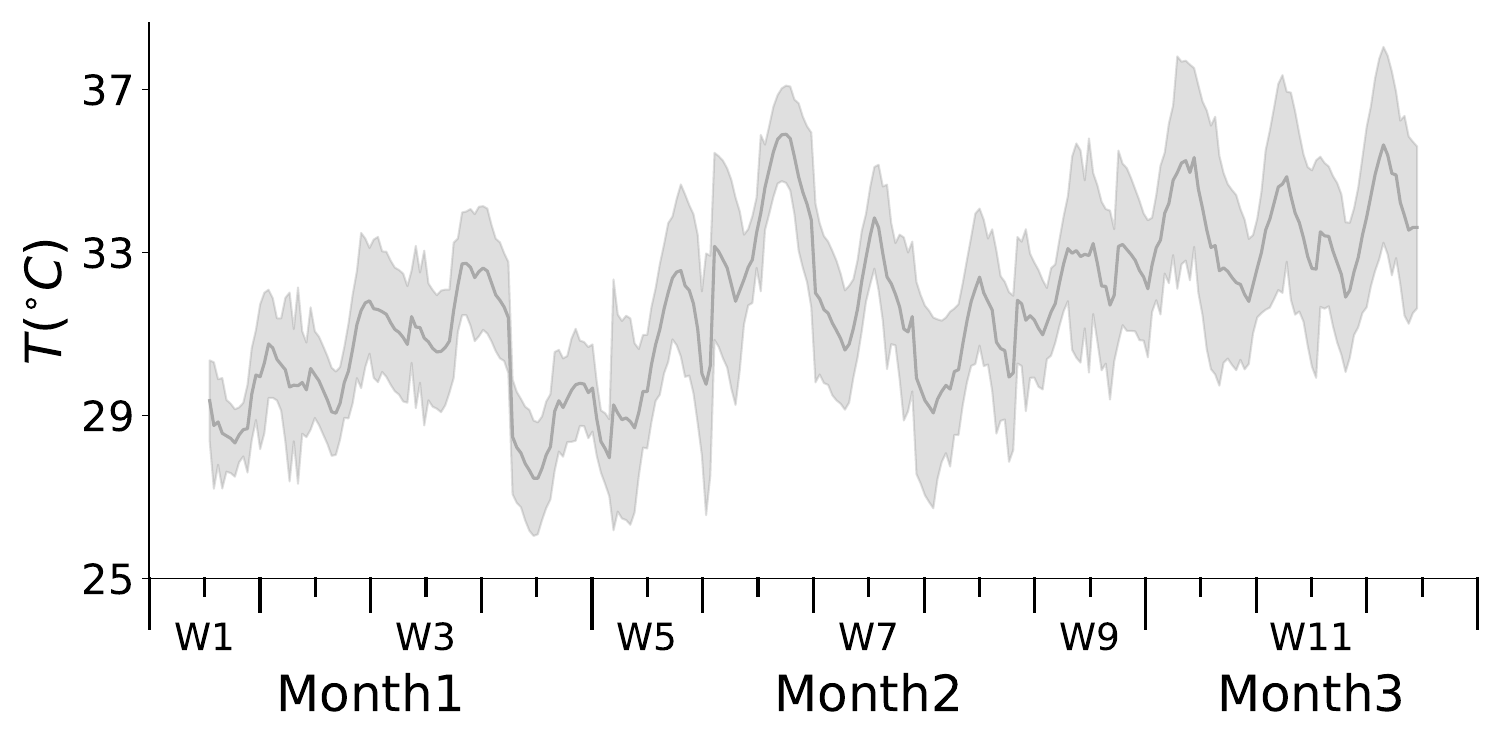}
		}
	    \subfloat[Humidity\label{fig:overall_hum}]{
	    	\includegraphics[width=0.45\columnwidth,keepaspectratio]{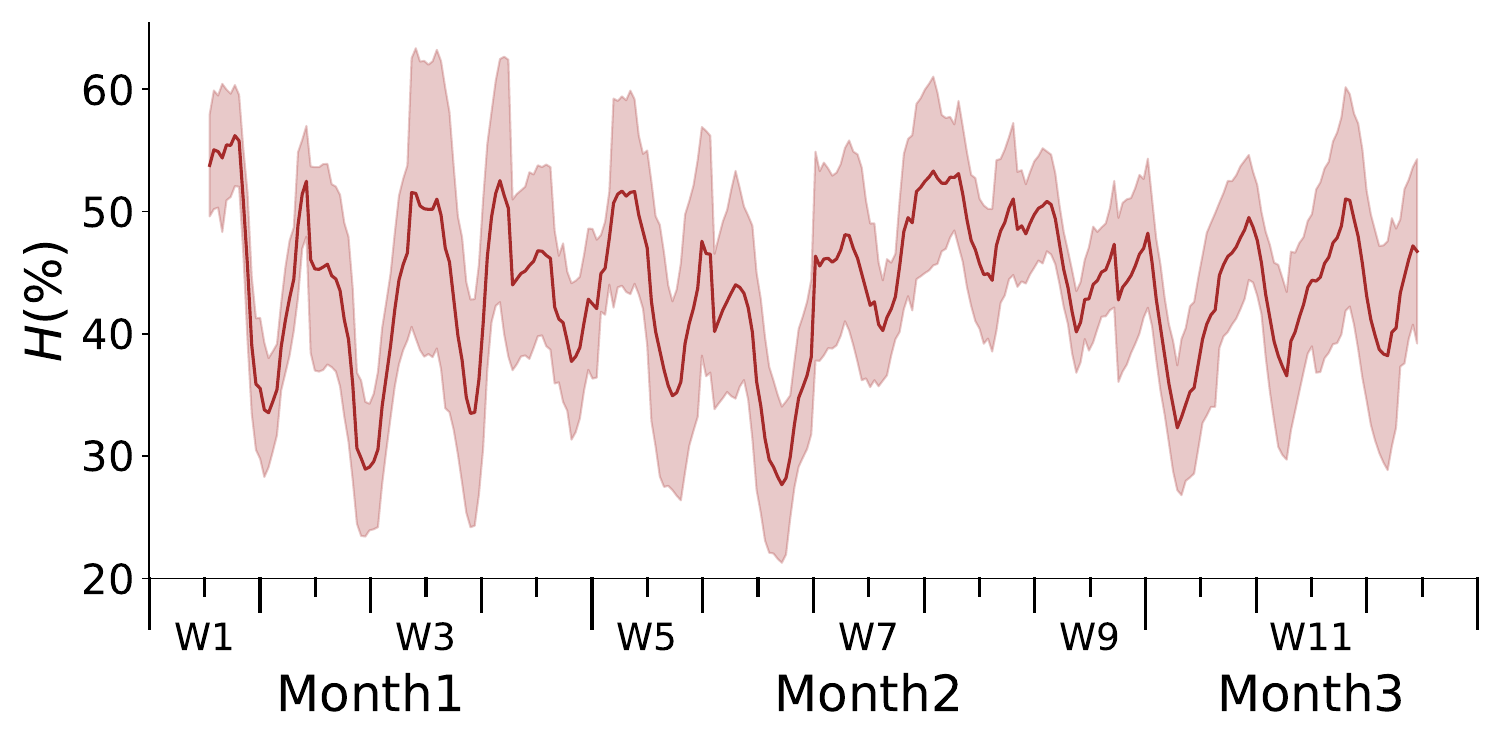}
	    }\
	\end{center}
	\caption{Daily indoor trends by week and month.}
	\label{fig:pattern}
\end{figure}

\subsection{Repetitive Patterns in the Data}
\figurename~\ref{fig:pattern} shows the weekly variation of daily VOC, CO\textsubscript{2} exposure, along with temperature and humidity change throughout the data collection period of three months. We observe a similar trend in the mean VOC exposure for the kitchen and bedroom as per \figurename~\ref{fig:kitchen_voc} and~\ref{fig:bedroom_voc}, which indicates that, in general, pollutants emitted from the kitchen are spread towards the bedrooms. Moreover, as the data is collected during the summertime, we observe a steady rise in temperature over the months as per \figurename~\ref{fig:overall_t}. As shown in \figurename~\ref{fig:overall_hum}, the overall humidity also increases during the second month of the data collection. The food items and fruits degrade quickly in high temperatures and humidity, releasing excessive VOCs; thus, we observe a rise in the mean VOC levels in the kitchens and bedrooms during the second month. Regarding CO\textsubscript{2} exposure, we observe a maximum peak in the kitchen during the first month when the temperature remains relatively comfortable, as shown in \figurename~\ref{fig:kitchen_co2}. The primary reason for such observation is that we are more sensitive towards temperature change (detail explanation and analysis in Section~\ref{sec:complex_pollution}), thus in comfortable temperatures, the kitchen exhaust fans are mostly turned off, resulting in poor ventilation for the emitted CO\textsubscript{2} (we observed this from the annotated labels as well). As the mean temperature increase over the months, we observe that the CO\textsubscript{2} peaks are reduced as the exhaust is turned on more frequently, providing much-needed ventilation. Interestingly, CO\textsubscript{2} in bedrooms do not significantly correlate with the kitchen, implying that CO\textsubscript{2} exposure is contained near the source, where VOC spread across the entire household. We next build upon these observations to analyze indoor pollutant behaviors for specific scenarios. 
\section{Indoor Pollution Dynamics}
\label{sec:pilot}
In this section, we further analyze the complex spatio-temporal distribution of indoor air pollutants. We also present evidence to design a comprehensive quantifier metric not limited to instantaneous pollution exposure but also considers the long-term accumulation of pollutants due to inadequate ventilation, uncertainty, and periodicity of pollutant concentration due to occupant activity.

\begin{figure}
	\captionsetup[subfigure]{}
	\begin{center}
		\subfloat[\label{fig:sc_house_f1}]{
			\includegraphics[width=0.42\columnwidth,keepaspectratio]{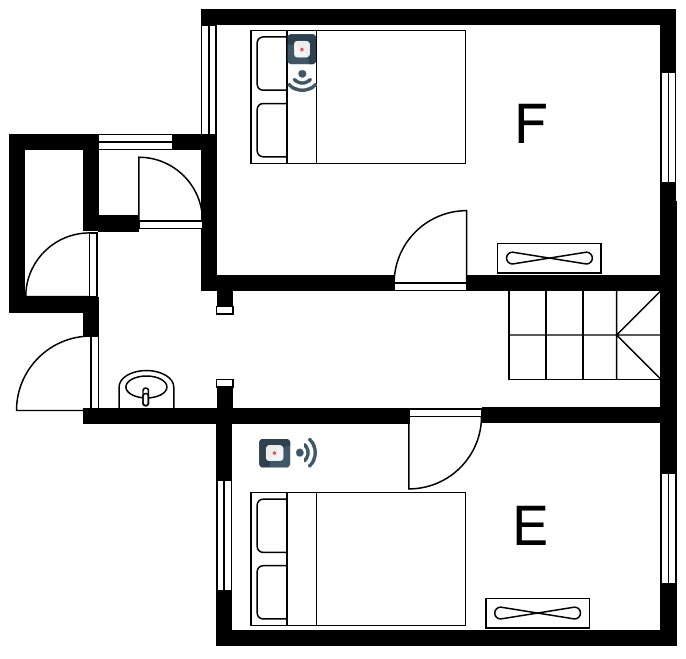}
		}
		\subfloat[\label{fig:gb_house}]{
			\includegraphics[width=0.48\columnwidth,keepaspectratio]{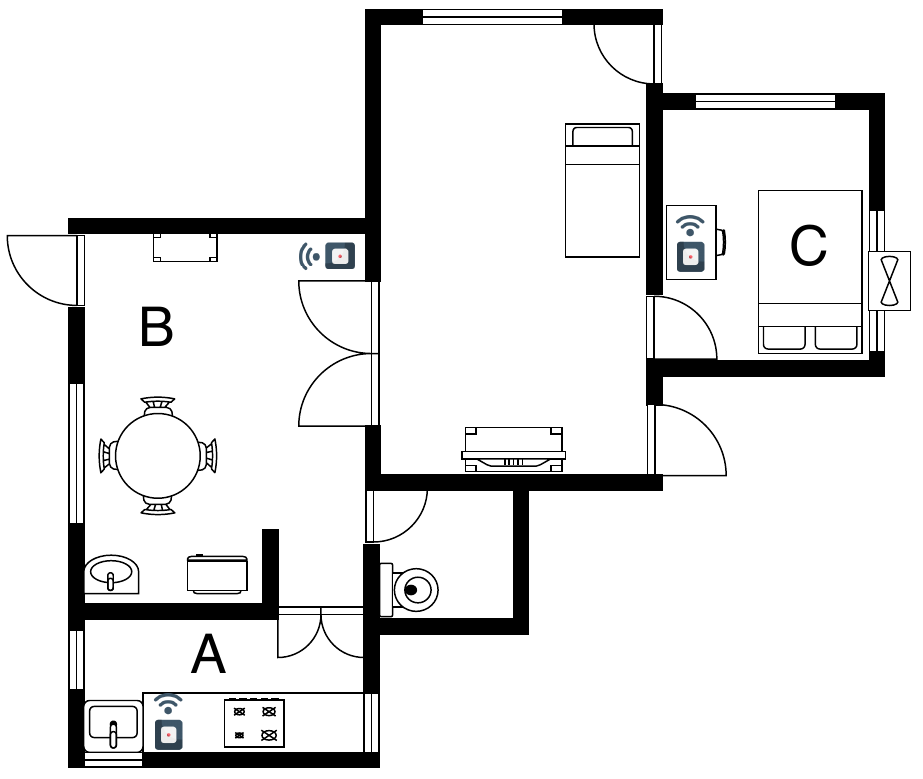}
		}
	\end{center}
	\caption{Floor plan (a) Household-2 Floor1 (H2-F1), (b) Household-3 (H3).}
	\label{fig:house_scf1_gb}
\end{figure}
\begin{figure}
	\captionsetup[subfigure]{}
	\begin{center}
		\subfloat[CO\textsubscript{2} -- H2\label{fig:sc_co2_time}]{
			\includegraphics[width=0.48\columnwidth,keepaspectratio]{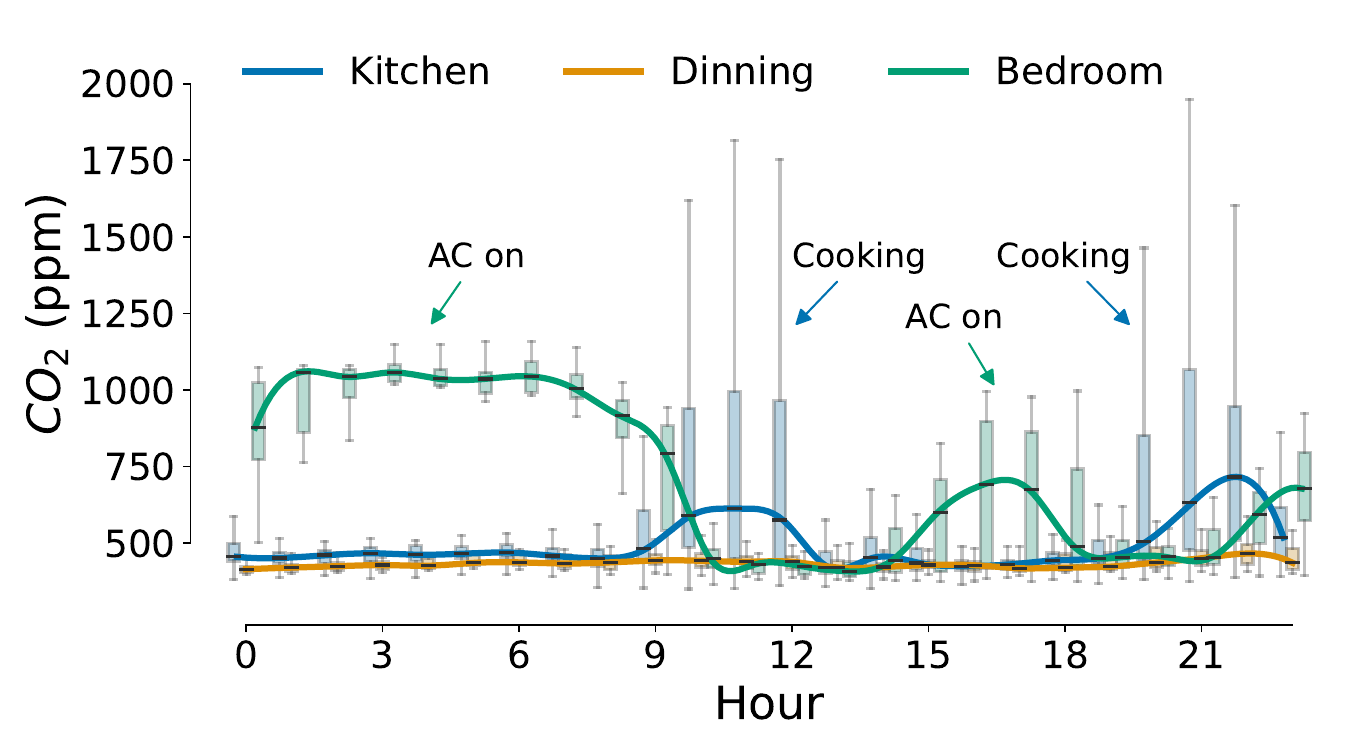}
		}
	    \subfloat[CO\textsubscript{2} -- H3\label{fig:gb_co2_time}]{
	    	\includegraphics[width=0.48\columnwidth,keepaspectratio]{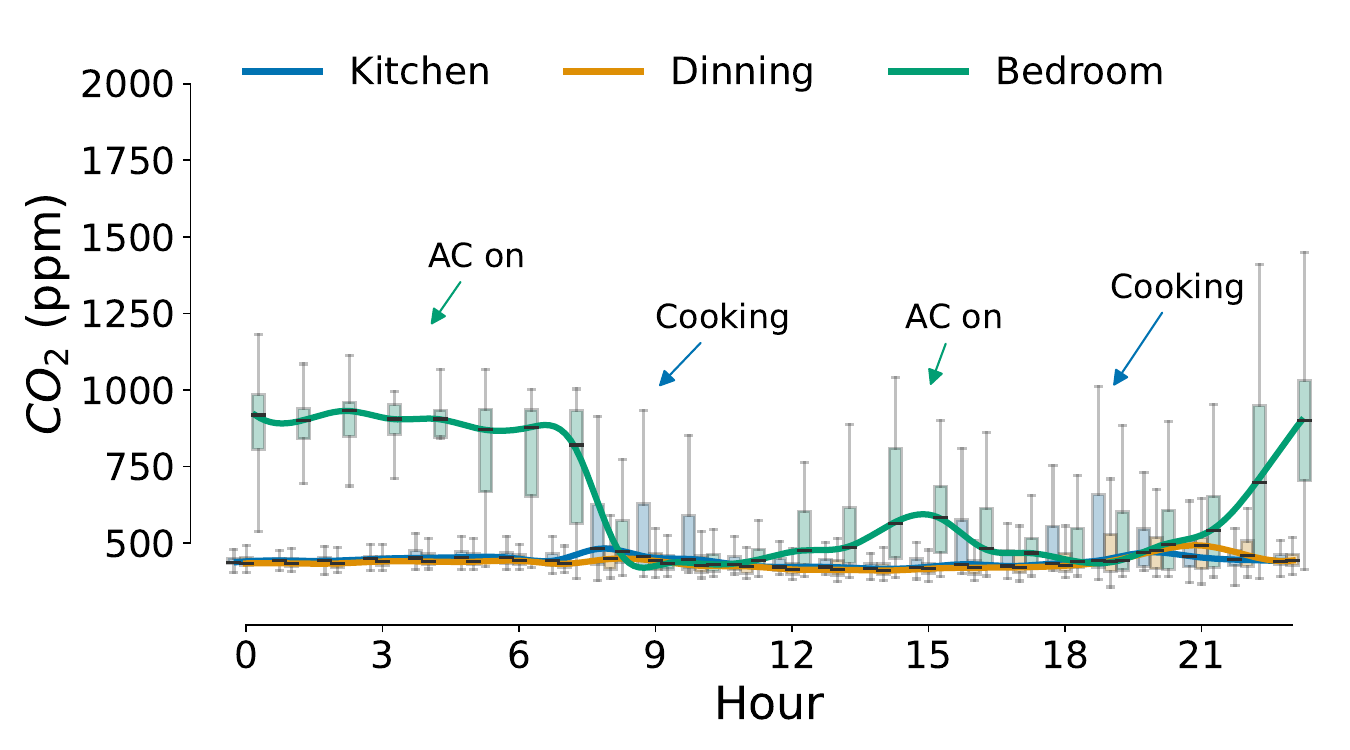}
	    }\
		\subfloat[VOC -- H2\label{fig:sc_voc_time}]{
			\includegraphics[width=0.48\columnwidth,keepaspectratio]{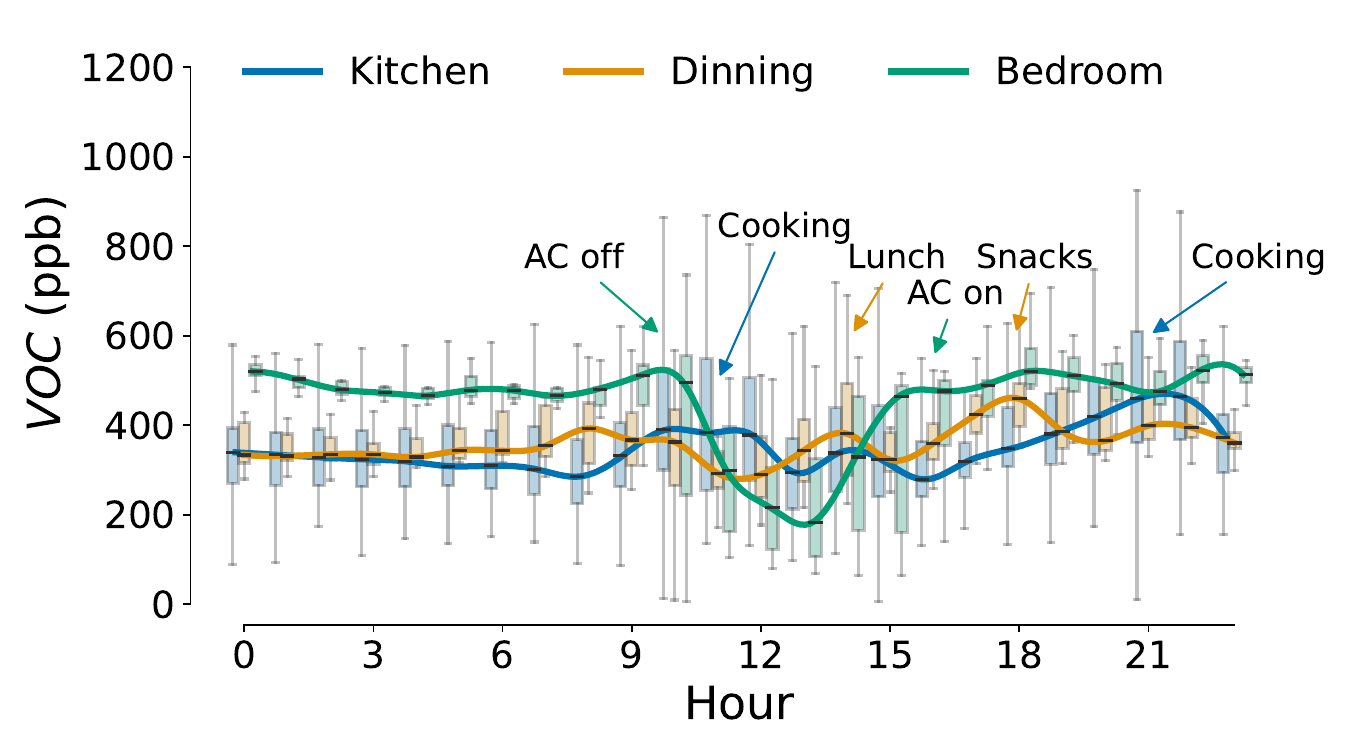}
		}
	    \subfloat[VOC -- H3\label{fig:gb_voc_time}]{
	    	\includegraphics[width=0.48\columnwidth,keepaspectratio]{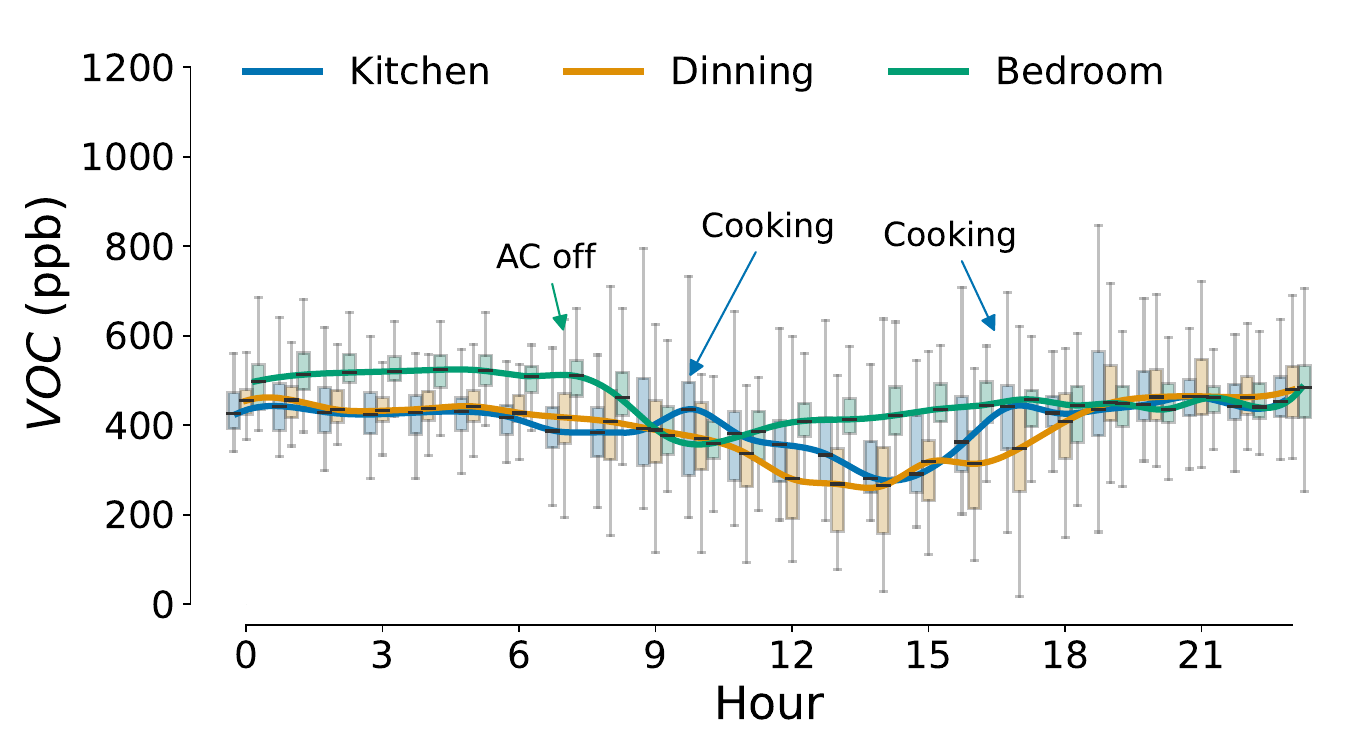}
	    }\
	\end{center}
	\caption{Pollutant distributions at different rooms of two households H2 and H3.}
	\label{fig:poll_time}
\end{figure}

% \begin{figure}
%     \centering
%     \includegraphics[width=0.6\columnwidth]{Figures/Diagrams/GB_marked.pdf}
%     \caption{Household-3 (H3)}
%     \label{fig:gb_house}
% \end{figure}

\subsection{Spatio-temporal Nature}
The daily activities of its occupants influence indoor environments; therefore, indoor pollutants follow a periodic pattern as our daily household activities. Specifically, different parts of the indoors act as pollution sources at different times of the day. As shown in \figurename~\ref{fig:poll_time}, we observe that the median CO\textsubscript{2} and VOC concentrations are significantly different across kitchen, dining, and bedroom for H2 and H3 (see the floor-plan in \figurename~\ref{fig:sc_house_marked} and~\ref{fig:house_scf1_gb}, respectively). 

Moreover, pollutants in an indoor area exhibit distinct periodic patterns based on the activities performed in that area. \figurename~\ref{fig:sc_co2_time} and~\ref{fig:gb_co2_time} show that in the kitchen, CO\textsubscript{2} is emitted during cooking, and with good ventilation (exhaust fans, open windows, etc.), it quickly descends to normal levels. However, for the bedroom, the median CO\textsubscript{2} levels are high (more than even the kitchen's peak CO\textsubscript{2}) throughout the night hours, mainly because of operating a split AC system. For better efficiency, split AC circulates the air internally~\cite{harby2019investigation}; however, such a technique allows the emitted pollutants to be accumulated over a long period. 

%Moreover, CO\textsubscript{2} in the dining region remains at normal levels and is slightly influenced by human gatherings during lunch and dinner.

VOC behaves similarly to CO\textsubscript{2} in the kitchen and bedroom and is impacted mainly by cooking and AC on/off, respectively. However, unlike CO\textsubscript{2}, VOC does not deplete rapidly even with good ventilation and lingers for an extended period, resulting in long-term exposure as shown in \figurename~\ref{fig:sc_voc_time} (see bedroom from 18:00 to 21:00). Moreover, VOC is naturally emitted from fruits, vegetables, left-over food; thus in both \figurename~\ref{fig:sc_voc_time} and~\ref{fig:gb_voc_time}, we see a steady increase of VOC in the kitchen from the evening hours until the kitchen is cleaned (see 22:00). Similarly, the dining place also observed an increase in the VOC during lunch, which gets descended after the dining was cleaned.

\begin{takeaway}{}{tw1}
 Different pollutants show different spatio-temporal behavior in indoor environments. While CO\textsubscript{2} and VOC both increase rapidly during activities like cooking, CO\textsubscript{2} depletes rapidly with good ventilation, whereas VOC can linger for an extended period. 
\end{takeaway}

\subsection{Inadequate Ventilation}
\begin{figure}
	\captionsetup[subfigure]{}
	\begin{center}
		\subfloat[Saturation of pollutants\label{fig:vent_all_time}]{
			\includegraphics[width=0.6\columnwidth,keepaspectratio]{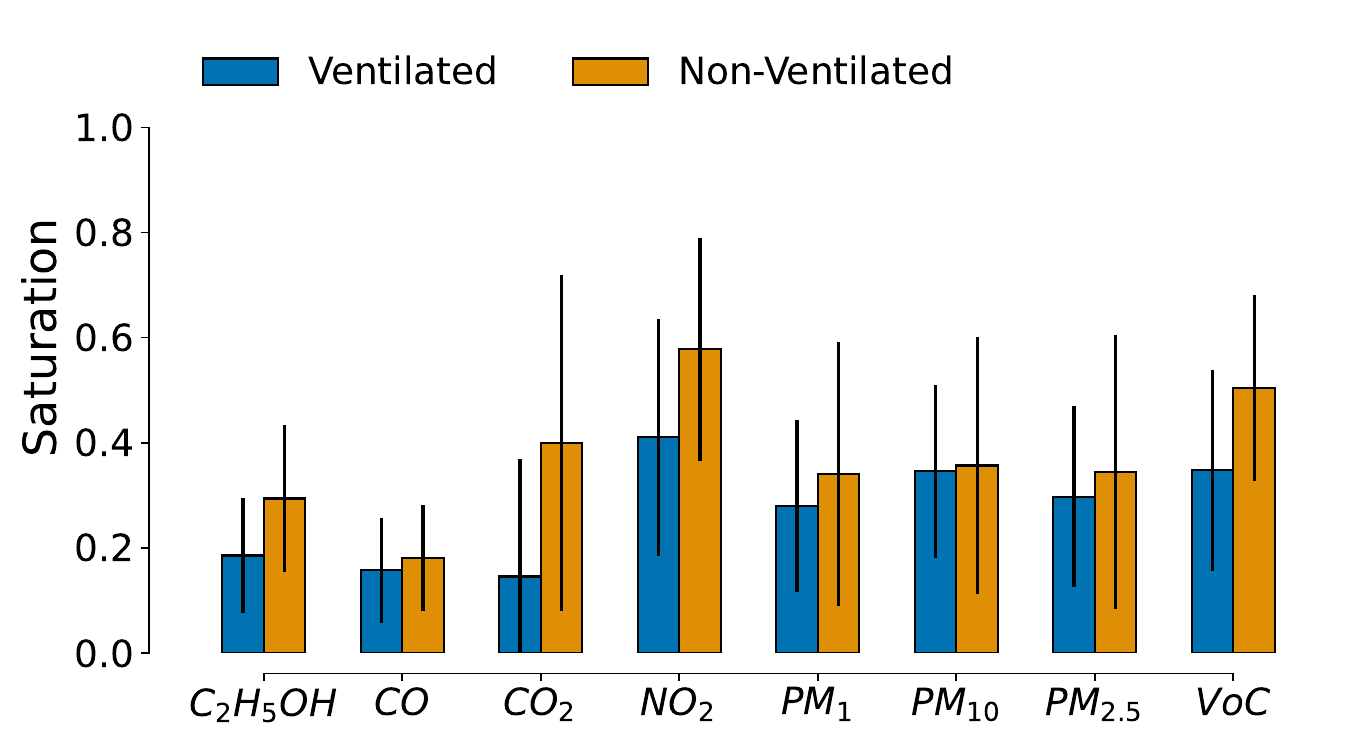}
		}\
		\subfloat[CO\textsubscript{2} concentration \label{fig:vent_co2_time}]{
			\includegraphics[width=0.48\columnwidth,keepaspectratio]{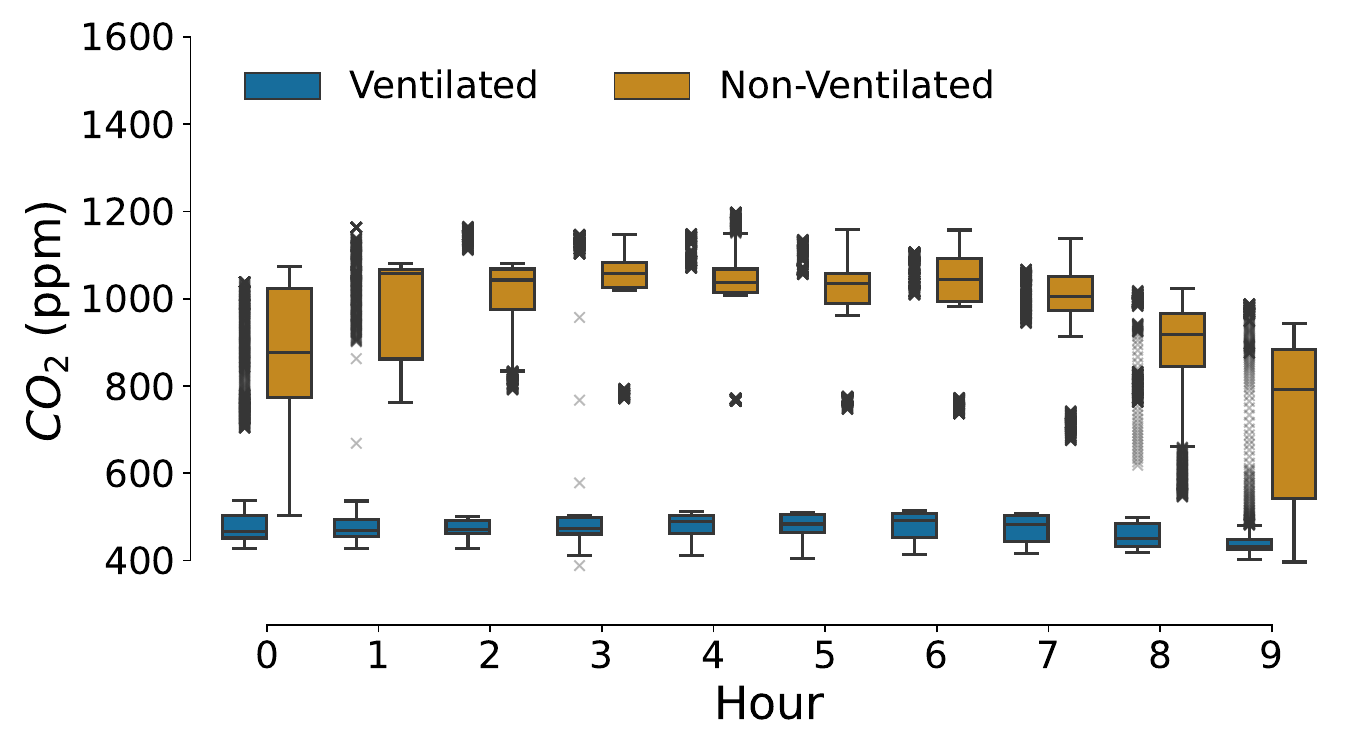}
		}
		\subfloat[VOC concentration\label{fig:vent_voc_time}]{
			\includegraphics[width=0.48\columnwidth,keepaspectratio]{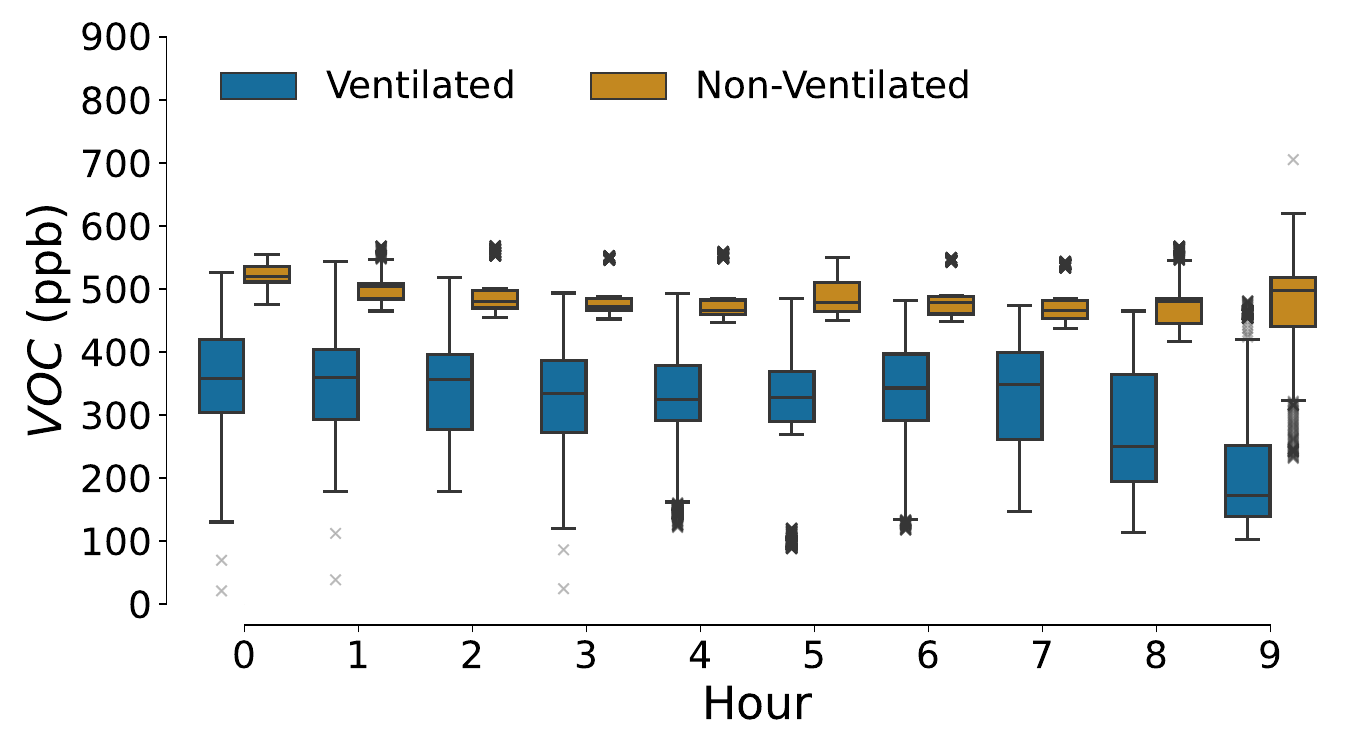}
		}\
	\end{center}
	\caption{Ventilated vs. non-ventilated scenario.}
	\label{fig:vent_time}
\end{figure}

\begin{figure}
	\captionsetup[subfigure]{}
	\begin{center}
			\subfloat[CO\textsubscript{2} concentration\label{fig:hm_co2_time}]{
				\includegraphics[width=0.48\columnwidth,keepaspectratio]{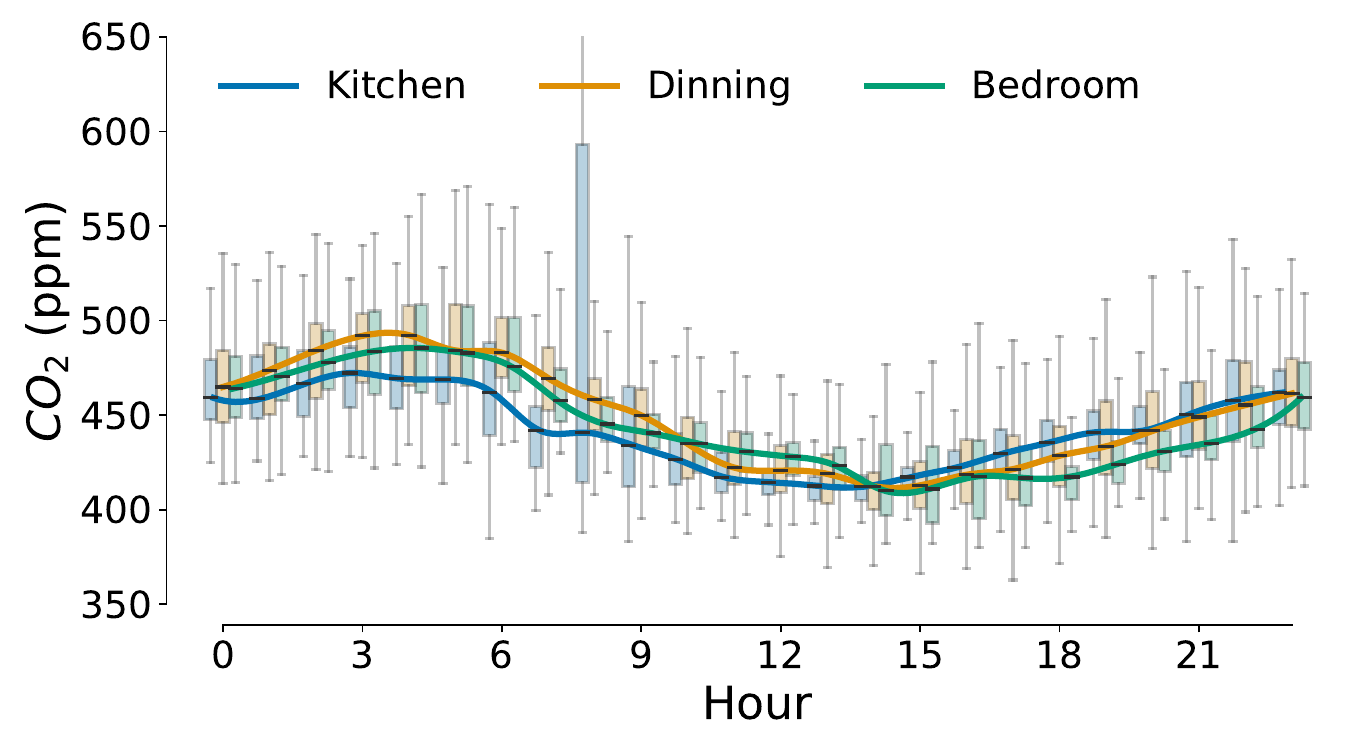}
			}
			\subfloat[VOC concentration\label{fig:hm_voc_time}]{
				\includegraphics[width=0.48\columnwidth,keepaspectratio]{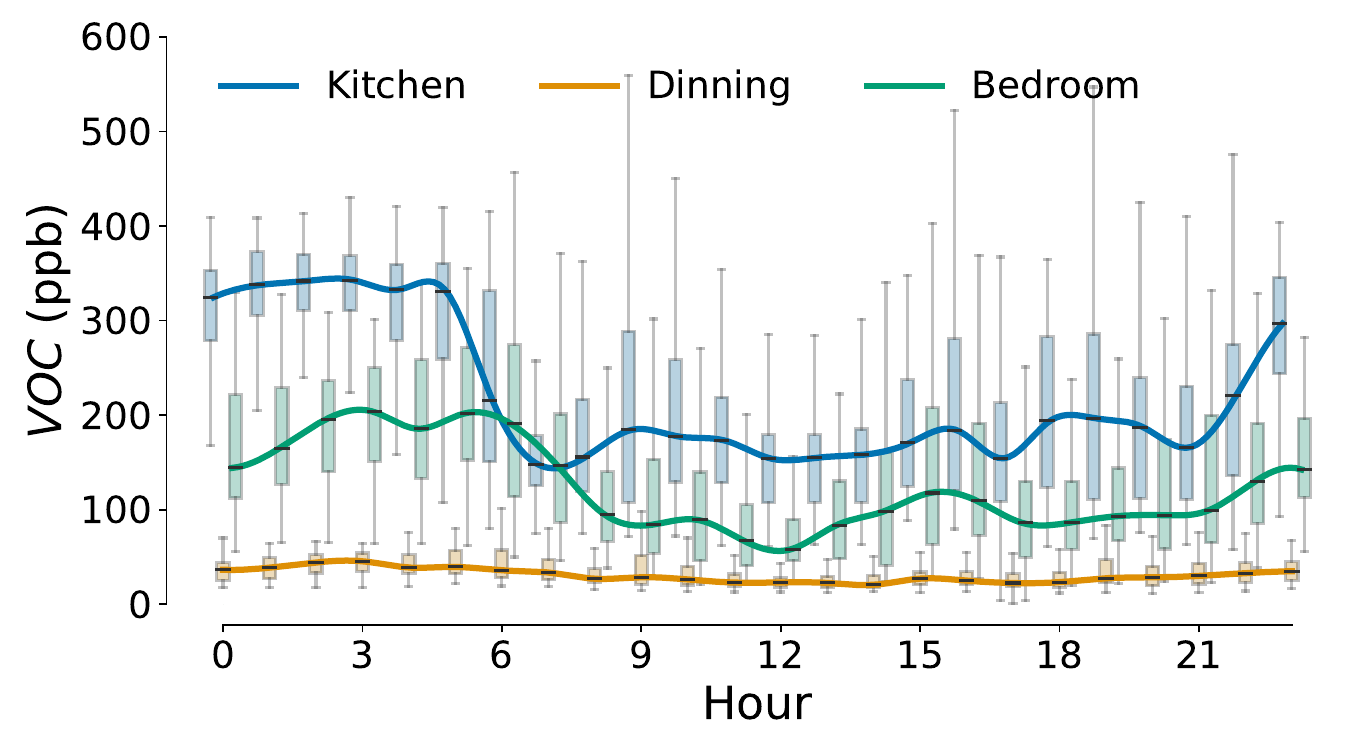}
			}\
	\end{center}
	\caption{Pollutants in relatively well ventilated H1.}
	\label{fig:vent_well}
\end{figure}

\figurename~\ref{fig:vent_all_time} depicts the degree of pollutant accumulation in non-ventilated versus ventilated scenarios for the bedroom (E) in H2-F1 (see floor plan in \figurename~\ref{fig:sc_house_f1}). We observe that each pollutant is impacted differently by the ventilation. In \figurename~\ref{fig:vent_co2_time} and~\ref{fig:vent_voc_time}, we observe the data distribution from midnight to early morning when pollutants are accumulated due to occupants sleeping in the room while the AC is turned on, keeping the windows closed. \figurename~\ref{fig:vent_co2_time} shows that the occupants experience on average two times CO\textsubscript{2} exposure due to poor ventilation. Similarly, VOC accumulates approximately $1.3$ times more strongly in lack of ventilation as depicted in \figurename~\ref{fig:vent_voc_time}. In contrast, we observe relatively less accumulation of pollutants in H1 (see floor plan in \figurename~\ref{fig:hm_house_marked}) due to the open slider windows around the building, even at night, providing adequate ventilation. As per \figurename~\ref{fig:vent_well}, we notice that the overall CO\textsubscript{2} and VOC distributions of H1 stay around healthy levels throughout the day.
\begin{takeaway}{}{tw2}
  Ventilation significantly impacts the pollutant contamination in a room. 
\end{takeaway}

\subsection{Complex Pollutant Dynamics}
\label{sec:complex_pollution}
\begin{figure}
	\centering
	\includegraphics[width=0.6\columnwidth]{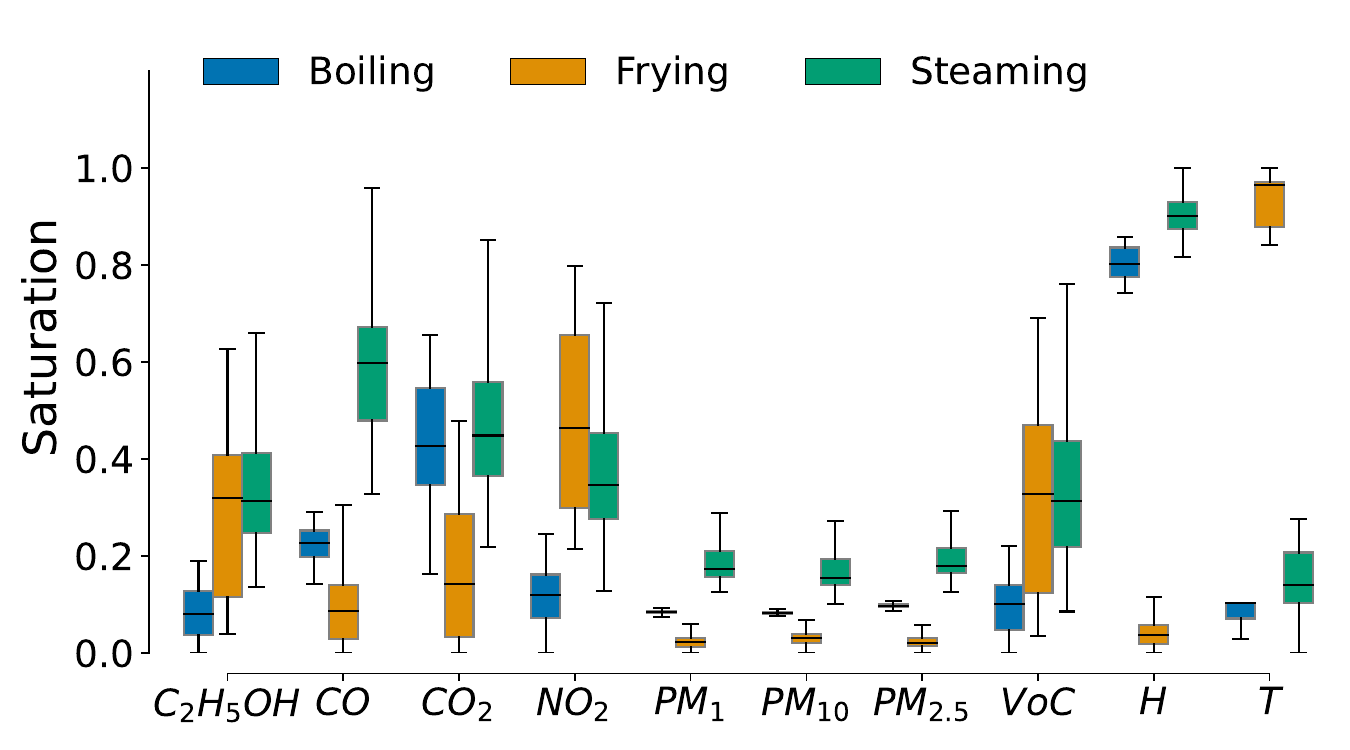}
	\caption{Saturation level of the pollutants, humidity, and temperature for different methods of cooking.}
	\label{fig:poll_cook}
\end{figure}

Pollutants can exhibit entirely different distributions based on how an activity is performed. Thus, the context of the activity like \textit{``What is being cooked''} or \textit{``Which detergent is used while cleaning the floor''} is more critical for characterizing which pollutants will majorly contaminate the indoor environment. To realize such complex pollutant dynamics, we observe three types of cooking activity, namely \textit{boiling}, \textit{frying}, and \textit{steaming}, which have significantly dissimilar pollutant signatures as shown in \figurename~\ref{fig:poll_cook}. 

In the case of \textit{boiling}, we can see an increase in the humidity, while most of the pollutants are dormant except CO and CO\textsubscript{2} as the kitchen's ventilation system is usually underused, resulting in accumulation of such gases. Whereas, \textit{frying} emits a lot of C\textsubscript{2}H\textsubscript{5}OH, NO\textsubscript{2}, VOC, and increases the temperature in the kitchen; thus, the exhaust fan is generally turned on, significantly lowering the concentration of CO, CO\textsubscript{2} and particulate matters (PM\textsubscript{x}). Unlike \textit{frying}, \textit{steaming} does not increase temperature significantly, leading to underutilization of the exhaust fan, as we observed from our dataset. However, unlike \textit{boiling}, \textit{steaming} emits lots of pollutants such as C\textsubscript{2}H\textsubscript{5}OH, NO\textsubscript{2}, PM\textsubscript{x}, VOC that accumulates throughout the activity. Moreover, due to lack of ventilation, CO and CO\textsubscript{2} also accumulate, resulting highest pollution exposure among the three cooking activities.

\begin{takeaway}{}{tw3}
The pollutants emitted and the general human response will vary depending on the activity. Humans are more sensitive to temperature change, so they know the need to turn on ventilation, but they can not feel or sense pollutants accumulating around them, resulting in unintentional exposure.
\end{takeaway}

% \begin{figure}
% 	\captionsetup[subfigure]{}
% 	\begin{center}
% 		\begin{minipage}{0.48\columnwidth}
% 			\subfloat[\label{fig:hm_house}]{
% 				\includegraphics[width=\columnwidth]{Figures/Diagrams/PK.png}
% 			}\
% 		\end{minipage} \hfil
% 		\begin{minipage}{0.47\columnwidth}
% 			\subfloat[\label{fig:hm_co2_time}]{
% 				\includegraphics[width=\columnwidth,keepaspectratio]{Figures/Motiv/hm_co2_time.pdf}
% 			}\
% 			\subfloat[\label{fig:hm_voc_time}]{
% 				\includegraphics[width=\columnwidth,keepaspectratio]{Figures/Motiv/hm_voc_time.pdf}
% 			}\
% 		\end{minipage}
% 	\end{center}
% 	\caption{Concentration of pollutants such as (b) CO\textsubscript{2}, (c) VOC in relatively well ventilated Deployment Site (a) S\textsubscript{3}}
% 	\label{fig:vent_well}
% \end{figure}

\subsection{Spatial Spread of Pollutants}
To realize the spread of different pollutants from the kitchen to other rooms, we consider three ventilation conditions in H1 (see \figurename~\ref{fig:hm_house_marked}) in the same \textit{cooking} circumstances based on the status of the kitchen exhaust fan and the ceiling fan in the dining, namely \textbf{ventilated} (exhaust on-ceiling off), \textbf{Naturally-ventilated} (exhaust off-ceiling off), and \textbf{pull inward} (exhaust off-ceiling on).

\begin{figure}
	\captionsetup[subfigure]{}
	\begin{center}
		\subfloat[Ventilated\label{fig:kit_best}]{
			\includegraphics[width=0.33\columnwidth,keepaspectratio]{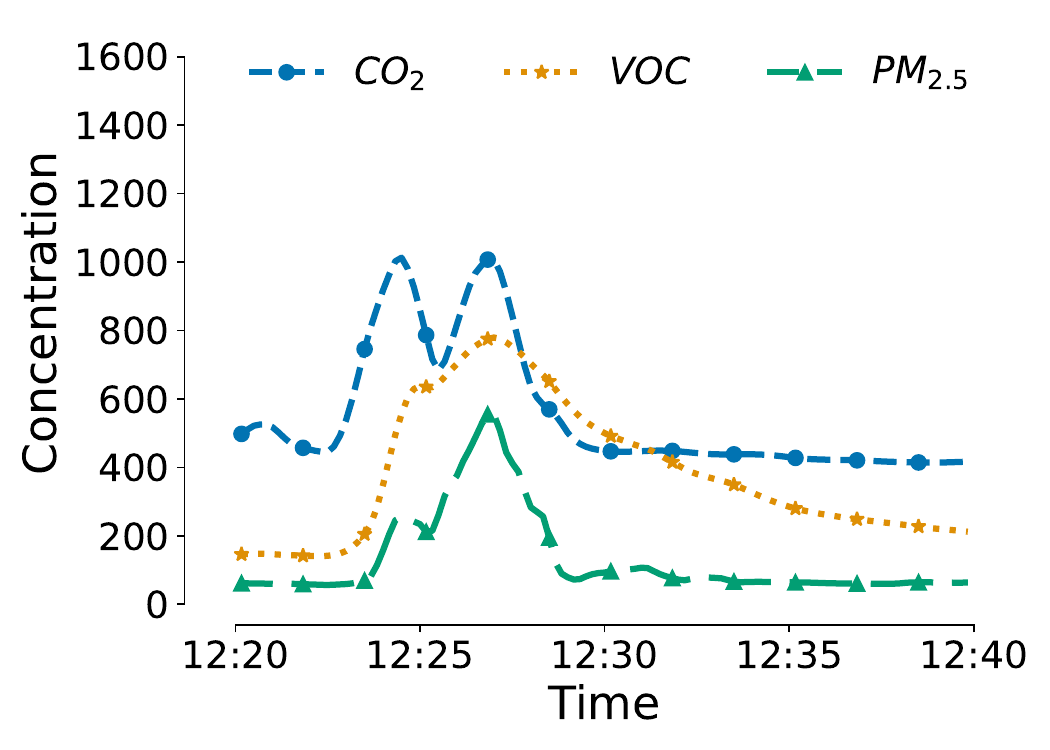}
		}
		\subfloat[Naturally-ventilated\label{fig:kit_medium}]{
			\includegraphics[width=0.33\columnwidth,keepaspectratio]{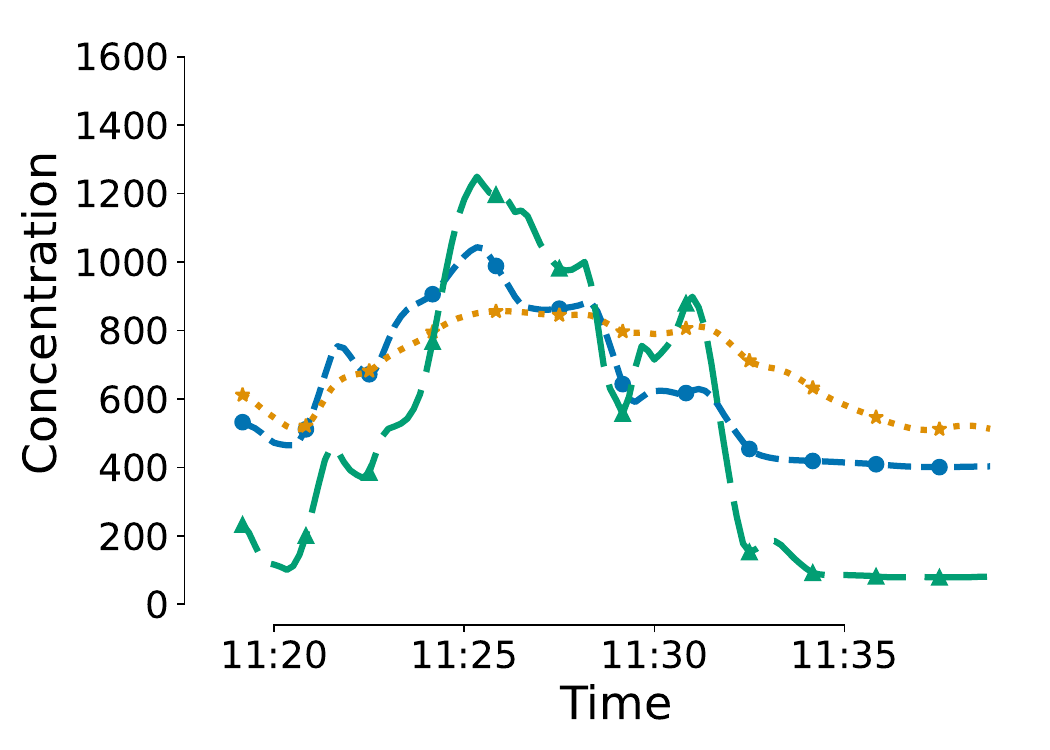}
		}
		\subfloat[Pull inward\label{fig:kit_worst}]{
			\includegraphics[width=0.33\columnwidth,keepaspectratio]{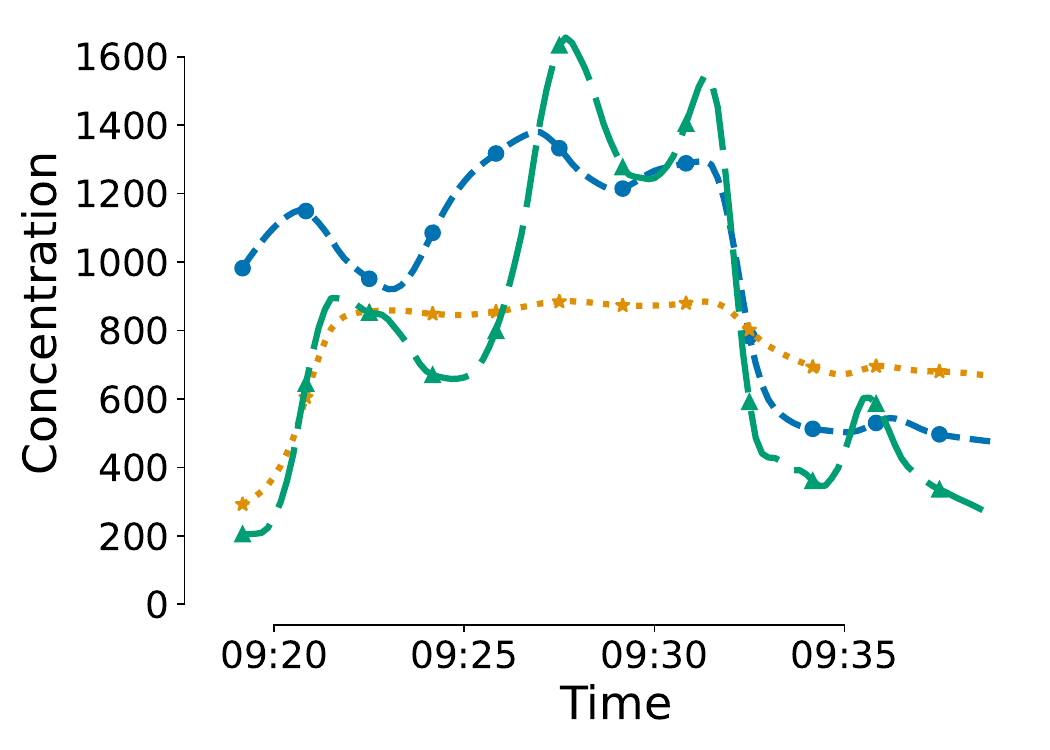}
		}\
		\subfloat[Ventilated\label{fig:rm_best}]{
			\includegraphics[width=0.33\columnwidth,keepaspectratio]{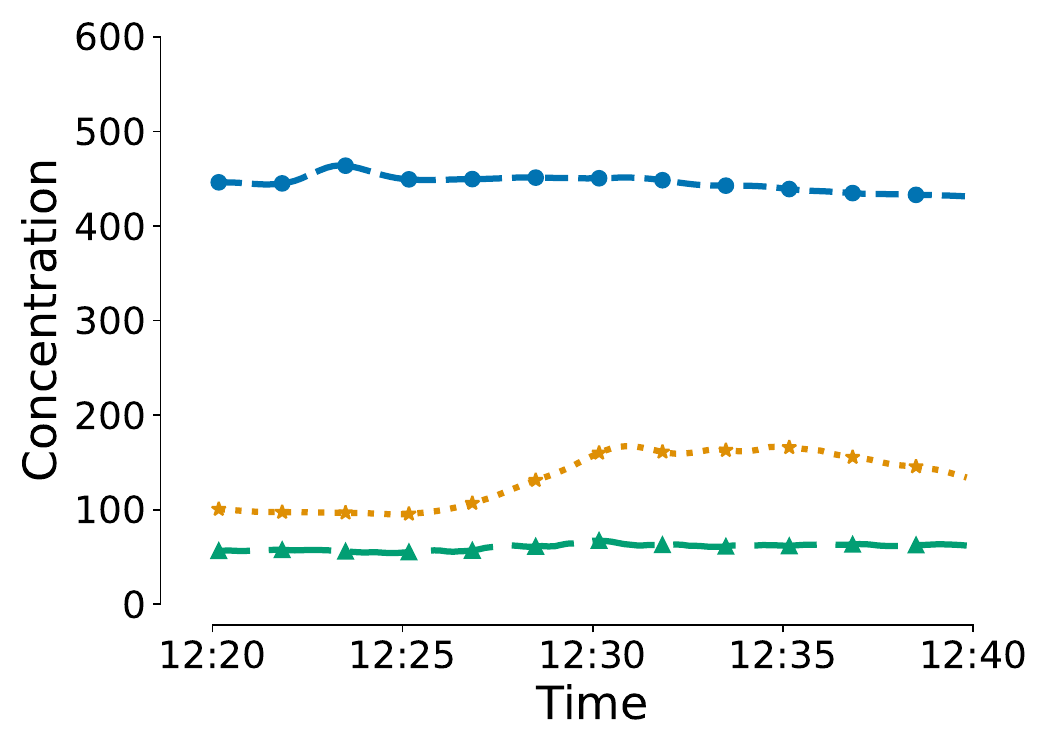}
		}
		\subfloat[Naturally-ventilated\label{fig:rm_medium}]{
			\includegraphics[width=0.33\columnwidth,keepaspectratio]{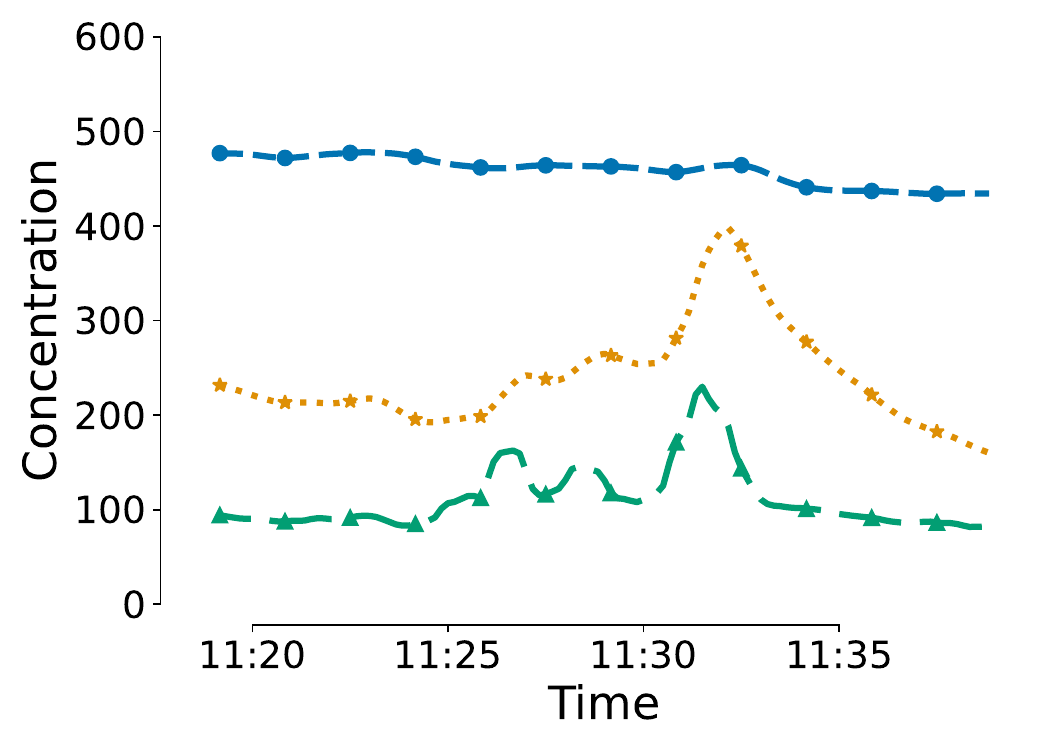}
		}
		\subfloat[Pull inward\label{fig:rm_worst}]{
			\includegraphics[width=0.33\columnwidth,keepaspectratio]{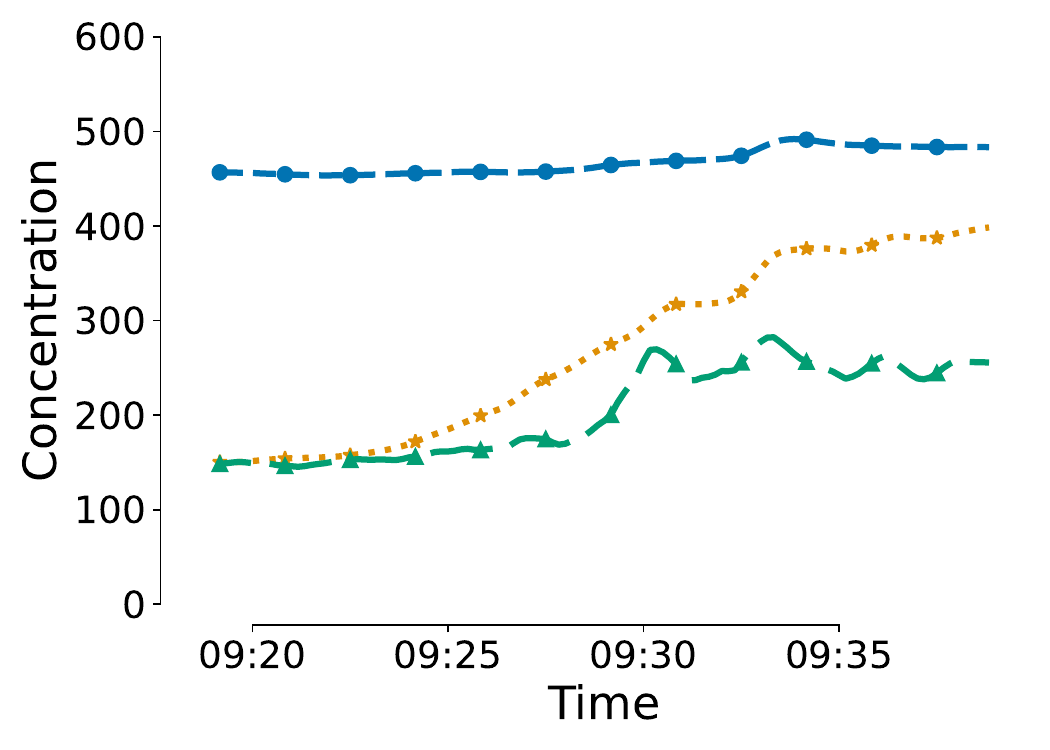}
		}\
		\subfloat[Ventilated\label{fig:din_best}]{
			\includegraphics[width=0.33\columnwidth,keepaspectratio]{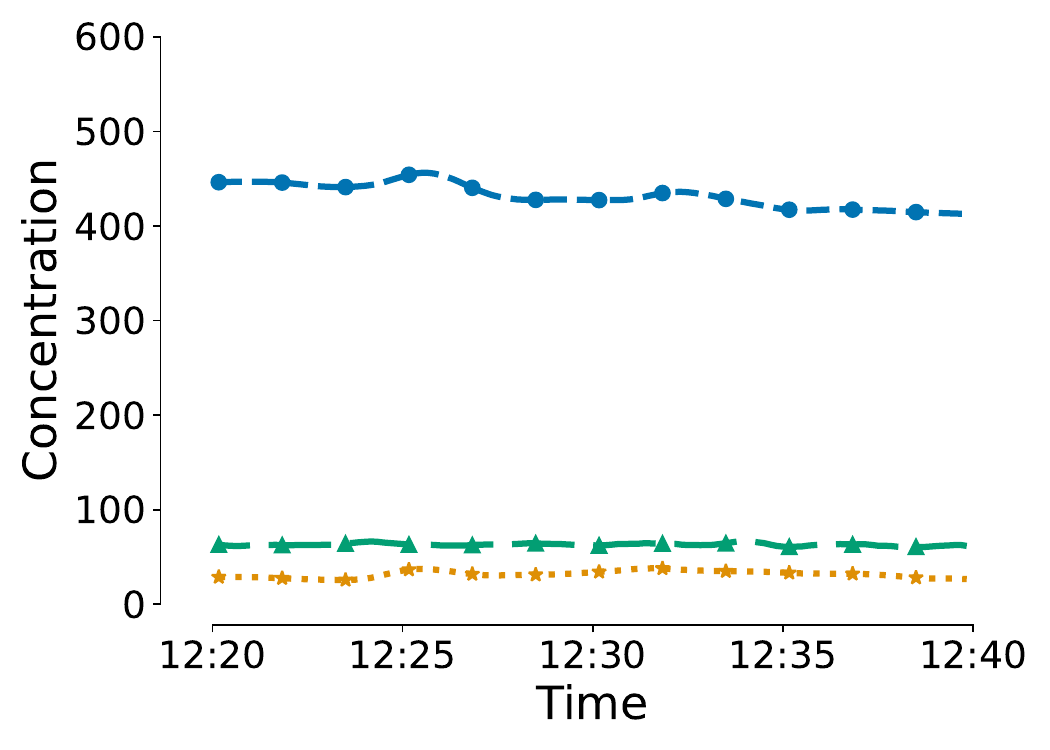}
		}
		\subfloat[Naturally-ventilated\label{fig:din_medium}]{
			\includegraphics[width=0.33\columnwidth,keepaspectratio]{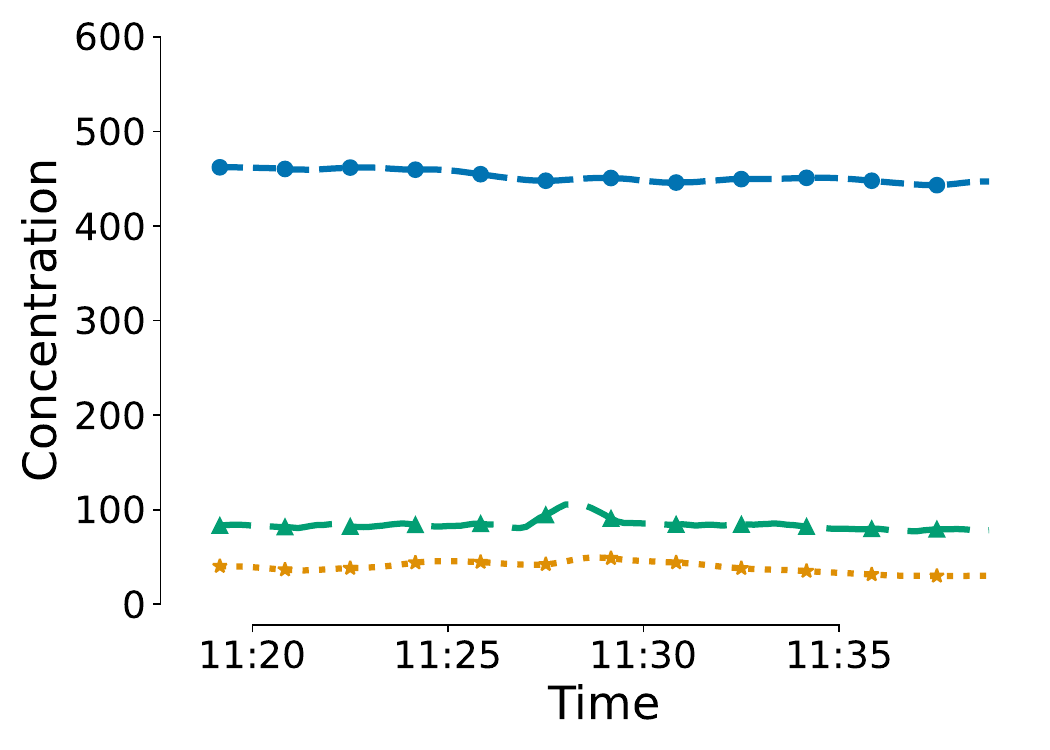}
		}
		\subfloat[Pull inward\label{fig:din_worst}]{
			\includegraphics[width=0.33\columnwidth,keepaspectratio]{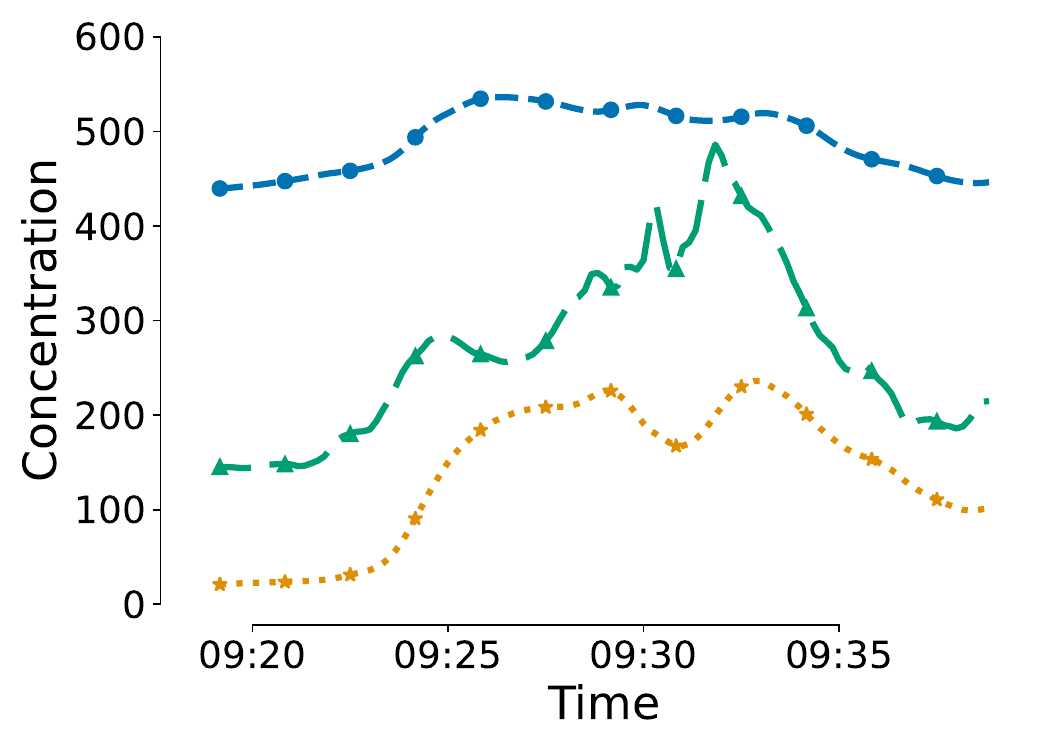}
		}\
	\end{center}
	\caption{Spread of pollutants in H1 due to cooking in (a)(b)(c) Kitchen, (d)(e)(f) Side by Room, (g)(h)(i) Dining. for different ventilation conditions.}
	\label{fig:spread}
\end{figure}

\noindent$\bullet\;\;\textbf{Ventilated}$: As shown in \figurename~\ref{fig:kit_best},~\ref{fig:rm_best},~\ref{fig:din_best}, despite lots of pollutants being released during cooking, the kitchen witness lowest levels among three cases as the kitchen exhaust is on, providing the adequate ventilation. Since the dining fan is off, pollutants are not pulled towards the side by the bedroom and dining, resulting in less impact in both areas.

\noindent$\bullet\;\;\textbf{Naturally-ventilated}$:
As shown in \figurename~\ref{fig:kit_medium},~\ref{fig:rm_medium},~\ref{fig:din_medium}, keeping both the exhaust and the dining fan off allows the pollutants to naturally spread further in the indoor environment. Within the next $10$ minutes of the started cooking, the side-by bedroom experiences a significant increase in the VOC and PM\textsubscript{2.5}. In addition, a slight increase in pollutant concentration was seen in the dining area. Most CO\textsubscript{2} gets ventilated through the kitchen's open window. Moderate ventilation is provided by keeping the exhaust and dining fans off, and the overall air quality is slightly degraded.

\noindent$\bullet\;\;\textbf{Pull inward}$:
As shown in \figurename~\ref{fig:kit_worst},~\ref{fig:rm_worst},~\ref{fig:din_worst}, the dining fan pulls pollutants inwards when the exhaust is off, resulting in maximum spread throughout the household. Interestingly, dining saw a sharp increase in pollutants, whereas side-by-side bedrooms gradually increased, and pollutants lingered for prolonged periods after cooking. Even with opened kitchen window, CO\textsubscript{2} is pulled into the dining. Therefore, keeping the dining fan on and the kitchen exhaust off will result in worst ventilation and adversely affect air quality.
\begin{takeaway}{}{tw4}
    Pollutants spread as a result of complex air circulation patterns. Different rooms may observe different patterns depending on how pollutants are pulled in. 
\end{takeaway}

\subsection{Can AQI Capture Such Dynamics?}
AQI is primarily designed considering the outdoor environment where pollutants do not accumulate and quickly spread over the surrounding regions based on wind flow and direction~\cite{gao2016mosaic,feng2017estimate,aula2022evaluation}. Moreover, AQI is measured for a specific pollutant (i.e., PM\textsubscript{2.5}, CO, O\textsubscript{3}, etc.), and the concentration range-bins are decided by the authorities of each country~\cite{lin2018calibrating,concas2021low}. Since outdoor pollutants are primarily influenced by weather, manufacturing, and road traffic, reducing pollution sources is the only way to improve the air~\cite{aula2022evaluation,ramachandran2019immersive,zheng2013u,cheng2014aircloud,liu2018third,montrucchio2020densely}. Therefore, AQI represents the average pollution exposure for a particular pollutant for $24$ hours. As observed above, indoor pollution varies rapidly, and major pollutants shift throughout the day depending on the occupants' activities. Thus, such a weak formulation is insufficient for indoor pollution because its variation depends on many contextual factors such as indoor structure, ventilation, and activity. 

\begin{figure}
	\captionsetup[subfigure]{}
	\begin{center}
		\subfloat[IAQI~\cite{mujan2021development}\label{fig:IAQI}]{
			\includegraphics[width=0.48\columnwidth,keepaspectratio]{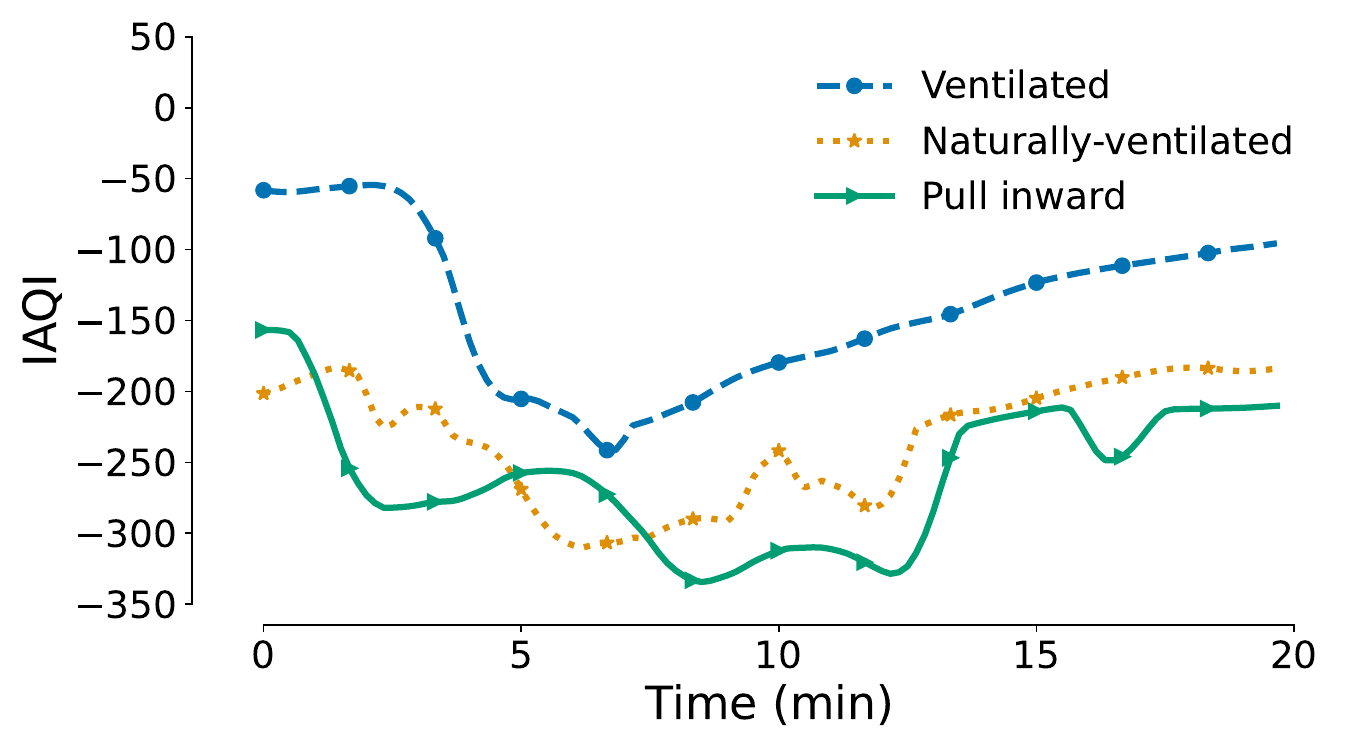}
		}
		\subfloat[Breeze IAQI~\cite{breeze}\label{fig:brz_IAQI}]{
			\includegraphics[width=0.48\columnwidth,keepaspectratio]{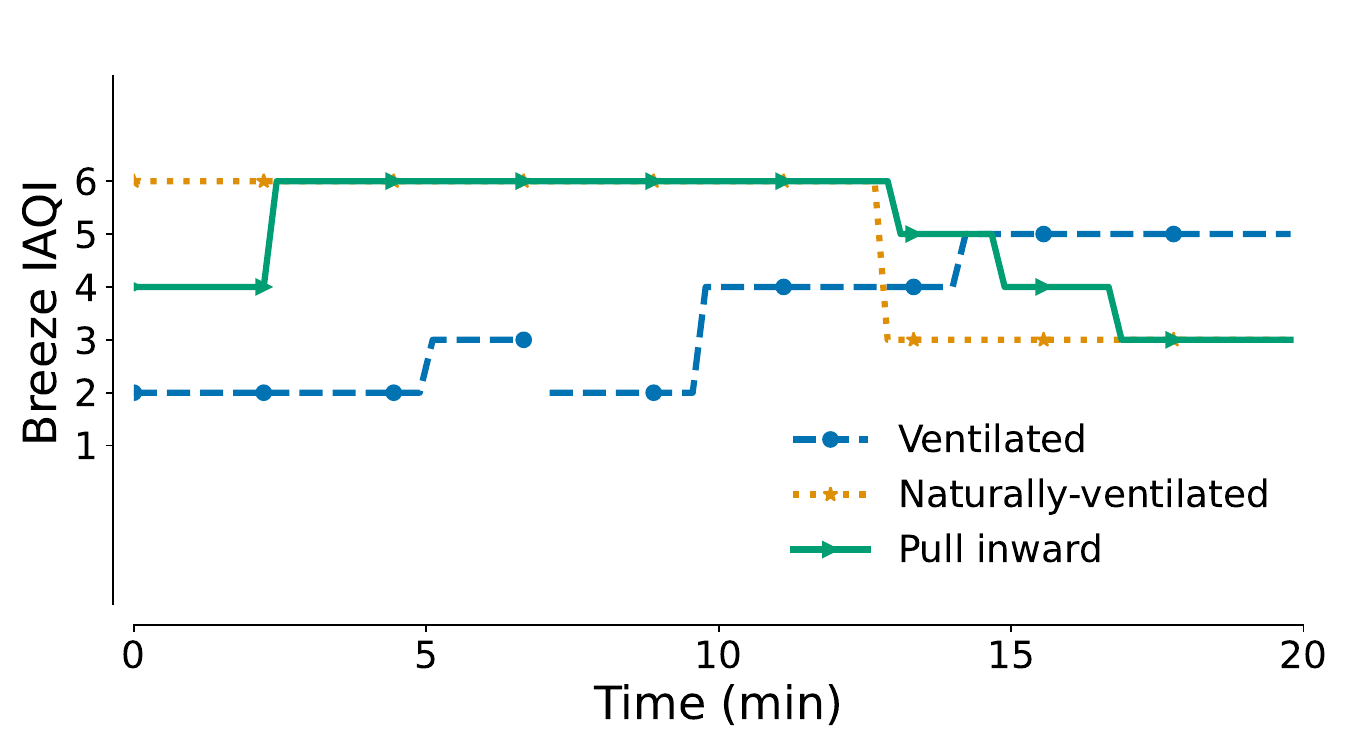}
		}\
	\end{center}
	\caption{Existing indoor air quality index.}
	\label{fig:base_metrics}
\end{figure}
Researchers have proposed Indoor Air Quality Index~\cite{mujan2021development} (IAQI), inspired by the AQI~\cite{pramanik2023aquamoho}. In addition, Breeze Technologies, an industry leader in air quality sensing, formulates actionable IAQI differently as their measure of indoor air quality. \figurename~\ref{fig:base_metrics} shows the IAQI and Breeze's IAQI, computed for all three ventilation scenarios from \figurename~\ref{fig:spread}. As shown in \figurename~\ref{fig:IAQI}, IAQI only considers local exposure in the kitchen and is completely unaware of the causal spread towards the bedroom and dining, thus underestimating the severity of poor ventilation. Whereas, \figurename~\ref{fig:brz_IAQI}, shows that Breeze's IAQI is saturated most of the time in case of naturally-ventilated and pull inward scenarios and unable to capture the difference in the spread and lingering of the pollutants. Thus, existing IAQI formulations are insufficient to characterize indoor complex pollution dynamics. Building upon this understanding, we formulate a metric to better capture indoor pollution dynamics under different scenarios.

\section{Healthy Home Index}
\label{sec:hhi}
The \textit{Healthy Home Index} (HHI) is a comprehensive measure of the quality of the indoor environment equipped with multiple sensing modules. HHI considers average exposure, statistical variations, and the spread of various pollutants over time to estimate a robust indicator of indoor health. Key constituent factors of the HHI are the following.

\noindent\textbf{Average level of pollutants:} The HHI considers several key pollutants that can affect indoor air quality, as we see in Figure~\ref{fig:vent_time}. These include PM2.5 and PM10 (fine particulates that can affect respiratory health), volatile organic compounds (VOCs, chemicals that can be emitted from a wide range of household activities such as cooking, eating, cleaning etc), carbon oxide (COx, high concentration of which can indicate poor ventilation), humidity levels, and temperature.   % (extreme humidity or temperatures can affect comfort and health).

\noindent\textbf{Statistical Variation:}  In addition to the average pollutant levels, the HHI incorporates the statistical properties of each pollutant as we observed in Section~\ref{sec:pilot}. It looks at the maximum and minimum concentrations observed, indicating the range of pollutant levels and their saturation, as observed in Figure~\ref{fig:poll_cook}. Moreover, it considers the standard deviation, which indicates the extent of fluctuation in pollutant levels and incorporates the rate of change in pollutant levels. This accounts for how swiftly pollutants contaminate a room (as we observe in Figure~\ref{fig:poll_time}). Furthermore, the HHI considers the number of times the level exceeded the unsafe threshold (Peak Count), the total duration of the peak (Peak Duration), and the duration for which a certain pollutant level remained above a safe threshold (Long Stay). As a result, these elements provide a more detailed picture of how pollutants linger and get trapped depending on various room structures, ventilation, and, accordingly, the inward airflow pattern (as observed in Figure~\ref{fig:spread}). 
%These statistical properties also indirectly indicate the ventilation of that space and the spatio-temporal variation of pollutants across multiple sensing modules.
\subsection{Detail of Statistical Properties}
\label{sec:stat_props}
% \SC{Do you compute these statistics within some window? What is the nature of this window -- sliding? How do you tune this window? [A window must be there as too old information might not be suitable]}
We compute several statistical properties over a sliding window of duration $\tau$ to capture abrupt changes, spread, lingering, etc., of the pollutants. The properties are as follows.

\noindent$\bullet\;\;$\textbf{Maximum and Minimum ($\mathbf{max}$, $\mathbf{min}$)}: These reflect the highest and lowest levels of pollutants recorded indoors. High maximum values could indicate very poor air quality episodes, potentially harmful to health. Very low minimum values suggest good ventilation. % and effective removal of pollutants.
    
\noindent$\bullet\;\;$\textbf{Standard Deviation ($\mathbf{std}$)}: This measures how much the pollutant levels fluctuate. A high standard deviation indicates large swings in pollutant levels, due to inconsistent ventilation or sporadic pollutant sources (like cooking). %Large fluctuations in pollutants could exacerbate health issues, particularly for respiratory patients.

\noindent$\bullet\;\;$\textbf{Rate of Change ($\mathbf{roc}$)}: This measures how quickly pollutant levels rise or fall. Rapid increases might occur if there's a sudden release of pollutants (e.g., burning food), while rapid decreases might indicate effective ventilation. Sudden spikes in pollutants could pose health risks, particularly for sensitive individuals. We consider the rate of change of pollutants for both raising ($roc_{raise}$) and falling ($roc_{fall}$) edges.

\noindent$\bullet\;\;$\textbf{Peak Count ($\mathbf{peak_c}$)}: This is the number of times pollutant levels exceeds a certain unsafe threshold. Multiple peaks could indicate recurring sources of pollutants or inconsistent ventilation, and frequent exposure to high pollution levels can harm health.

\noindent$\bullet\;\;$\textbf{Peak Duration ($\mathbf{peak_{\Delta}}$)}: This measures the total time that pollutant levels were above a certain unsafe threshold. Longer spikes in pollutant concentration could mean a higher risk of health effects and point to issues with ventilation or persistent pollutant sources.

\noindent$\bullet\;\;$\textbf{Long Stay ($\mathbf{\Delta_{exc}}$)}: This represents the duration of moderate pollution levels above the safe threshold. Extended periods of moderate pollution indicate poor air quality for prolonged periods, which can harm health. It also suggests inadequate ventilation or persistent sources of pollutants.

%These statistical properties provide insight into how pollutant levels change over time, which can be as important as average pollutant levels to understand indoor healthiness. They can help identify issues like inconsistent ventilation or sporadic pollutant sources and provide a more nuanced view of potential health risks.

\subsection{Formulation of HHI}
HHI considers an adaptive and collaborative approach by consulting with multiple instances of the sensing module, whenever available, deployed in different strategic locations of a household. In the case of one sensing module, HHI estimates air quality around that sensor. When multiple sensing modules are deployed in different household rooms, HHI correlates the pollutant readings of individual modules to provide an overall picture of air quality across rooms.

Let the set of sensing modules $\mathcal{D}=\{d_i|i=1,2,\dots N\}$, where $N$ is the number of modules, and the set of pollutants and comfort factors $\mathcal{F}=\{$$CO_2$, $VOC$, $PM_{2.5}$, $PM_{10}$, $T$, $H$$\}$. Moreover, the set of statistical properties over $\tau$\footnote{We took $\tau=10$ minutes as per empirical observations.} minute sliding window of the pollutants $\mathcal{S}=\{$$min$, $max$, $std$, $roc_{raise}$, $roc_{fall}$, $peak_{c}$, $peak_{\Delta}$, $\Delta_{exc}$$\}$, as described in Section~\ref{sec:stat_props}. Based on the understanding of Section~\ref{sec:pilot}, we formulate HHI as the weighted sum of two components, normalized within [0, 1000], as shown in Equation~\eqref{eq:hhi}. The components resemble average pollution exposure ($\mathcal{C}_1$ ), and statistical variations and spread of pollutants ($\mathcal{C}_2$), in Equation~\eqref{eq:c1} and Equation~\eqref{eq:c2}, respectively. The product operation in the second component ($\mathcal{C}_2$) computes cross-correlation between statistical properties of the modules, representing the spread of pollutants throughout an indoor space.
\begin{gather}
    \label{eq:hhi}
    HHI=(1-\Phi(\lambda_1 \cdot \mathcal{C}_1 + \lambda_2 \cdot \mathcal{C}_2))\cdot 1000 \\
    \label{eq:c1}
    \mathcal{C}_1 =\frac{1}{(|\mathcal{F}|\cdot |\mathcal{D}|)} \sum_{f \in \mathcal{F}} \sum_{d \in \mathcal{D}} \Phi \left( \frac{1}{\tau} \sum_{i=1}^{\tau} f_d^{(i)} \right)\\
    \label{eq:c2}
    \mathcal{C}_2=\frac{1}{(|\mathcal{F}|\cdot |\mathcal{S}|)} \sum_{f \in \mathcal{F}} \sum_{\psi \in \mathcal{S}} \prod_{d \in \mathcal{D}} \Phi \left(\psi \left(\{f_d^{(i)}| i=1,2\dots \tau\}\right)\right) \\
    \text{Here,}~\Phi (\mathcal{A}) = \left\{\frac{max(\mathcal{A})-a_i}{max(\mathcal{A})-min(\mathcal{A})} | a_i \in \mathcal{A}\right\} \notag \\ 
    \text{for}~\mathcal{A}= \{a_1,a_2,\dots a_n\} \notag
\end{gather}
Note that, $f_d^{(i)}$ is the $i^{th}$ sample of pollutant $f\in\mathcal{F}$  of sensing module $d\in\mathcal{D}$. As shown in Eqn.~\eqref{eq:hhi}, HHI varies on a scale from 0 (worst) to 1000 (best). Thus, lower HHI values indicate a greater level of pollutants and more variation, suggesting poorer overall home health. The HHI comprehensively assesses indoor environment quality, considering the levels of key pollutants and their variation over time. This can help identify potential issues, understand the effectiveness of countermeasures to improve air quality, and compare the health of different indoor environments.
\section{Implementation and Evaluation}
\label{sec:eval}
We use the \ourmethod{} IoT platform (Section~\ref{sec:dalton}) for seamlessly collecting indoor pollutant data from $28$ deployment sites (Section~\ref{sec:datacollection}) to compute HHI and evaluate it in comparison with IAQI. \tablename~\ref{tab:ovl_spec} highlights the sensing module's system specification and operational bounds. The detail follows. 

%We first discuss the \ourmethod{} IoT Backbone architecture in detail that has been used to collect this data and subsequently analyze the performance and impact of HHI on indoor pollution measurement. 

\begin{table*}[!t]
	\centering
	\scriptsize
	\caption{Overall Specifications of DALTON Sensing Module}
	\label{tab:ovl_spec}
	\begin{tabular}{|l|l|l|l|l|l|c|c|c|c|c|} 
		\cline{1-3}\cline{5-11}
		\multicolumn{3}{|c|}{\textbf{System Specification}}                                                              & \multicolumn{1}{c|}{\textbf{}} & \multicolumn{2}{c|}{\multirow{2}{*}{\textbf{Sensor}}} & \multicolumn{5}{c|}{\textbf{Operational Details}}                                                                                                                                                                                                                                                                                                                                                         \\ 
		\cline{1-3}\cline{7-11}
		\multicolumn{2}{|l|}{Microprocessor}   & \begin{tabular}[c]{@{}l@{}}Xtensa®32-bit LX6\\Clock 80\textasciitilde{}240 MHz\end{tabular} & \multicolumn{1}{c|}{\textbf{}} & \multicolumn{2}{c|}{}                                 & \textbf{Range}                               & \textbf{Resolution}  & \textbf{Error Margin}                                                                                                                   & \begin{tabular}[c]{@{}c@{}}\textbf{Response}\\\textbf{Time}\end{tabular} & \begin{tabular}[c]{@{}c@{}}\textbf{Operational }\\\textbf{Temp \& RH}\end{tabular}                               \\ 
		\cline{1-3}\cline{5-11}
		\multirow{2}{*}{Memory} & ROM          & 448 KB                                                                  &                                & \multirow{3}{*}{DUST}   & PM\textsubscript{x}                         & 0\textasciitilde{}500 $\mu g/m^3$                    & 1                    & \multicolumn{1}{l|}{\begin{tabular}[c]{@{}l@{}}$\pm$ 10 $\mu g/m^3$ @0\textasciitilde{}100 $\mu g/m^3$\\$\pm$ 10\% @100\textasciitilde{}500 $\mu g/m^3$\end{tabular}} & \multirow{3}{*}{$\leq$10 s}                                                   & \multirow{3}{*}{\begin{tabular}[c]{@{}c@{}}-10\textasciitilde{}60 \textdegree C\\0\textasciitilde{}99\%\end{tabular}}  \\ 
		\cline{2-3}\cline{6-9}
		& SRAM         & 520 KB                                                                  &                                &                         & RH                          & 0\textasciitilde{}99 \%                      & \multirow{3}{*}{0.1} & \multicolumn{1}{l|}{$\pm$ 2\%}                                                                                                             &                                                                          &                                                                                                                \\ 
		\cline{1-3}\cline{6-7}\cline{9-9}
		\multicolumn{2}{|l|}{Connectivity}     & Wi-Fi 2.4GHz                                                            &                                &                         & T                           & -20\textasciitilde{}99 \textdegree C                 &                      & \multicolumn{1}{l|}{$\pm$ 0.5 \textdegree C}                                                                                                       &                                                                          &                                                                                                                \\ 
		\cline{1-3}\cline{5-7}\cline{9-11}
		\multicolumn{2}{|l|}{Scan Rate (Hz)}    & 1                                                                       &                                & \multirow{4}{*}{MCGS}   & NO\textsubscript{2}                         & 0.1\textasciitilde{}10 ppm                   &                      & \multirow{4}{*}{--}                                                                                                                      & \multirow{3}{*}{$\leq$30 s}                                                   & \multirow{6}{*}{\begin{tabular}[c]{@{}c@{}}-10\textasciitilde{}50 \textdegree C\\0\textasciitilde{}95\%\end{tabular}}  \\ 
		\cline{1-3}\cline{6-8}
		\multicolumn{2}{|l|}{Max Power (W)}        & 3.55                                                                       &                                &                         & C\textsubscript{2}H\textsubscript{5}OH                      & \multirow{2}{*}{1\textasciitilde{}500 ppm}   & \multirow{2}{*}{1}   &                                                                                                                                         &                                                                          &                                                                                                                \\ 
		\cline{1-3}\cline{6-6}
		\multicolumn{2}{|l|}{Max Current (mA)}     & 760                                                                     &                                &                         & VOC                         &                                              &                      &                                                                                                                                         &                                                                          &                                                                                                                \\ 
		\cline{1-3}\cline{6-8}\cline{10-10}
		\multicolumn{2}{|l|}{Dimensions(mm)}   & 112 $\times$ 112 $\times$ 55                                                          &                                &                         & CO                          & 5-5000 ppm                                   & 0.5                  &                                                                                                                                         & $\leq$10 s                                                                    &                                                                                                                \\ 
		\cline{1-3}\cline{5-10}
		\multicolumn{2}{|l|}{Weight (g)}       & 160                                                                     &                                & \multirow{2}{*}{MH-Z16} & \multirow{2}{*}{CO\textsubscript{2}}        & \multirow{2}{*}{0\textasciitilde{}10000 ppm} & \multirow{2}{*}{1}   & \multirow{2}{*}{$\pm$ 100ppm +6\%value}                                                                                                    & \multirow{2}{*}{$\leq$30 s}                                                   &                                                                                                                \\ 
		\cline{1-3}
		\multicolumn{2}{|l|}{Power Adapter}    & DC (5V, 15W)                                                            &                                &                         &                             &                                              &                      &                                                                                                                                         &                                                                          &                                                                                                                \\
		\cline{1-3}\cline{5-11}
	\end{tabular}
\end{table*}

\subsection{\ourmethod{} IoT Backbone}
\begin{figure}[!ht]
	\centering
	\includegraphics[width=0.9\columnwidth]{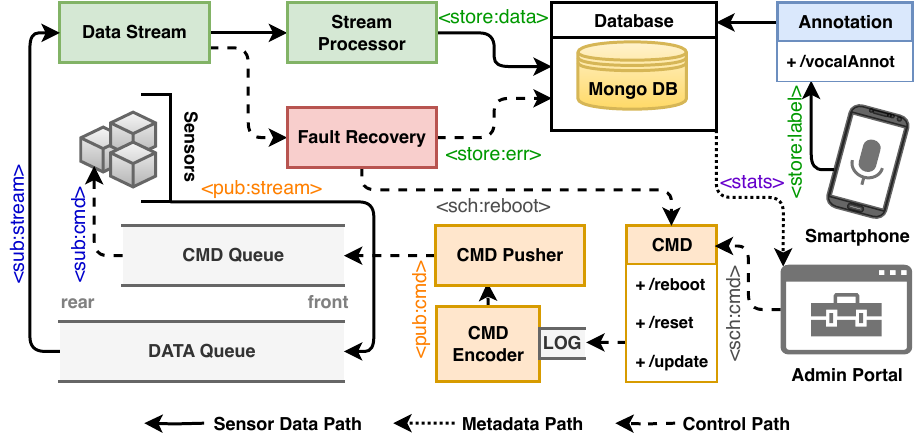}
	\caption{\ourmethod{} IoT Backbone Architecture.}
	\label{fig:backend}
\end{figure}

To handle multiple real-time data streams from a large set of sensing modules, we design the IoT backbone (shown in \figurename~\ref{fig:backend}) with multiple modular micro-services individually responsible for singular tasks. Further, we use publisher-subscriber based IP-agnostic design using MQTT\footnote{Eclipse Mosquitto (\url{https://mosquitto.org/})} to uniquely identify each device and access the data from there over an asynchronous communication channel.  

%The benefit of such a micro-service architecture is that it allows real-time data analytics and stream processing, which reduces the possibility of single points of failure. Below are the design choices and micro-services of the IoT backbone.

% \subsubsection{IP Agnostic Design}
% In real-world, large-scale IoT sensing networks with multiple local area networks, the public IP addresses of each module are not always accessible because of restrictions imposed by ISPs and limited expertise by end users. We chose a publisher-subscriber-based IP agnostic approach (through a micro-service described in the next subsection) where any sensing module can be uniquely identified. In addition, sensing modules and micro-services of the IoT backbone can establish an asynchronous communication channel. The design simplifies the initial setup for the end user and allows supervisory control over each module, allowing remote debugging, and reconfiguration.

%\subsubsection{Micro-services} 
In addition, the IoT backbone is comprised of several micro-services: \textbf{ (1) Queuing} service hosts an MQTT broker that is utilized as the message broker that manages the DATA queue and the command (CMD) queue as shown in \figurename~\ref{fig:backend} and ensures one-time FIFO delivery in the underline asynchronous wireless channel. \textbf{(2) Database} service hosts a MongoDB server that keeps track of four collections, namely data, error log, users, and annotation, to facilitate storage for the corresponding micro-services of the IoT backbone. \textbf{(3) Data Stream Processor} subscribes to the mixed data stream of the queuing service and splits it into individual streams from each sensing module. \textbf{(4) Command Pusher} exposes endpoints for remotely executing commands such as reboot, reset, reconfiguration, etc., in the sensing modules. \textbf{(5) Fault Recovery} is triggered by the stream processor upon detecting any anomaly, and it determines the suitable recovery action (i.e., reboot) for the specific device. \textbf{(6) Annotation} service exposes secured endpoints for tagging indoor activities for each occupant via a simple Android application named \textit{VocalAnnot} (details in Section~\ref{sec:datacollection}). \textbf{(7) Admin Portal} with a role-based access control, provides an interface for executing commands remotely, listing live modules that are sending data, viewing error logs, and recovery actions to debug any infrastructural or hardware issues (i.e., unstable power supply, damaged sensor, etc.). 

%In summary, \ourmethod{} is a very extensible IoT platform comprised of an  in-house sensing module and a robust IoT backbone to facilitate plug-and-play end-user experience. Next, we describe the field study with the developed platform for data collection.
\subsection{Hardware/Software Configurations}
%\subsubsection{Software setup}
Hardware details of the sensing module are given in Table~\ref{tab:ovl_spec}. The IoT backbone is hosted in Microsoft Azure on a Standard B2s Virtual Machine instance ($2$ CPU cores, $4$GB primary memory, $30$GB storage, with Ubuntu 20.04). We have used Python v3.9.1 and Flask v2.2.2 library to implement the IoT backbone. For handling the data and command stream, we have utilized Eclipse Mosquitto, an open-source lightweight (EPL/EDL licensed) message broker that implements the MQTT protocol version 5.0. Moreover, we have hosted a MongoDB 6.0 database to store the data. We have used Docker v20.10.23 to containerize and deploy the services over the cloud. Furthermore, Arduino IDE v2.1.0 uploads C++ programs to the sensing module.

%\subsubsection{Hardware setup}

%Moreover, we used an M1 iMac with $16$GB primary memory running MacOS v13.3.1 with kernel version 22.4.0 to analyze and experiment with the collected data.

\subsection{Evaluation metric}
To measure the similarity between average pollution exposure and the spread of pollutants with HHI, we have utilized normalized cross-correlation and the time lag between the change of sensor measurements and the HHI response.
\subsubsection{Cross-Correlation ($\mathcal{X}_r$)} It is a measure of the similarity of two signals as a function of displacement of one to the other. Cross-correlation for 1D signals is computed as the dot product of the sliding window of one signal with the other signal. Let we have two signals $x$ and $y$, having length of $|x|$ and $|y|$ ($|x| \leq |y|$), respectively. Then cross-correlation between these signals is expressed as shown in Eqn.~\eqref{eq:X_r}.
\begin{equation}
    \label{eq:X_r}
    \mathcal{X}_r(x,y)=\left\{ \sum_{l=0}^{|x|-1} x_l \cdot y^{pad}_{(l-k+|x|-1)},~\forall_{k \in \left[-|x|+1,|y|-1\right]}\right\}
\end{equation}
Here, $y^{pad}$ is derived with zero-padding ($|x|-1$ both side) signal $y$. Thus $\mathcal{X}_r$ has a total length of $(|x|+|y|-2)$.

\subsubsection{Time lag ($k$)} The variable $k$ in Eqn.~\eqref{eq:X_r} represents time lag between signal $x$ and $y$. Negative and positive $k$ value means leading and lagging cross-correlation. Higher $\mathcal{X}_r$ value with a positive time lag $k$ means signals $x$ and $y$ show similar (highly correlated) patterns after $k$ unit time. Notably, zero lag means the signals are not correlated. 

%Whereas, if the cross-correlation is maximized at $k=0$, both signals $x$ and $y$ vary.

%time delay in packet transmission through WiFi
%Power Usage by each module

\subsection{Evaluation}
This section describes the properties of HHI and shows that it is a comprehensive measure of both pollution exposure and spread throughout the indoors. Moreover, we establish that HHI highly correlates to IAQI when only one sensing module is deployed, thus indicating that it provides no less information than IAQI while capturing the indoor dynamics with multiple sensors' data. Lastly, we touch upon the actionable HHI levels and their indications in terms of pollutants.

\begin{figure}
	\captionsetup[subfigure]{}
	\begin{center}
		\subfloat[Ventilated\label{fig:hhi_best}]{
			\includegraphics[width=0.33\columnwidth,keepaspectratio]{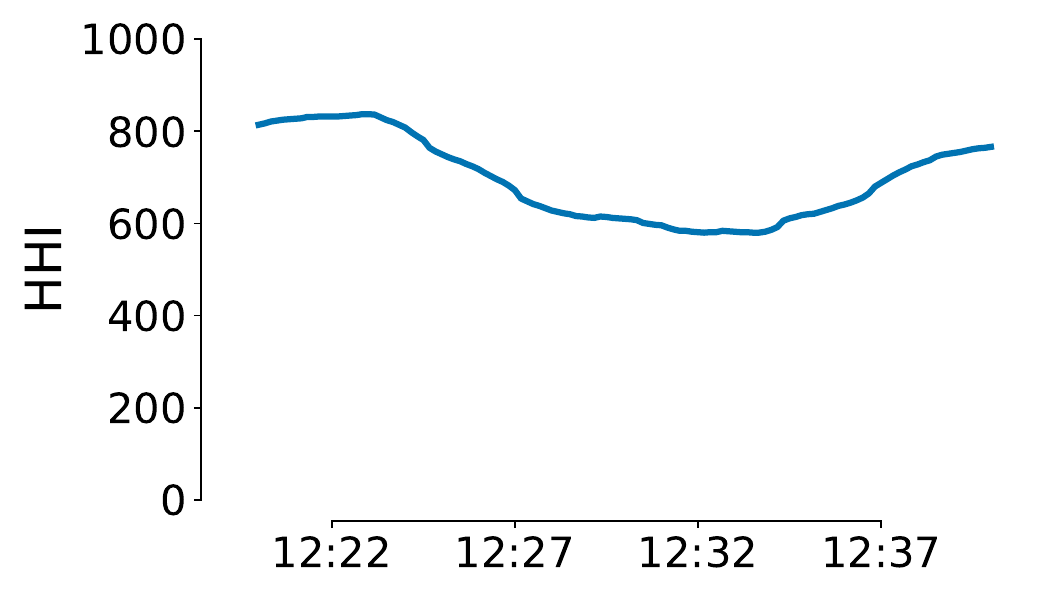}
		}
            \subfloat[Naturally-ventilated\label{fig:hhi_mod}]{
			\includegraphics[width=0.33\columnwidth,keepaspectratio]{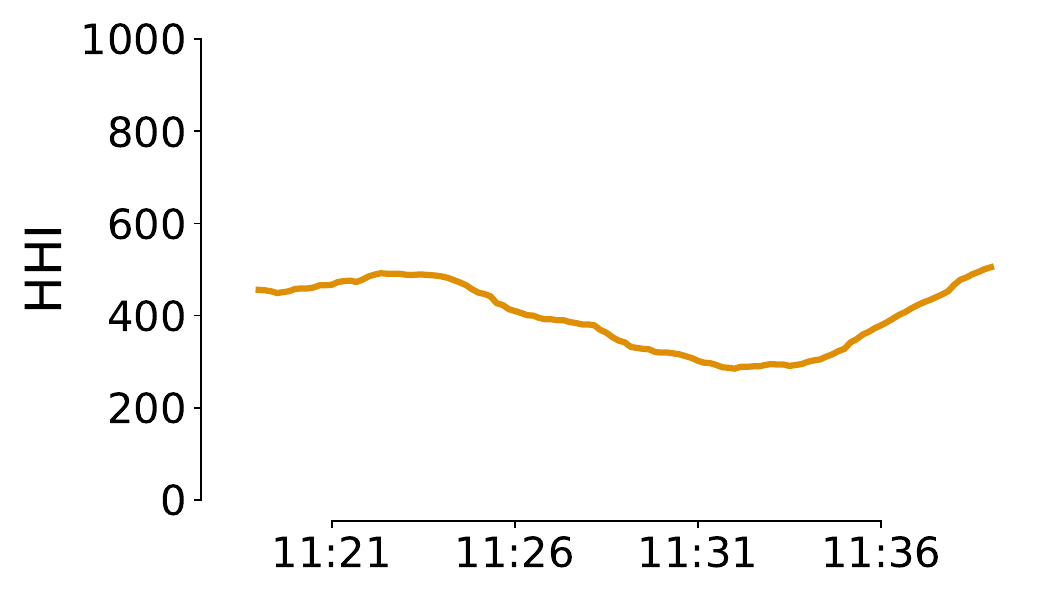}
		}
            \subfloat[Pull inward\label{fig:hhi_worst}]{
			\includegraphics[width=0.33\columnwidth,keepaspectratio]{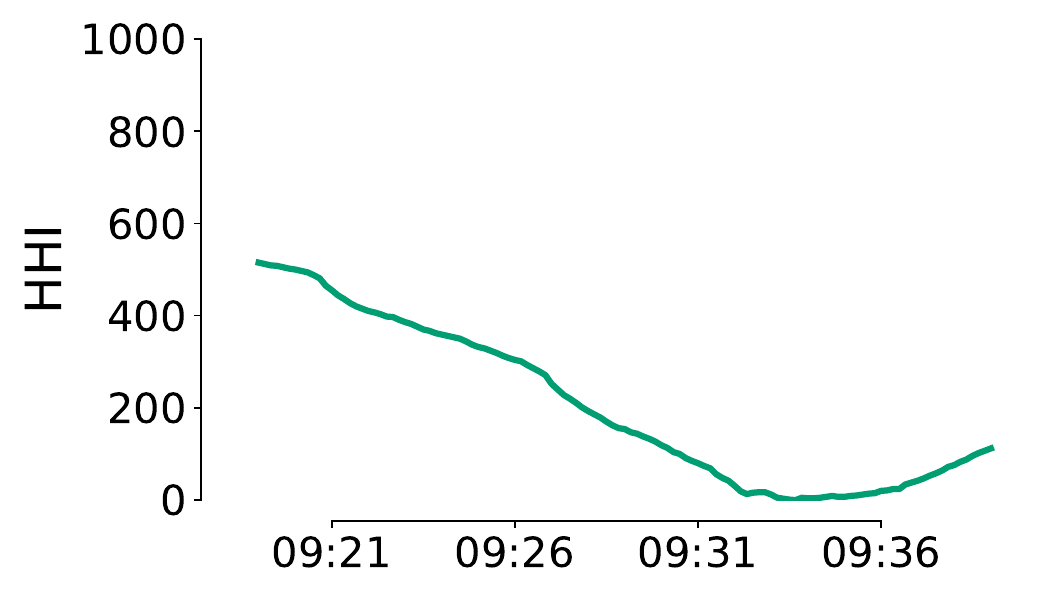}
		}\
		\subfloat[Cross-correlation\label{fig:cross_corr_hhi}]{
			\includegraphics[width=0.49\columnwidth,keepaspectratio]{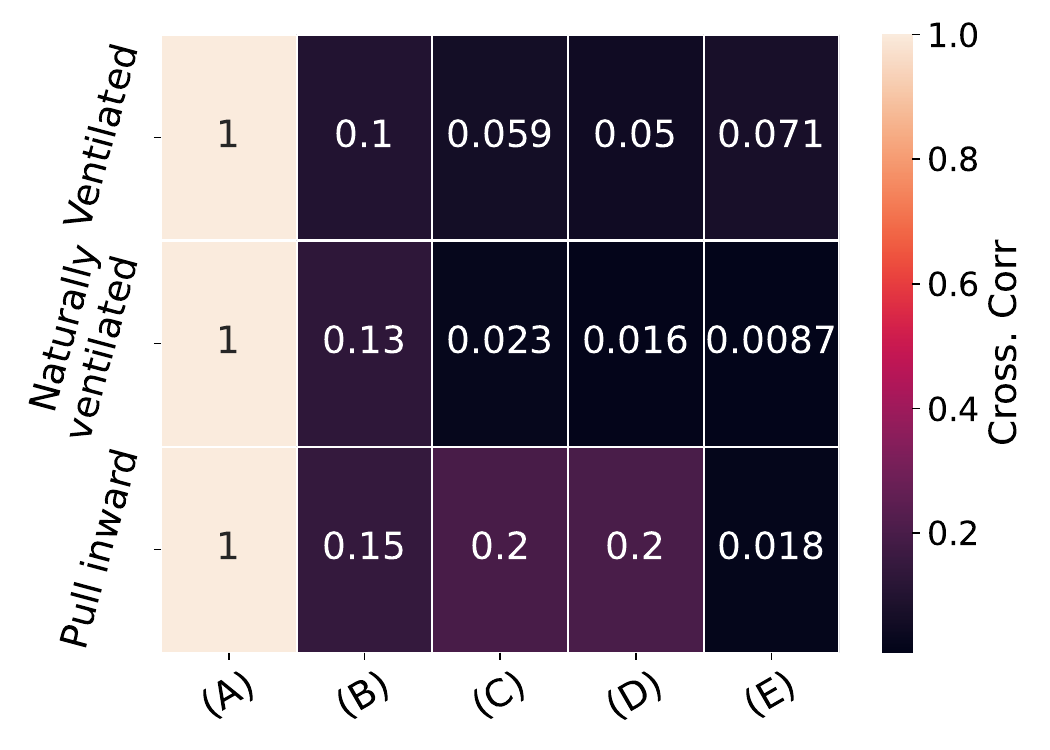}
		}
		\subfloat[Time lag\label{fig:lag_hhi}]{
			\includegraphics[width=0.49\columnwidth,keepaspectratio]{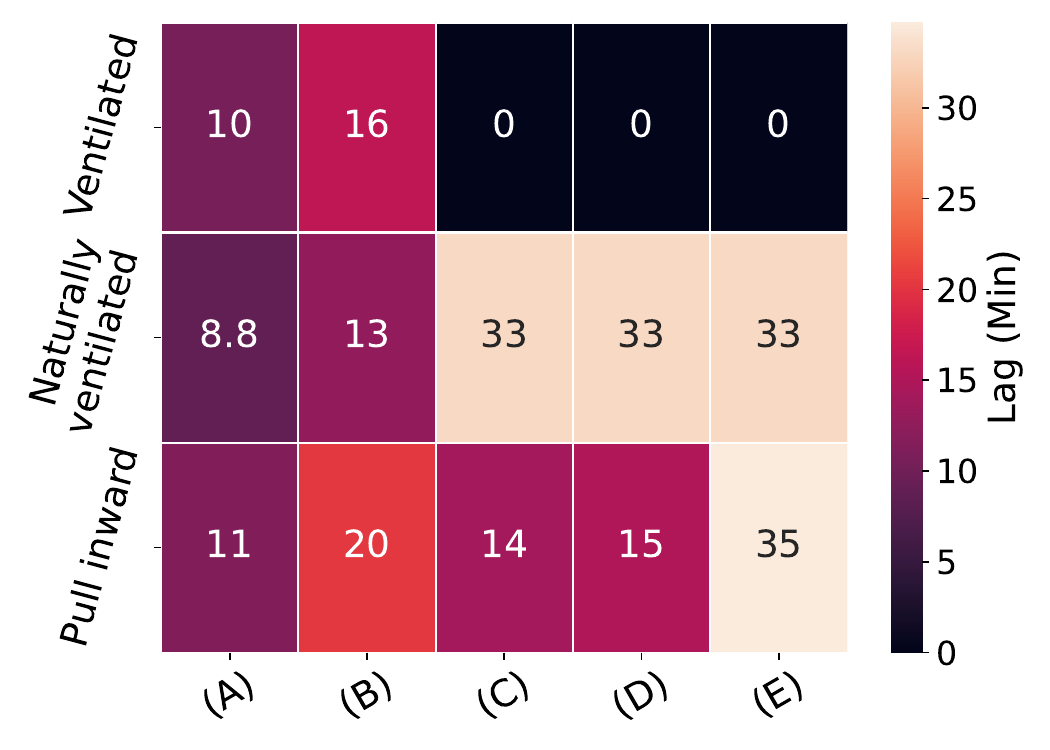}
		}\
	\end{center}
	\caption{Variation of HHI with sensing modules in H1 (A--E are different locations as shown in \figurename~\ref{fig:hm_house_marked}).}
	\label{fig:hhi_case}
\end{figure}

\subsubsection{Pollution Exposure}
HHI is computed over a period based on the average pollution exposure at each deployed sensing module. Therefore, unlike IAQI and Breeze IAQI, it is not limited to local observations and captures an overall estimate of exposure from indoors. From \figurename~\ref{fig:spread}, we observe that naturally-ventilated and pull-inward scenarios have similar effects on the kitchen as the exhaust fan is off for both cases. However, they impact the nearby room differently. While IAQI and Breeze IAQI cannot capture such impact and underestimate the spread of pollutants, HHI shows a steeper decline in air quality as shown in \figurename~\ref{fig:hhi_worst}.

\subsubsection{Spatial Spread}
Revisiting \figurename~\ref{fig:spread}, we observe that in household H1, different degrees of ventilation either contain the pollutants near the source (kitchen, when ventilated), only pollute nearby rooms (side room of kitchen, when naturally-ventilated), or spread the pollutants (when pulled inward by the ceiling fan in dining). \figurename~\ref{fig:base_metrics} shows that IAQI and AQI metrics from breeze tech do not capture such spatio-temporal properties of pollutants, whereas HHI provides a better indication to estimate the spread and lingering of pollutants in different parts of the indoor, as shown in \figurename~\ref{fig:hhi_case}. We can see a gradual decrease in HHI in \figurename~\ref{fig:hhi_best},~\ref{fig:hhi_mod},~\ref{fig:hhi_worst} with poorer ventilation. HHI captures such events when the kitchen exhaust is off, and the ceiling fan in the dining room pulls pollutants, resulting in lingering and trapping pollutants in some parts of H1, showing a lower air quality as shown in \figurename~~\ref{fig:hhi_worst}, even if the primary source of pollution, the kitchen, is not emitting.

Moreover, \figurename~\ref{fig:cross_corr_hhi} shows the cross-correlation between HHI and the pollutant measures from the deployed sensors (averaged and normalized over different pollutants) in H1. \figurename~\ref{fig:lag_hhi} highlights the time lag for maximum correlation. We observe that HHI shows maximum correlation with the kitchen pollutants and then follows the pollution spread patterns. Based on the pollution spread according to ventilation, different rooms correlate with HHI at different time lag as shown in \figurename~\ref{fig:lag_hhi}. Note that for the best ventilation (exhaust on, fan off), other rooms except the kitchen and its adjacent rooms remain mostly unaffected (zero lag). 

% \begin{figure}
%     \centering
%     \includegraphics[width=0.8\columnwidth,keepaspectratio]{Figures/result/HHI/lg_stay_hhi.pdf}
%     \caption{Variation of HHI with long stay of pollutants}
%     \label{fig:linger}
% \end{figure}

\begin{figure}
	\captionsetup[subfigure]{}
	\begin{center}
                \subfloat[Long Stay\label{fig:linger}]{
				\includegraphics[width=0.33\columnwidth,keepaspectratio]{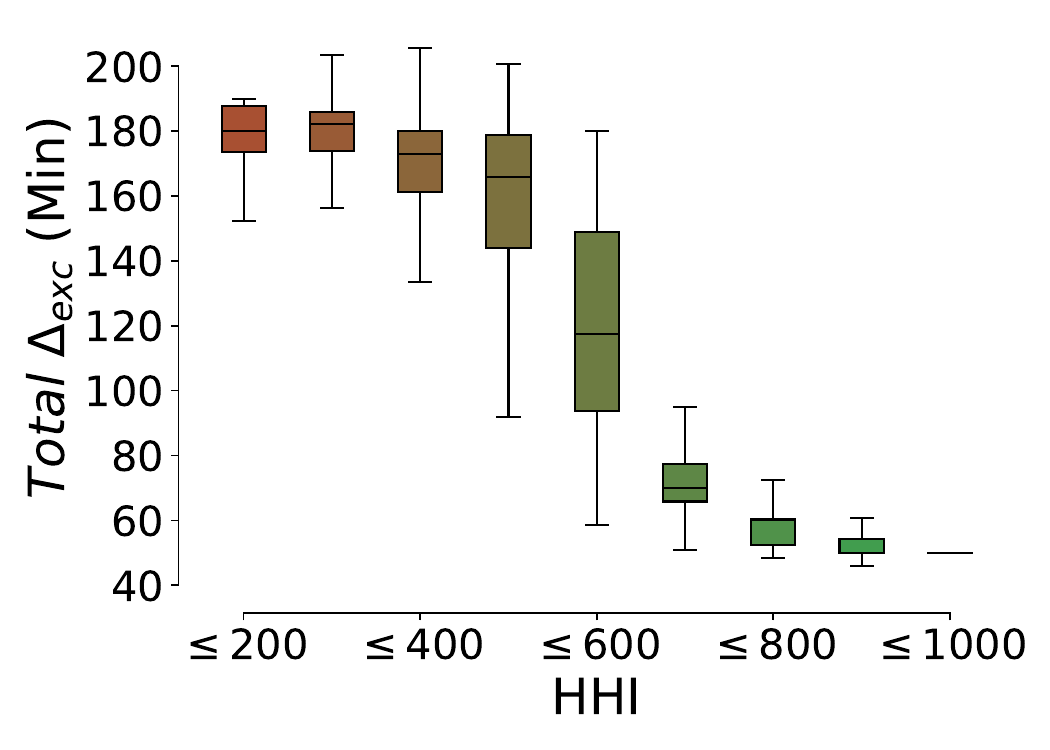}
			}
		      \subfloat[Peak Count\label{fig:hhi_pc}]{
				\includegraphics[width=0.33\columnwidth,keepaspectratio]{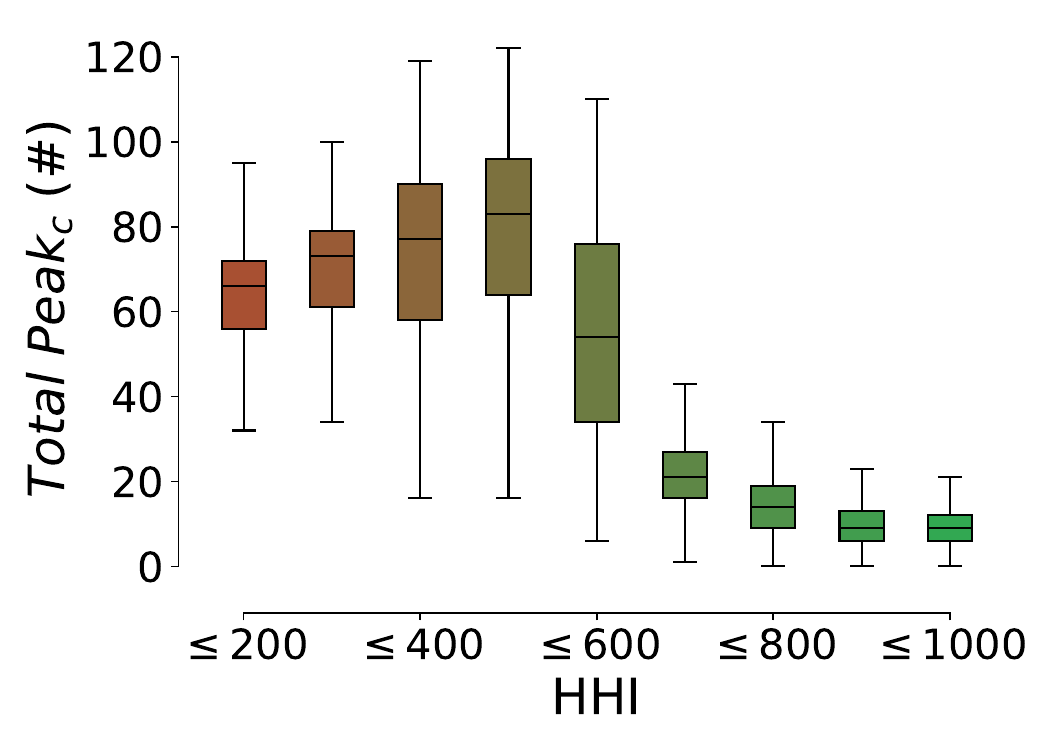}
			}
			\subfloat[Peak Duration\label{fig:hhi_pd}]{
				\includegraphics[width=0.33\columnwidth,keepaspectratio]{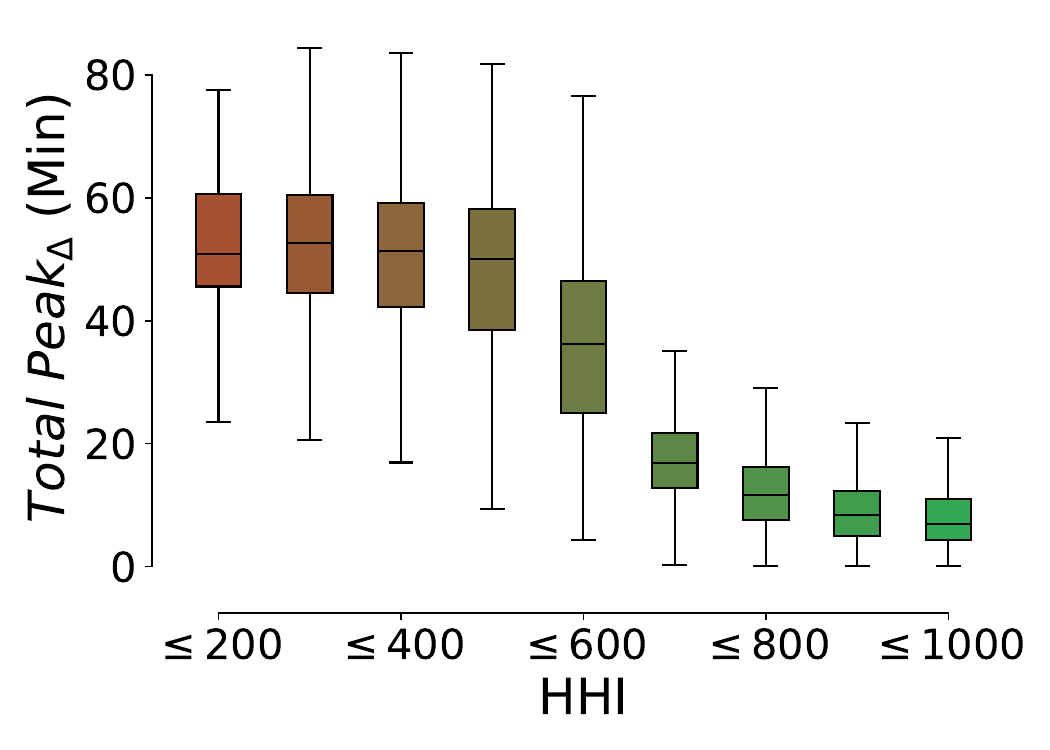}
			}
        \end{center}
	\caption{Variation of HHI with pollution properties.}
	\label{fig:linger_peaks}
\end{figure}

\subsubsection{Long stay of Pollutants}
Moderate-level pollution exposure for an extended period harshly impacts our health. Even though an instantaneous measure of pollutant concentration shows tolerable air quality, we formulate HHI to capture long-term moderate-level pollution exposure to represent true air quality. In \figurename~\ref{fig:linger}, we show that more than $100$ minutes of moderate pollution exposure across all the sensing modules results in $<600$ HHI values. As the pollutants linger long, the HHI value further decreases and indicates unhealthy indoor air quality. %and indicate the flaw in the indoor ventilation system by representing poor 

\subsubsection{Abrupt Increase in Pollutants}
Indoor spaces also experience a sudden increase in pollution due to activities such as cleaning, cooking, gathering, etc. Abrupt changes in pollutant concentration resemble failure of the ventilation system in case of a major pollution event and indicate the spread and lingering of pollutants unless ventilated readily. HHI signifies poor air quality with more peaks in measured pollutants, considering peak duration. From \figurename~\ref{fig:hhi_pc},~\ref{fig:hhi_pd}, we observe that HHI decreases as the duration or the number of such abrupt high pollution events increases. Such HHI response triggers necessary ventilation or warns the occupant about frequent short-term exposure to bad indoor air.

\begin{figure}[]
	\captionsetup[subfigure]{}
	\begin{center}

            \begin{minipage}{0.63\columnwidth}
                \subfloat[IAQI is subset of HHI\label{fig:iaqi_hhi_divs}]{
    			\includegraphics[width=\columnwidth,keepaspectratio]{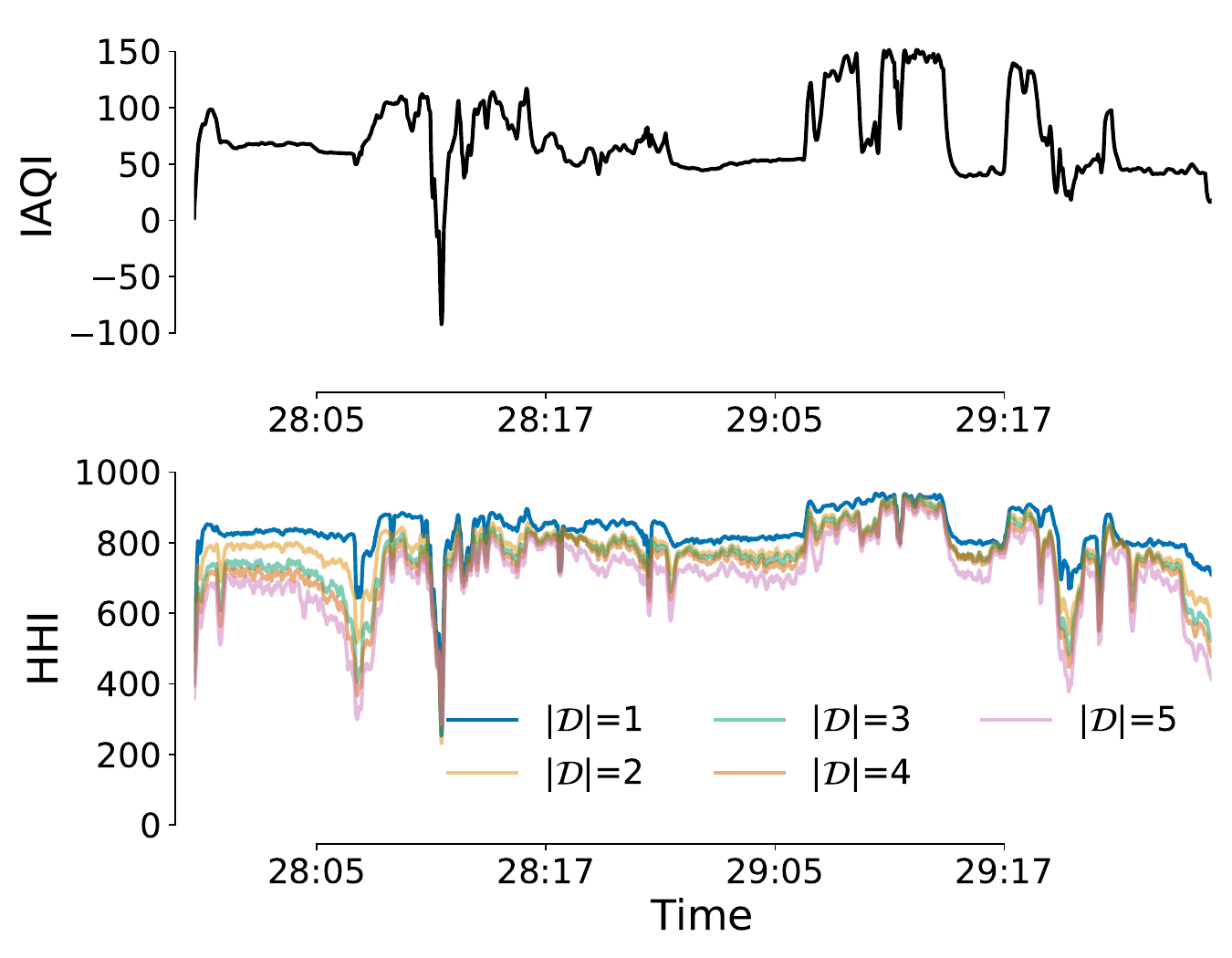}
    		}
            \end{minipage}\hfil
            \begin{minipage}{0.32\columnwidth}
                \subfloat[Cross-correlation\label{fig:cross_corr_iaqi_hhi}]{
    			\includegraphics[width=\columnwidth,keepaspectratio]{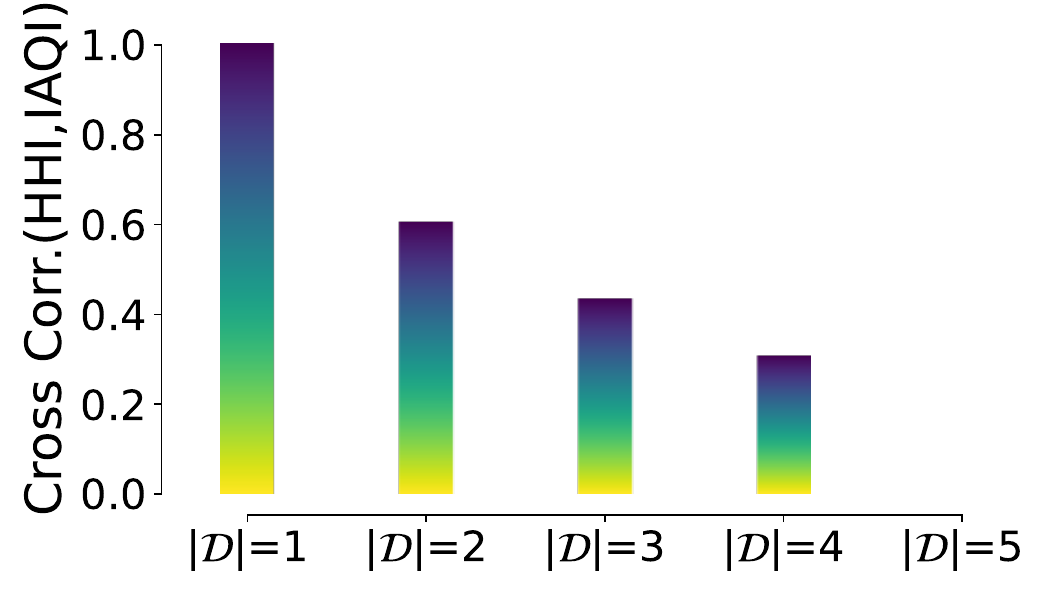}
    		}\
    		\subfloat[Distribution change\label{fig:hhi_divs}]{
    			\includegraphics[width=\columnwidth,keepaspectratio]{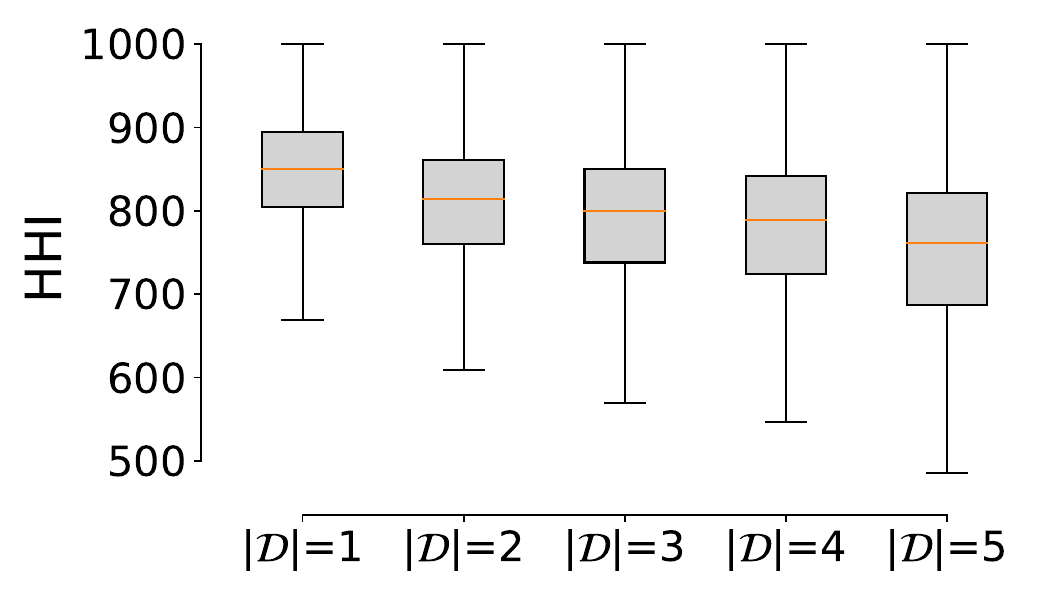}
    		}\
            \end{minipage}

	\end{center}
	\caption{HHI with a varying number of modules.}
	\label{fig:more_divs}
  %\vspace{-8mm}
\end{figure}

\subsubsection{Varying number of Sensing Modules}
Here, we analyze the change in computed HHI values by varying the number of deployed sensing modules in the indoor environment. IAQI is primarily developed to capture local exposure, undermining the spatial spread of the pollutants; with one sensing module, HHI behaves close to IAQI as shown in \figurename~\ref{fig:iaqi_hhi_divs}. We observe that multiple sensing modules bring in necessary pollutant information from indoors such that HHI emerges as a superior metric in capturing spatial spread and average pollution exposure. \figurename~\ref{fig:cross_corr_iaqi_hhi}, shows relative cross-correlation between HHI and IAQI with varying numbers of sensing modules to $|\mathcal{D}|$=5 sensing modules. Moreover, \figurename~\ref{fig:hhi_divs} shows the distributional change in HHI values with an increasing number of modules. Based on the figure, multiple modules provide better estimates of poorer air quality by collecting exposure data from various locations. % and combining them to estimate the spatial spread of the pollutants.
%Despite that advantage, HHI works as IAQI equivalent metric and shows maximal relative cross-correlation as per \figurename~\ref{fig:cross_corr_iaqi_hhi}. However,

\begin{figure}
	\captionsetup[subfigure]{}
	\begin{center}
		      \subfloat[CO\textsubscript{2} concentration\label{fig:co2_safe}]{
				\includegraphics[width=0.33\columnwidth,keepaspectratio]{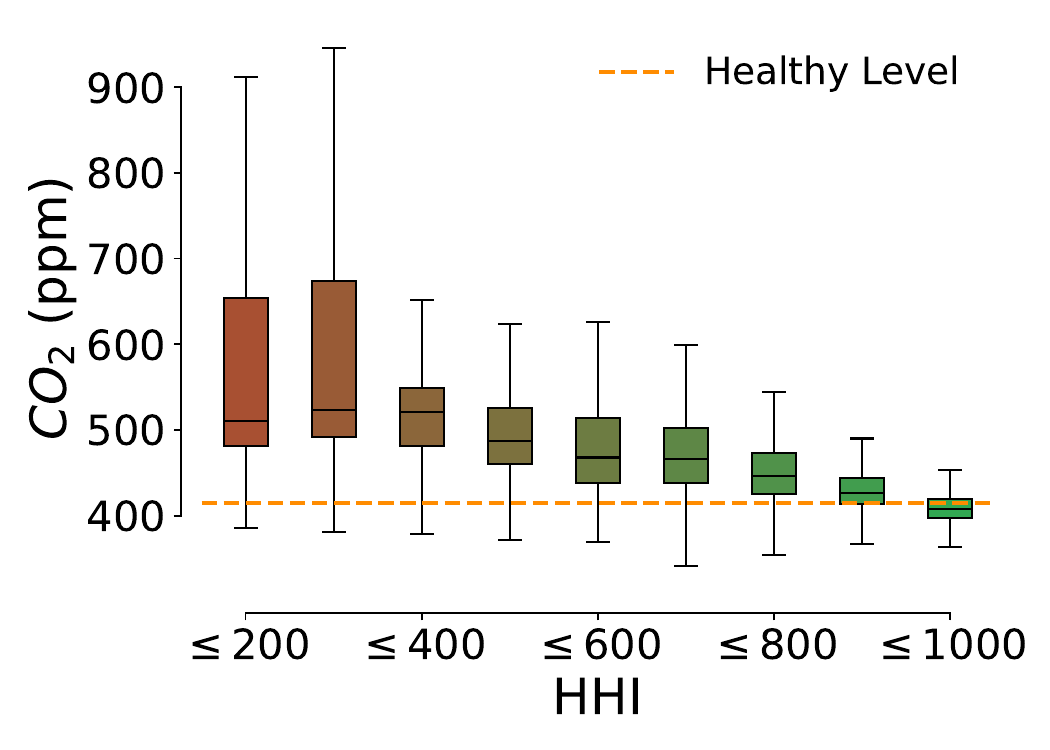}
			}
			\subfloat[VOC concentration\label{fig:voc_safe}]{
				\includegraphics[width=0.33\columnwidth,keepaspectratio]{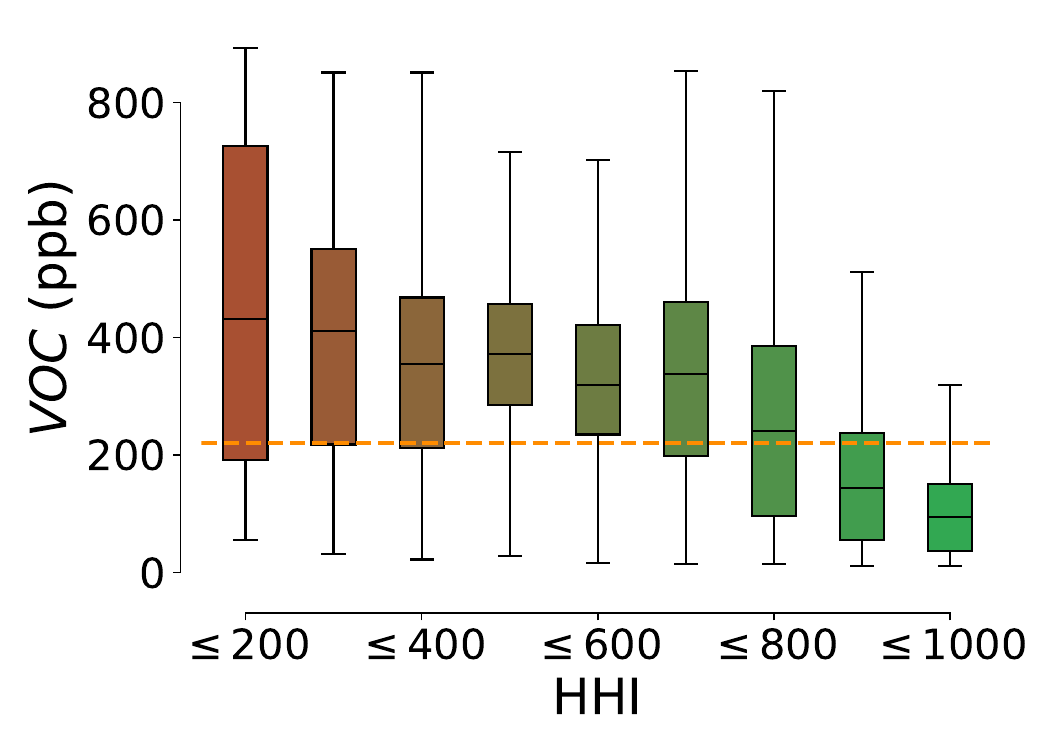}
			}
                \subfloat[PM\textsubscript{2.5} concentration\label{fig:pm2_5_safe}]{
				\includegraphics[width=0.33\columnwidth,keepaspectratio]{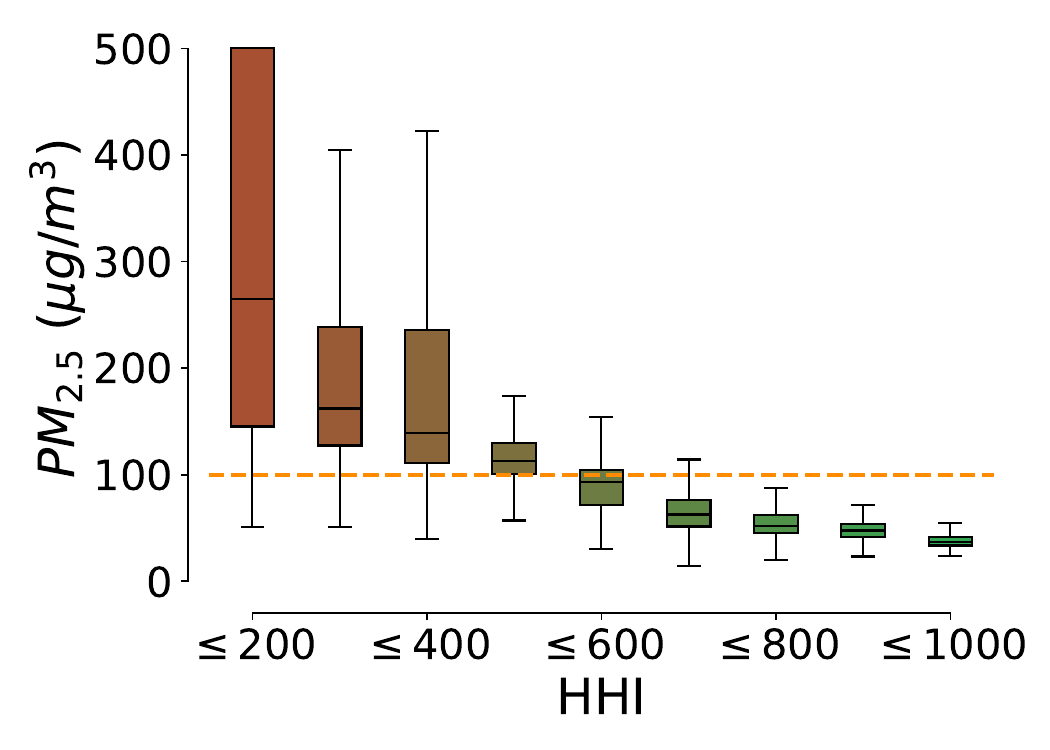}
			}\
			\subfloat[PM\textsubscript{10} concentration\label{fig:pm10_safe}]{
				\includegraphics[width=0.33\columnwidth,keepaspectratio]{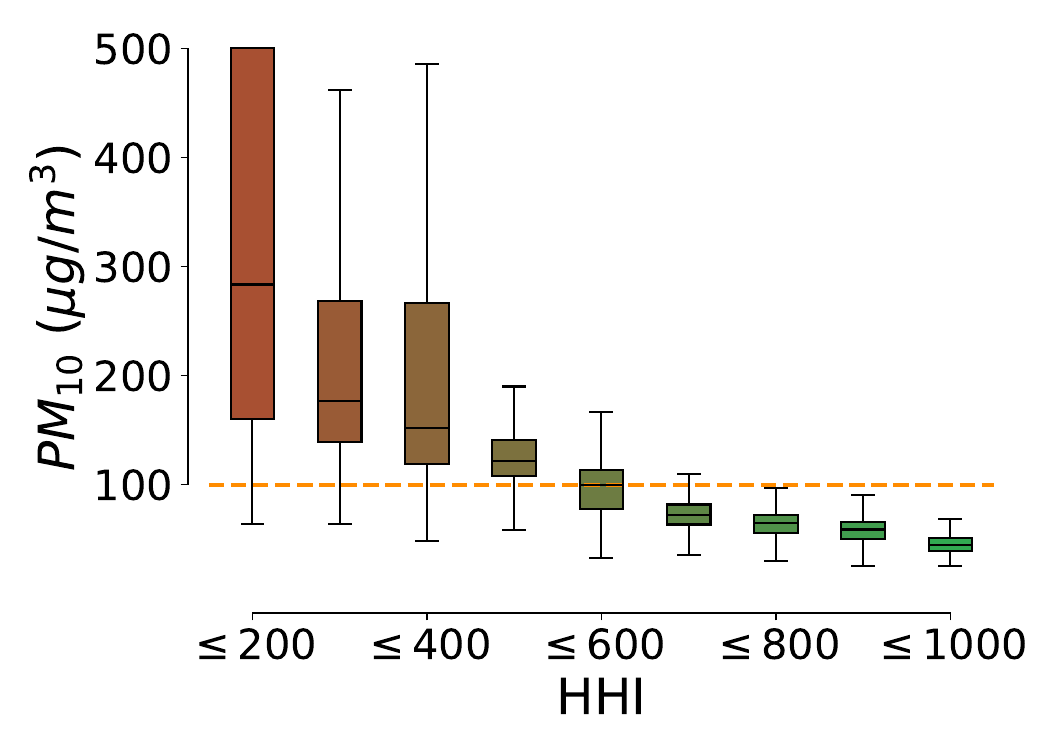}
			}
                \subfloat[Temperature\label{fig:t_safe}]{
				\includegraphics[width=0.33\columnwidth,keepaspectratio]{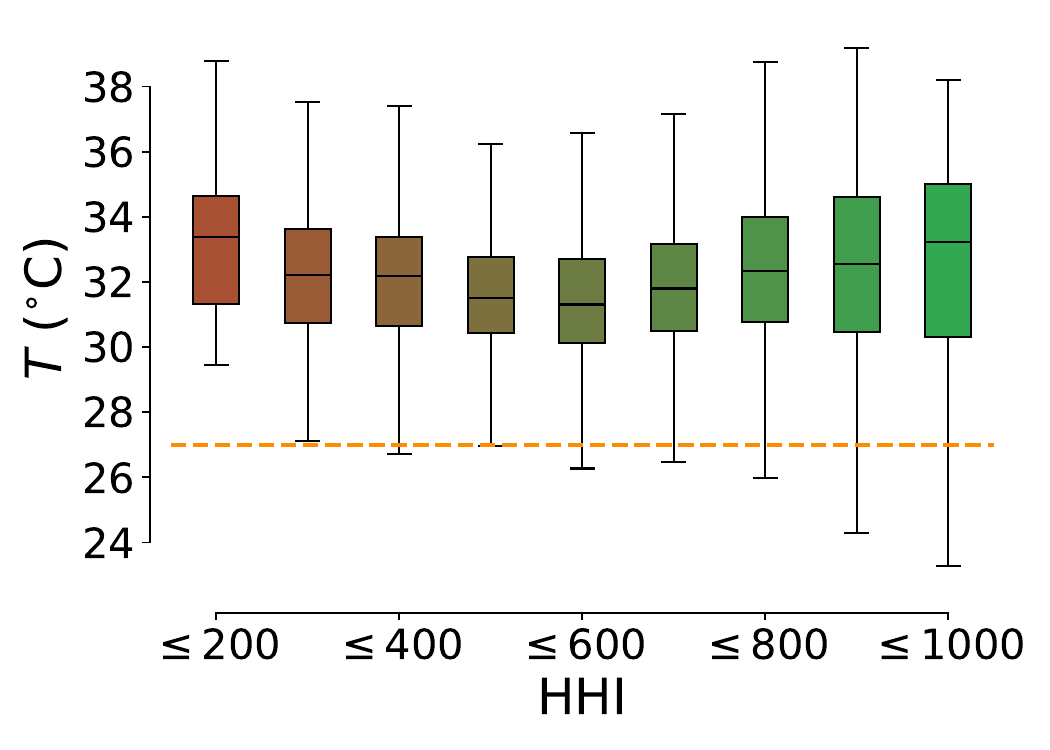}
			}
			\subfloat[Humidity\label{fig:h_safe}]{
				\includegraphics[width=0.33\columnwidth,keepaspectratio]{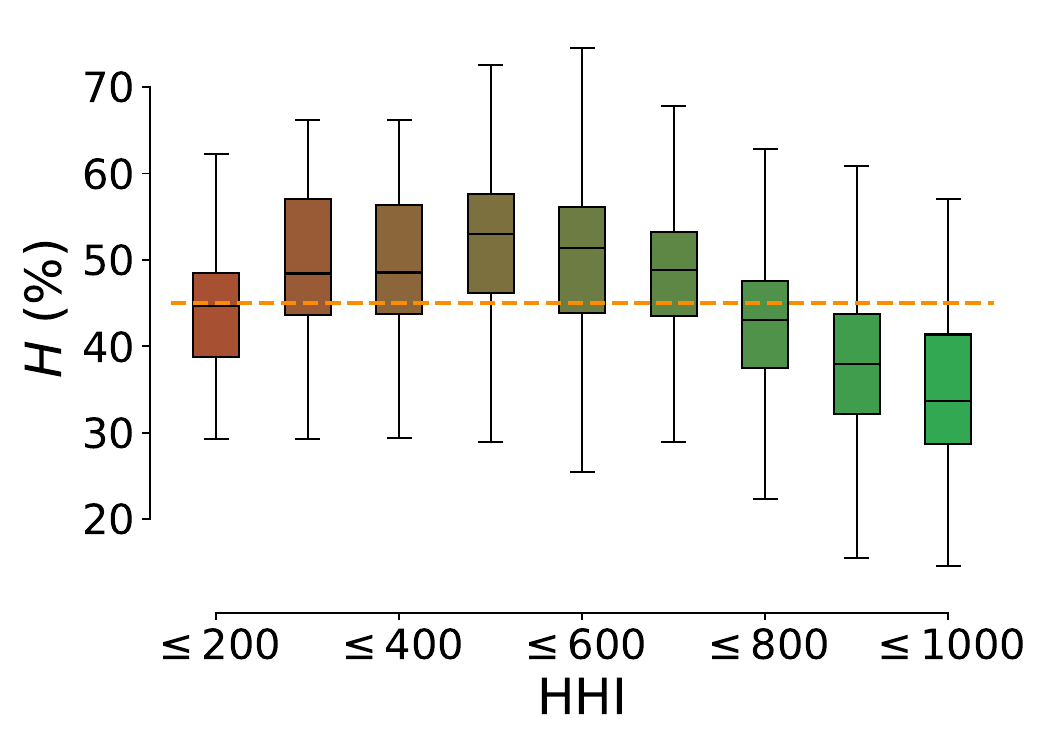}
			}
	\end{center}
	\caption{HHI with varying indoor air quality.}
	\label{fig:safety}
  %\vspace{-5mm}
\end{figure}

\subsubsection{Safety Analysis and Actionable HHI}
Here we analyze the impact of different pollutants over the computed HHI value as shown in \figurename~\ref{fig:safety}. Moreover, depending on the permissible concentrations of each pollutant, we come up with intuitions on permissible levels or safety thresholds. Below are the key observations: 

%\begin{itemize}
\noindent$\bullet\;\;$\textbf{CO\textsubscript{2}}: As shown in \figurename~\ref{fig:co2_safe}, the healthy level (orange dashed line in the figure) of CO\textsubscript{2} is 415 ppm. HHI $\leq$ 700 helps to trigger an alert to the occupant after a significant decrease in air quality. HHI $\leq$ 400 needs ventilation.

\noindent$\bullet\;\;$\textbf{VOC}: Similarly, healthy levels of VOC is 220 ppb. beyond which HHI is below 700 and can help alert occupants. Moreover, HHI $\leq$ 400 is unhealthy and requires countermeasure as shown in \figurename~\ref{fig:voc_safe}.

\noindent$\bullet\;\;$\textbf{Particulate matter}: Both PM\textsubscript{2.5} and PM\textsubscript{10}, remain within the healthy level till 700 HHI as shown in \figurename~\ref{fig:pm2_5_safe},~\ref{fig:pm10_safe}. But, HHI $\leq$ 400 needs the immediate attention of the occupant.

\noindent$\bullet\;\;$\textbf{Temperature and Humidity}: HHI value above 700 means moderate temperature with lower humidity. In comparison, a decrease in HHI resembles an increase in moisture and temperature, shown in \figurename~\ref{fig:t_safe},~\ref{fig:h_safe}.

Thus, HHI is grouped into three broad categories based on the impact of the indoor air quality on the occupant's health: (i) \textbf{Healthy} (1000 $\geq$ HHI $\geq$ 700), (ii) \textbf{Alert} (700 $>$ HHI $\geq$ 400), and (iii) \textbf{Actionable} (400 $>$ HHI $\geq$ 0).

\begin{figure}
	\captionsetup[subfigure]{}
	\begin{center}
                \begin{minipage}{0.60\columnwidth}
                    \subfloat[Varying $\lambda_1$ and $\lambda_2$ in HHI \label{fig:lambda_vary}]{
    				\includegraphics[width=\columnwidth,keepaspectratio]{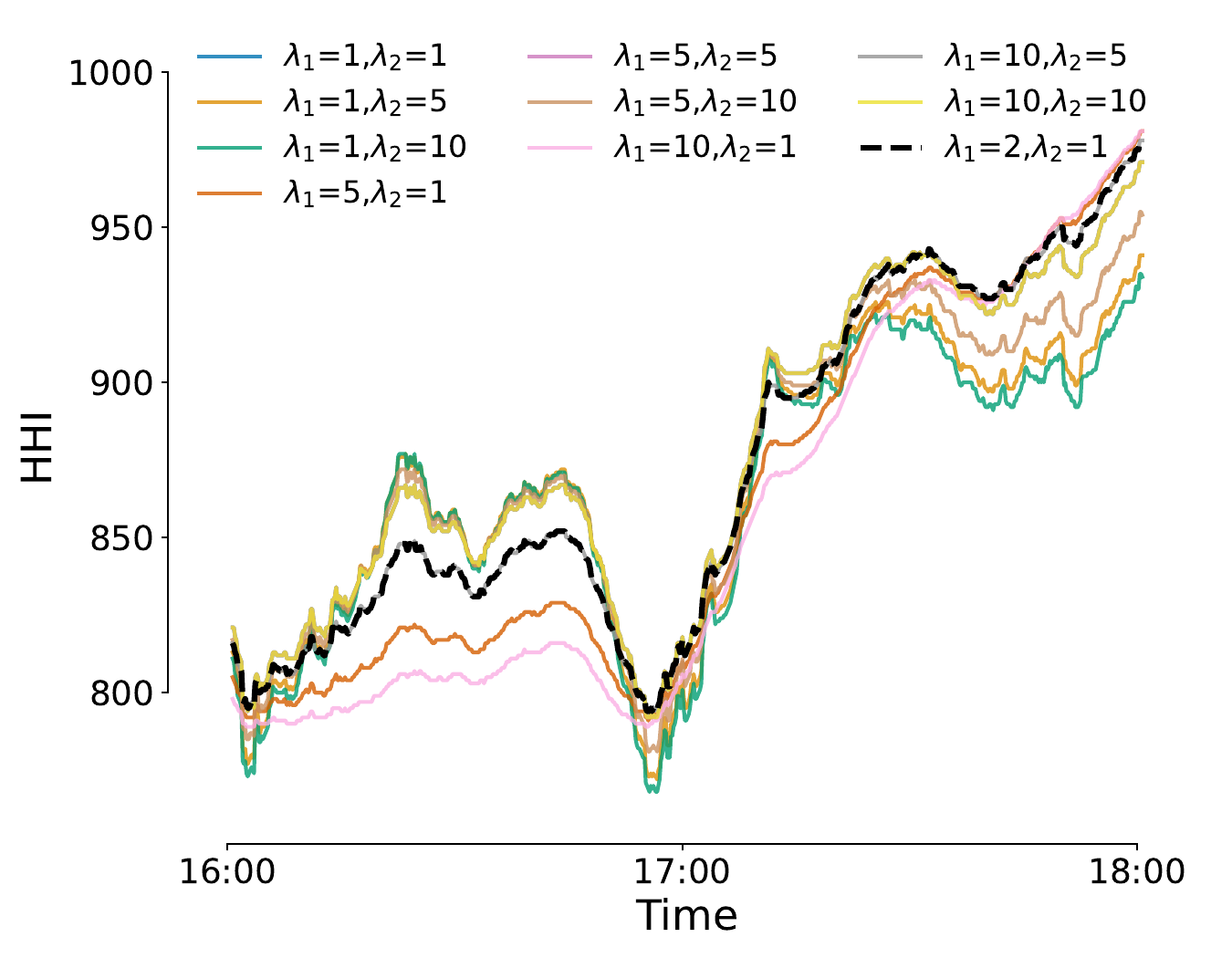}
    			}
                \end{minipage}\hfil
                \begin{minipage}{0.30\columnwidth}
                    \subfloat[CO\textsubscript{2} concentration \label{fig:co2_lambda}]{
    		          \includegraphics[width=\columnwidth,keepaspectratio]{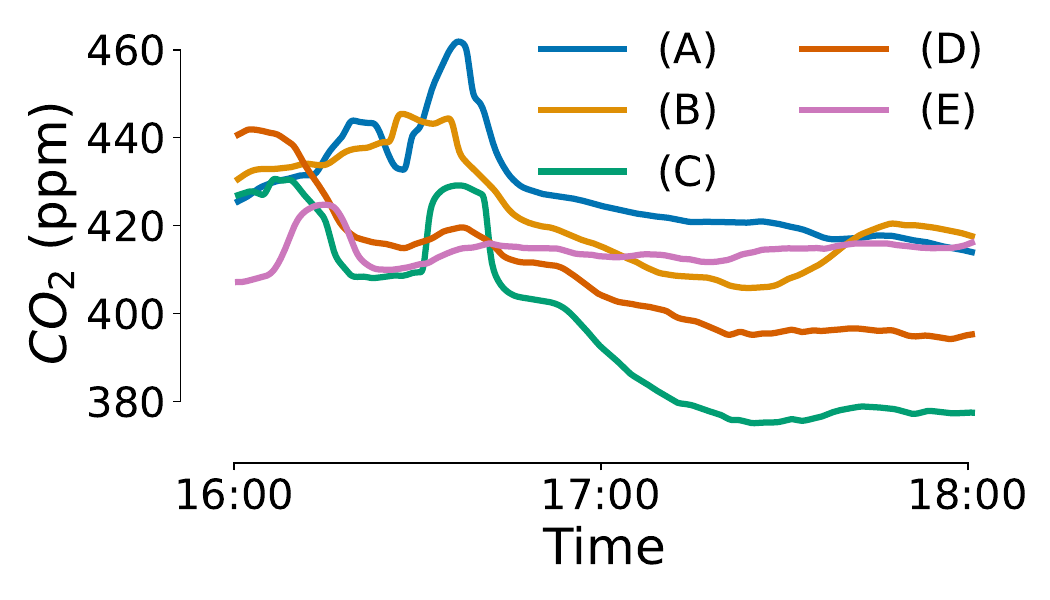}
    			}\
    			\subfloat[VOC concentration \label{fig:voc_lambda}]{
    				\includegraphics[width=\columnwidth,keepaspectratio]{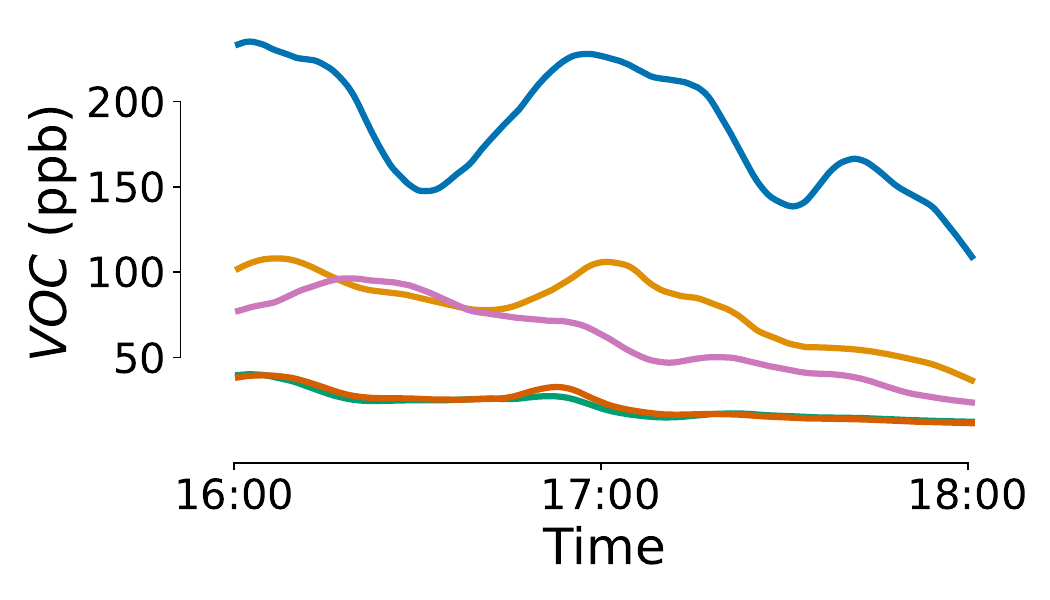}
    			}\
       %              \subfloat[\label{fig:pm2_5_lambda}]{
    			% 	\includegraphics[width=\columnwidth,keepaspectratio]{Figures/result/ablation/PMS2_5_lambda_vary.pdf}
    			% }\
                \end{minipage}
	\end{center}
	\caption{Impact of $\lambda_1$ and $\lambda_2$ on HHI in H1.}
	\label{fig:vary_l1_l2}
 %\vspace{-3mm}
\end{figure}

% \begin{figure}
% 	\captionsetup[subfigure]{}
% 	\begin{center}
% 		      \subfloat[\label{fig:lambda_vary}]{
% 				\includegraphics[width=0.9\columnwidth]{Figures/result/ablation/lambda_vary.pdf}
% 			}\
% 		      \subfloat[\label{fig:co2_lambda}]{
% 				\includegraphics[width=0.3\columnwidth,keepaspectratio]{Figures/result/ablation/CO2_lambda_vary.pdf}
% 			}
% 			\subfloat[\label{fig:voc_lambda}]{
% 				\includegraphics[width=0.3\columnwidth,keepaspectratio]{Figures/result/ablation/VoC_lambda_vary.pdf}
% 			}
%                 \subfloat[\label{fig:pm2_5_lambda}]{
% 				\includegraphics[width=0.3\columnwidth,keepaspectratio]{Figures/result/ablation/PMS2_5_lambda_vary.pdf}
% 			}\
% 			\subfloat[\label{fig:pm10_lambda}]{
% 				\includegraphics[width=0.3\columnwidth,keepaspectratio]{Figures/result/ablation/PMS10_lambda_vary.pdf}
% 			}
%                 \subfloat[\label{fig:t_lambda}]{
% 				\includegraphics[width=0.3\columnwidth,keepaspectratio]{Figures/result/ablation/T_lambda_vary.pdf}
% 			}
% 			\subfloat[\label{fig:h_lambda}]{
% 				\includegraphics[width=0.3\columnwidth,keepaspectratio]{Figures/result/ablation/H_lambda_vary.pdf}
% 			}
% 	\end{center}
% 	\caption{HHI values in H1 with (a) varying $\lambda_1$ and $\lambda_2$ under changing indoor conditions (b) CO\textsubscript{2}, (c) VOC, (d) PM\textsubscript{2.5}, (e) PM\textsubscript{10}, (f) Temperature, (g) Humidity}
% 	\label{fig:vary_l1_l2}
% \end{figure}

\subsection{Ablation Study}
To understand the impact of $\lambda_1$ and $\lambda_2$ on the HHI computation, we vary them from 1 to 10 as shown in \figurename~\ref{fig:vary_l1_l2} and observe against the recorded pollutants. As HHI has two components that resemble average pollution exposure and the spread of pollutants, we observe degraded HHI in both scenarios. As shown in the figures, with an increase in CO\textsubscript{2} around 16:15, HHI is impacted. Similarly, HHI depletes around 17:00 as VOC increases. Subsequently, HHI falls after 17:45 with the increase in variance among pollutants. Hence, with extreme $\lambda_1$ and $\lambda_2$ values, HHI either prefer average pollution or their variation as observed from \figurename~\ref{fig:lambda_vary}. We took $\lambda_1$=2 and $\lambda_2$=1 (black dashed line) marked in \figurename~\ref{fig:lambda_vary} as it results in a balanced HHI curve. % that is unbiased towards exposure or variance of pollutants in the indoor space.

\section{Related Works}
\label{sec:relatedwork}
The literature on indoor air quality can be categorized into the distinct groups described below.\\
\noindent$\bullet\;\;$\textbf{Analysis of Indoor Pollution Dynamics:}
Researchers have conducted several field studies to understand pollution dynamics under different circumstances. \cite{mujan2021development} has deployed their in-house pollution sensing module to two of their office buildings and one educational building with $69$ and $56$ participants, respectively. They observed correlations between indoor environmental quality (IEQ) and occupant comfort levels. Several other works~\cite{temprano2020indoor,daisey2003indoor,gao2022understanding} have analyzed pollution dynamics of environments like a classroom. In the work~\cite{gao2020n}, they have established a relationship between indoor air quality and students' concentration levels during lectures. A few works in the literature~\cite{hussain2023indoor,kumar2016real,brunelli2014povomon} have explored the challenges of deploying real-time sensors in urban buildings. Further, Mendoza \textit{et al.}~\cite{mendoza2021long} have analyzed the impact of outdoor pollutants on indoor air quality. Few researchers have conducted field studies to estimate cookstoves' impact on indoor air~\cite{chartier2017comparative} during cooking and accordingly modeled the spatial pollution distribution for households~\cite{putri2022spatial}. However, these works are limited in scale and the diversity of indoor spaces, whereas we performed our study for three months over $28$ locations of different indoor types.

\noindent$\bullet\;\;$\textbf{Continuous Monitoring of Indoor Air Quality:}
Several works have developed real-time pollution overlays as well as data management systems over the wireless sensing infrastructure in educational~\cite{brazauskas2021real,brazauskas2021data} and office buildings~\cite{mujan2021development,ji2019indoor} or rural areas~\cite{li2022field}. In~\cite{zhu2022dynamic,yang2021distributed,ding2019octopus}, the authors have considered adjusting the ventilation rate in a controlled setup. These works primarily focus on specific environments, therefore, are constrained by their experimental spaces. In contrast, our HHI-based approach is a holistic way to quantify air quality in unconstrained indoor spaces. %(i.e., residential houses, kitchens, restaurants, etc.).

\noindent$\bullet\;\;$\textbf{Impact of Activities on Indoor Air Quality:}
Studies have shown that occupants' activities and major events, such as lunch breaks, meetings, etc., influence indoor air quality significantly. Fang \textit{et al.}~\cite{fang2016airsense} proposed a machine learning-based approach to detect occupant activities like cooking, smoking, and spraying in small apartments based on sensing the air pollutants. In~\cite{verma2021racer}, the authors have inferred limited indoor activities such as cooking, window opening or closing, corridor walking, etc., from air quality sensing. The primary shortcoming in the above works is a small set of considered activities. To address this, we use our intuitive Android app (VocalAnnot) to record all possible indoor activities with the help of our participants. %, enabling them to add voice annotations in real-time.
%, in which participants can easily annotate indoor activities in real-time with voice commands.\\ %\cite{verma2021sensert} have developed an IoT framework to stream real-time data from the sensors deployed in the two department buildings in Cambridge University. In~\cite{gao2022understanding}, the authors have further taken recurring surveys from $29$ participants, and in a subsequent work~\cite{gao2020n}, they have established a relationship between indoor air quality and concentration levels of students during lectures. In~\cite{gao2022understanding}, the authors have further taken recurring surveys from $29$ participants, and in a subsequent 
%\cite{kumar2016real,brunelli2014povomon}, have explored the challenges in deploying real-time sensors in urban buildings.  %(.F\cite{kumar2016real,brunelli2014povomon}, have explored the challenges in deploying real-time sensors in urban buildings.

\noindent$\bullet\;\;$\textbf{Competing Pollution Monitoring Platforms:} Recent studies on indoor air quality have proposed several competing systems. In~\cite{maag2018w,jiang2011maqs,zhong2020hilo}, the authors have introduced mobile sensing modules that can be attached to the occupant's body to estimate personalized exposure. Works like~\cite{kim2013inair,cheng2014aircloud,qin2023system,zhong2021complexity,snow2019performance,yang2021distributed,sun2022c} have deployed sensors across buildings to visualize, analyze, and forecast pollution. A few studies~\cite{moore2018managing,zhong2020hilo,kim2020awareness,ingelrest2010sensorscope} considered user interactions and comfort level annotation to associate with pollution dynamics. However, such studies are limited to the specific homogeneous environment with a small set of user annotations and do not consider effects like linger, spread, pull inward, etc., on the pollutants depending on the complex indoor dynamics.
\section{Conclusion}
In this paper, we design, develop, and deploy the \ourmethod{} platform to conduct a three-month-long study in $28$ deployment sites with various indoor types like labs, classrooms, households, studio apartments, food canteen, etc., to understand indoor pollution dynamics. Through this large-scale study, we uncover fascinating trends in indoor pollution, including its spread, lingering effect, and the impact of activities, ventilation, and room layout on air quality. These observations inspire the development of the \textit{Healthy Home Index} (HHI), an adaptive metric that holistically captures true indoor air quality. Extensive evaluations across various indoor setups demonstrate the scalability of HHI, outperforming baseline metrics in accurately reflecting adequate pollution exposure. Overall, our research significantly contributes to the understanding and assessment of indoor air quality, offering the \ourmethod{} platform and the HHI as valuable tools for monitoring and evaluating indoor environments, empowering individuals, researchers, and policymakers in creating healthier indoor living spaces of the future. %Categorized into Healthy, Alert, and Actionable thresholds, HHI prioritizes the well-being of occupants.
% \input{Sections/acknowledge}

%References
\bibliographystyle{ACM-Reference-Format}
\bibliography{reference}

\end{document}